\documentstyle[12pt,epsf]{article}
\input{amssym.def}
\input{amssym.tex}
\input{epsf}
\catcode`\@=11
\newcount\hour
\newcount\minute
\newtoks\amorpm
\hour=\time\divide\hour by60
\minute=\time{\multiply\hour by60 \global\advance\minute by-\hour}
\edef\standardtime{{\ifnum\hour<12 \global\amorpm={am}%
        \else\global\amorpm={pm}\advance\hour by-12 \fi
        \ifnum\hour=0 \hour=12 \fi
        \number\hour:\ifnum\minute<10 0\fi\number\minute\the\amorpm}}
\edef\militarytime{\number\hour:\ifnum\minute<10 0\fi\number\minute}
\def\draftlabel#1{{\@bsphack\if@filesw {\let\thepage\relax
   \xdef\@gtempa{\write\@auxout{\string
      \newlabel{#1}{{\@currentlabel}{\thepage}}}}}\@gtempa
   \if@nobreak \ifvmode\nobreak\fi\fi\fi\@esphack}
        \gdef\@eqnlabel{#1}}
\def\@eqnlabel{}
\def\@vacuum{}
\def\draftmarginnote#1{\marginpar{\raggedright\scriptsize\tt#1}}
\def\draft{\oddsidemargin -.2truein
        \def\@oddfoot{\sl preliminary draft \hfil
        \rm\thepage\hfil\sl\today\quad\militarytime}
        \let\@evenfoot\@oddfoot \overfullrule 3pt
        \let\label=\draftlabel
        \let\marginnote=\draftmarginnote
   \def\@eqnnum{(\theequation)\rlap{\kern\marginparsep\tt\@eqnlabel}%
\global\let\@eqnlabel\@vacuum}  }
\relax
\def\sqr#1#2{{\vcenter{\vbox{\hrule height.#2pt
        \hbox{\vrule width.#2pt height#1pt \kern#1pt
           \vrule width.#2pt}
        \hrule height.#2pt}}}}
\def\square{\mathchoice\sqr64\sqr64\sqr{2.1}3\sqr{1.5}3}
\def\lsim{{\displaystyle
{{\raise-8pt\hbox{$ <$}}
\atop{\raise5pt\hbox{$\sim$}}}}}
\def\gsim{{\displaystyle
{{\raise-8pt\hbox{$ >$}}
\atop{\raise5pt\hbox{$\sim$}}}}}
\def\slsim{{\displaystyle
{{\raise-8pt\hbox{$\scriptstyle <$}}
\atop{\raise5pt\hbox{$\scriptstyle \sim$}}}}}
\def\sgsim{{\displaystyle
{{\raise-8pt\hbox{$\scriptstyle  >$}}
\atop{\raise5pt\hbox{$\scriptstyle \sim$}}}}}
\newcommand{\sump}[0]{\sum_{(h,g)}\!{\raise 4pt \hbox{$'$}}\,}
\def\a{\alpha}
\def\b{\beta}
\def\g{\gamma}
\def\d{\delta}
\def\e{\epsilon}
\def\m{\mu}
\def\n{\nu}
\def\t{\tau}
\def\p{\pi}

\def\r{\rho}
\def\th{\vartheta}

\def\s{\sigma}
\def\l{\lambda}

\def\et{\eta}

\def\Ga{\Gamma}

\def\bw{\bar{w}}
\def\bv{\bar{v}}
\def\bq{\bar{q}}

\def\br{\bar{r}}
\def\by{\bar{y}}
\def\bs{\bar{s}}

\def\tf{\tilde{f}}

\def\K{{\cal K}}

\def\cV{{\cal V}}

\def\cA{{\cal A}}

\def\cO{{\cal O}}

\def\hE{\hat{E}}
\def\hG{\hat{G}}
\def\hB{\hat{B}}

\def\Z{Z\!\!\! Z}
\def\one{{\mathchoice {\rm 1\mskip-4mu l} {\rm 1\mskip-4mu}
{\rm 1\mskip-4.5mu l} {\rm 1\mskip-5mu l}}}

\def\pa{\partial}

\def\ve{\vert}

\def\ra{\rightarrow}

\def\ti{\times}

\def\ba{\bar{a}}

\def\ch#1#2{\left( #1 \atop #2 \right)}

\def\un{\underline}
\def\ni{\noindent}
\def\nl{\newline}
\def\limit#1#2{\smash { \mathop{#1} \limits_{#2} }  }

\def\thefootnote{\fnsymbol{footnote}}
\def\be{\begin{equation}}
\def\ee{\end{equation}}
\def\bs{\begin{subequations}}
\def\es{\end{subequations}}
\def\ben{\begin{enumerate}}
\def\een{\end{enumerate}}
\def\ba{\begin{eqnarray}}
\def\ea{\end{eqnarray}}

\def\vs{\vskip}

\def\rd{{\rm d}}

\def\ni{\noindent}
\def\nl{\newline}
\def\ed{\end{document}}

\def\bibtem#1{\bibitem{#1} }

\def\sp{\quad, \quad}

\def\pe{\, .}
\def\co{\, ,}

\def\Ga{\Gamma}
\def\p{\pi}
\def\t{\tau}
\def\rd{{\rm d}}
\def\n{\nu}
\def\et{\eta}

\def\ed{\end{document}}
\newtoks\@stequation
\def\subequations{\refstepcounter{equation}%
  \edef\@savedequation{\the\c@equation}%
  \@stequation=\expandafter{\theequation}
  \edef\@savedtheequation{\the\@stequation}
  \edef\oldtheequation{\theequation}%
  \setcounter{equation}{0}%
  \def\theequation{\oldtheequation\alph{equation}}}

\def\endsubequations{\setcounter{equation}{\@savedequation}%
  \@stequation=\expandafter{\@savedtheequation}%
  \edef\theequation{\the\@stequation}\global\@ignoretrue
  \vspace*{-12pt} \\}
\def\thefootnote{\fnsymbol{footnote}}
\def\bea{\begin{eqnarray}}
\def\eea{\end{eqnarray}}
\def\be{\begin{equation}}
\def\ee{\end{equation}}
\def\bs{\begin{subequations}}
\def\es{\end{subequations}}
\newskip\humongous \humongous=0pt plus 1000pt minus 1000pt
\def\caja{\mathsurround=0pt}
\def\eqalign#1{\,\vcenter{\openup1\jot \caja
        \ialign{\strut \hfil$\displaystyle{##}$&$
        \displaystyle{{}##}$\hfil\crcr#1\crcr}}\,}
\newif\ifdtup

\def\un{\underline}
\def\limit#1#2{\smash { \mathop{#1} \limits_{#2} }  }

\def\thebibliography#1{%
\vskip 0.5cm \centerline{\bf References}
\list{%
[\arabic{enumi}]}{\settowidth\labelwidth{[#1]}
\leftmargin\labelwidth
\advance\leftmargin\labelsep
\usecounter{enumi}}
\def\newblock{\hskip .11em plus .33em minus .07em}
\sloppy\clubpenalty4000\widowpenalty4000
\sfcode`\.=1000\relax}

\renewcommand{\theequation}{\arabic{section}.\arabic{equation}}

\begin{document}

\begin{titlepage}
\begin{flushright}
CERN-TH/97-233\\
hep-th/9709058 \\
\end{flushright}
\begin{centering}
\vspace{.3in}
\boldmath
{\large\bf Heterotic/Type-I Duality in $D<10$ Dimensions, \\ 
Threshold Corrections
and D-Instantons} \\
\unboldmath
\vspace{1.1 cm}
E. KIRITSIS and
N.A. OBERS
\vskip 1cm
{\it Theory Division, CERN$^{\ \dagger}$}\\
{\it CH-1211, Geneva 23, Switzerland}\\
\medskip
\vspace{1.1cm}
{\bf Abstract}\\
\end{centering}
\vspace{.1in}

We continue our study of heterotic/type-I
 duality in $D<10$ dimensions. We consider the heterotic and type-I
theories compactified on tori to lower dimensions. We calculate
the  special (``BPS-saturated'') ${\cal F}^4$ and
  ${\cal R}^4$ terms in the
effective one-loop heterotic action.
These terms  are expected to be non-perturbatively exact for $D>4$.

The heterotic result is compared with the associated type-I result.
In $D<9$ dimensions, the type-I theory has instanton corrections due to
D1 instantons. In $D=8$ we use heterotic/type-I duality to give a
simple
prescription
of the D-instanton calculation on the type-I side.
We allow arbitrary Wilson lines and show that the D1-instanton
determinant
is the affine character-valued elliptic genus evaluated at the induced
complex structure of the D1-brane world-volume.
The instanton result has an expansion in terms of Hecke operators that
suggests an interpretation in terms of an $SO(N)$ matrix model of the
D1-brane.
The total result can be written in terms of generalized
prepotentials, 
revealing an underlying holomorphic structure.

In $D<8$ we calculate again the heterotic perturbative thresholds and
show that they agree with the D1-instanton calculation using the rules
derived in $ D=8$.

\vspace{.9cm}
\begin{flushleft}
CERN-TH/97-233  \\
September 1997 \\
\end{flushleft}
\vskip 2cm
\hrule width 6.7cm \vskip.1mm{\small \small \small
$^\dagger$ e-mail addresses: kiritsis,obers@mail.cern.ch. }
\end{titlepage}
\newpage
\setcounter{footnote}{0}
\renewcommand{\thefootnote}{\arabic{footnote}}

\setcounter{section}{0}

\section{Introduction and Results}

 D-brane solitons and instantons are a key element of all
non-perturbative duality conjectures.
While  solitons have been studied vigorously, the attention
paid to  instantons has been  lesser and more recent:
it includes work on the point-like D-instanton of type IIB
\cite{Po}--\cite{rt}, on the resolution of the type-IIA 
conifold
singularity by  Euclidean 2-branes \cite{BBS}--\cite{O}, and on
non-perturbative effects associated with  Euclidean 5-branes
\cite{W}--\cite{6}. Here we will look at a  simpler case, that  of
Euclidean D-strings present in type-I $SO(32)$ string
theory:
these are physically less interesting, since they  are mapped
by strong/weak-coupling dualities
  to standard world-sheet instanton effects on the
type-IIB, respectively heterotic side. Our motivation is however
different: we would like
to gain a better understanding of the rules of semi-classical
D-instanton calculations, which  could prove useful in more interesting
contexts. We will  at the same time  elucidate some
subtleties of the above duality maps, when applied below the
critical dimension.

There have been  many qualitative checks of various non-perturbative
dualities, but so far quantitative checks are scarce.
In order to do a tractable quantitative test of a non-perturbative
duality
we need to carefully choose the quantity to be computed.
Since usually a weak coupling computation has to be compared with a
strong coupling one, one has to choose a quantity whose strong coupling
computation
can also be done at weak coupling.
Such quantities are very special and generally turn out to be terms in
the
effective action that obtain loop contributions from BPS states only.
They are also special from the supersymmetry point of view, since the
dependence of their couplings on moduli must satisfy certain
holomorphicity
or harmonicity  conditions.
Moreover, when supersymmetry commutes with the loop expansion, they get
perturbative corrections from a single order in perturbation theory.
Such terms also have special properties concerning instanton
corrections
to their effective couplings.
In particular, they obtain corrections only from instantons that
leave some part
of the original supersymmetry unbroken.
Sometimes, such terms are directly linked to anomalies.

For ground states with $N=2$ supersymmetry\footnote{We count the
supersymmetries using four-dimensional language (in units of four
supercharges).}, the two-derivative
terms in the effective action have the properties mentioned above. All
the information
about the two-derivative effective action is contained in a
prepotential which is holomorphic in the vector-moduli, and another one
which contains the hypermultiplet moduli.
Moreover, there is a tower of higher-derivative terms \cite{fg} that
also have
such special properties, and their action can be written as an F-term.
The simplest such bosonic term is the $R^2$ term.

In the case of $N=4$ supersymmetry, the two-derivative effective action
does not receive any corrections, either perturbative or
non-perturbative.
The higher-derivative terms that have the special properties mentioned
above
are, among others, the four-derivative $F^4$ and $R^2$ terms, the
six-derivative $F^2 R^2$ terms and the eight-derivative $R^4$ terms
\cite{BaKi}\footnote{The analysis of \cite{BV} strongly indicates
that there is also an
infinite tower of such terms, as in the $N=2$ case, which are special.}. 
In this paper we will focus on such terms in vacua with $N=4$ 
supersymmetry.

In \cite{bk2} the relevant heterotic as well as some  type-I    
one-loop thresholds were calculated.
In $D=9$ no instanton corrections are expected and the two sides could
be
matched in perturbation theory.
The thresholds of the irreducible terms, $tr R^4$, $tr F^4$ obtain only
one-loop contributions on both sides.
Via the duality map, the heterotic result for the factorizable terms
$(tr F^2)^2$, $(trR^2)^2$, $trF^2trR^2$ were shown to contain terms
that come from
higher genus ($\chi=-1,-2$) on the type-I side.
These are contact (boundary) terms
on the type-I side and their appearance was motivated.
Their presence is associated with the (mild) non-holomorphicity of the
elliptic genus on the heterotic side, while they are related to the
different
structure of supersymmetry  on the type-I side.
World-sheet contact terms are responsible for this non-holomorphicity
on the heterotic side.
It was  shown that the  one-loop (non-contact) terms matched on both
sides.
This worked  because the winding sum in the heterotic
side can be traded for unfolding the torus fundamental domain to a
strip, which is
the relevant annulus fundamental domain on the type-I side.
It is crucial for this that no windings appear in the type-I theory.
This is essentially the old trick used in finite temperature
string theory, which maps a case with windings and a torus fundamental
domain
to a case without windings and an annulus domain.

The $D=8$ case was further considered, where D1-brane instanton
corrections are expected on
the type-I
side.
The Wilson lines were set to zero and the heterotic thresholds were
calculated as functions of the two-torus
moduli $T$, $U$.
Using the heterotic/type-I duality map, the heterotic result
was separated into perturbative and non-perturbative type-I  parts.
The perturbative part depends only on $U$ and has a structure similar 
to that in $D=9$.
The non-contact terms were again shown to agree with a
one-loop calculation on the
type-I side.
The non-perturbative part was given an elegant interpretation in terms
of
D1-brane instantons.
The relevant configurations turn out to be a single Euclidean D1-brane
wrapped (holomorphically) in all possible ways around the two-torus.
Wrapped configurations related by large diffeomorphisms
of the D1-brane world-sheet should be considered equivalent and not
be summed over.
Multiple D1-branes at a non-zero distance do not contribute, because of  
zero modes.
However, configurations that factorize into several independently
wrapped (overlapping) D1-branes should also be included.
This is necessary for restoring
the $SL(2,Z)_T$ $T$-duality symmetry.
The necessity of including independent wrapped D1-branes can be
interpreted
(in the Minkowski case)
as the presence of bound states at threshold.

By directly evaluating the classical D1-brane world-sheet action
(which is known independently) the exponential terms
$e^{2\pi iT}$ of the heterotic result were reproduced.
Most interestingly, the fluctuation determinant
turned out to be, not unexpectedly, the heterotic elliptic genus
evaluated at the complex structure modulus of the wrapped D1-brane.

In this paper we continue and generalize the analysis of \cite{bk2}.
In $D=8$ we turn on all possible moduli, the $T,U$ torus moduli as well
as the 16 complex Wilson lines, $y^i$.
We again evaluate the heterotic perturbative thresholds for the
gravitational terms $tr R^4$ and $(trR^2)^2$.
The piece that is non-perturbative on the type-I side is shown to be
given again by D1-instantons.
The fluctuation determinant is again holomorphic and is given by the
affine character-valued heterotic elliptic genus.
We show that the full threshold correction can be written in
terms
of generalized holomorphic  prepotentials indicating a hitherto unknown
holomorphic structure of these higher-derivative terms in the context
of $D=8$, $N=1$ supergravity.
The existence of such prepotentials is shown to be intimately related
to the presence of the two-torus.
Differential identities satisfied by the torus lattice sum translate
into existence conditions of prepotentials.

The instanton results can be expressed in terms of Hecke operators.
As pointed out in \cite{talk}, it is in this form that they should be 
derivable from a D1-matrix model.

We further compactify both theories to $D<8$. The heterotic threshold
is
perturbative for $D>4$. We evaluate it and subsequently show that it
translates into perturbative type-I contributions as well as
D1-instanton
corrections, where now the world-volume of the D1-brane (with $T^2$
topology) is mapped supersymmetrically in all possible ways into
$T^{10-D}$.
The one-loop determinant around the instanton is again given by the
heterotic elliptic genus evaluated at the induced complex structure on
the world-volume of the Euclidean D1-brane.

The structure of the paper is the following.
In Section 2 we present some general remarks on perturbative and
non-perturbative corrections for the special terms in the effective
action
in the presence of $N=4$ spacetime supersymmetry.
In Section 3 we discuss the form of one-loop thresholds for the
relevant $R^4$
and $F^4$ terms and their relation to the elliptic genus.
In Section 4 we present the calculation of the $D=8$ heterotic
thresholds, while these are further discussed in Section 5, along with
supersymmetric recursion relations and generalized prepotentials.
The  corresponding D1-brane instanton interpretation on the type-I side
is
given in Section 6. The case with non-zero Wilson lines and its
D1-brane
interpretation is given in Section 7. Section 8 discusses
toroidal compactifications of the heterotic string to lower dimensions
and
the corresponding D1-brane interpretation.
Finally,  Section 9  contains further remarks and
directions.
In Appendix \ref{modf} we present useful facts about modular forms and
various
modular covariant derivatives. In Appendix \ref{htd} we give the
duality map
of heterotic/type-I duality in less than ten dimensions.
In Appendix \ref{elg4} we outline the calculation of one-loop threshold
corrections for general heterotic $N=4$ ground states.
In Appendix \ref{lpr} we list various useful properties of the (2,2)
lattice.
In Appendix \ref{1li}  we evaluate the integrals relevant for the
heterotic
threshold calculation in $D=8$.
In Appendix \ref{lte}  we derive the large volume expansion of the
heterotic
thresholds.
In Appendix \ref{rrp} we discuss recursion relations satisfied by
heterotic
thresholds and how these translate into the existence of generalized
prepotentials. Finally, in Appendix \ref{htt} we calculate the one-loop
heterotic thresholds for
toroidal compactifications to $D <8$.

\section{The Setup and Some General Remarks}
\setcounter{equation}{0}

The effective action for $F^4$, $R^4$, and $R^2F^2$ terms in an $N=4$
theory
can receive
corrections
that are either perturbative or non-perturbative.
Of course, the distinction between perturbative and non-perturbative
corrections depends on a given string theory one starts with.
Perturbative corrections in one description can contain
non-perturbative
contributions when translated in a dual description in terms of a
different string theory.
When, however,  such terms obtain one-loop contributions in a given
description, then these contributions are proportional to a supertrace of the
helicity to the fourth power\footnote{In $N=2$ ground states the
supertrace
of the helicity squared is obtained instead.}\cite{BaKi}.
Since the helicity supertraces are essentially indices to which only
short  BPS multiplets contribute \cite{BaKi,lec}, the one-loop
contribution
to such terms is due to BPS states only.
The appropriate helicity supertraces count essentially the numbers of
``unpaired" BPS multiplets. It is only these that are protected from
renormalization
and can provide reliable information in strong coupling regions.
In fact, calling the helicity supertraces indices is more
than an analogy.
In our context, unpaired BPS states in lower dimensions are intimately
connected with the chiral asymmetry (conventional index) of the
ten-dimensional
theory.
It is well known that the ten-dimensional elliptic genus is the stringy
generalization of the Dirac index \cite{Windey,ellwit}.
Projecting the elliptic genus on
physical states
in ten dimensions gives precisely the massless states, responsible
for anomalies.
In lower dimensions, BPS states are determined uniquely by the elliptic
genus,
as well as the compact manifold data (in our case the toroidal lattice
sum).
Moreover, the amplitudes that only have BPS contributions
are governed by the ten-dimensional elliptic genus and
its covariant derivatives as will be shown later on in this paper.
It would be interesting to generalize in a model-independent way
the relationship of standard indices and helicity supertraces giving
rise to the elliptic genus.

For several four- or six-dimensional ground states with $N=2,4$ 
supersymmetry, there is a trio of dual descriptions corresponding to a
type-II, heterotic and type-I (open) description.
In the type-II description the special terms described above seem to
obtain
perturbative contributions from a single order in perturbation theory.
This order is proportional to the number of fields appearing in such a
term if it belongs to the gravitational sector.
Moreover, these different loop-order contributions satisfy recursion
relations
\cite{fg}.
In the heterotic description such terms seem to obtain perturbative
contributions only at one loop.
Successful comparisons of such corrections have been made \cite{ffg}
between heterotic/type-II $N=2$ dual pairs.

The case of the type-I  duals is more special. One of the reasons is
that supersymmetry in type-I theory does not ``commute" with the genus
expansion.
This can easily be seen by observing that, for example, the
Green--Schwarz anomaly
term
$B\wedge F^4$ appears at one loop while the CP-even term $F^4$ appears
at the disk level.
However, the two are related by supersymmetry \cite{roo}.
Since supersymmetry is essential in duality, we would expect subtleties
in comparing the type-I with the heterotic string past the tree level.
Already in \cite{ABFPT} a comparison was made between $N=2$ heterotic and
type-I vacua in four dimensions, using the techniques and results of \cite{BF}.
It was shown that the duality map has to
be modified since on the type-I side there are one-loop corrections to
the
Einstein term that modify the passage to the Einstein frame where
dual theories are compared. Moreover, similar comparisons in $N=2$ 
ground states have been
made for the higher F-terms \cite{FI}.
In \cite{bk2} it was shown that even for $N=4$ ground states such
subtleties
arise and have to be resolved.

On the heterotic side we consider compactifications of the
ten-dimensional
heterotic string on a torus down to $D<10$ non-compact dimensions.
In heterotic perturbation theory, the $R^2$ term appears only at tree
level
and does not get further perturbative corrections.
To argue about non-perturbative corrections, we will have to identify
the appropriate instantons that could contribute.
Since the $R^2$ term is of a special kind, only maximal
supersymmetric
instantons can contribute, and in the heterotic string this is the
Euclidean five-brane.
In a toroidal compactification, an instanton correction from the
five-brane can arise if its six-dimensional Euclidean
world-sheet can wrap
(supersymmetrically) around a compact six-torus.
We would thus conclude that there are no perturbative or  
non-perturbative
corrections to the $R^2$ term for $D>4$.
At $D=4$ we expect instanton corrections and these were calculated
using heterotic/type-II duality in \cite{HM,6} although a direct
five-brane calculation is still lacking.

The $R^4$, $R^2F^2$ and $F^4$ terms do get one-loop contributions.
So far, we have been vague concerning the tensor structure of such
terms.
Here, however, we will be more precise \cite{roo,Tseytlin,BaKi}.
There are three types of $R^4$ terms in ten dimensions: $t_8(trR^2)^2$,
$t_8trR^4$ and $(t_8t_8-\e_{10}\e_{10}/8)R^4$, where $t_8$ is the
standard eight-index tensor \cite{GSW}
and $\e_{10}$ is the ten-dimensional totally antisymmetric $\e$ symbol.
The precise expressions can be found for example in \cite{Tseytlin}.
There are also the $t_8 trR^2trF^2$, $t_8trF^4$ and $t_8(trF^2)^2$
terms.
These different structures can be completed in supersymmetric
invariants
\cite{roo,Tseytlin}.
The bosonic parts of these invariants are as follows:
\bs
\be
J_0=\left(t_8t_8-{1\over
8}\e_{10}\e_{10}\right)R^4\;\;\;,\;\;\;I_1=t_8trF^4-{1\over
4}\e_{10}BtrF^4\label{1}\ee
\be
I_2=t_8(trF^2)^2-{1\over 4}\e_{10}
B(trF^2)^2
\;\;\;,\;\;\;I_3=t_8trR^4-{1\over 4}\e_{10}BtrR^4
\label{2}\ee
\be
I_4=t_8(trR^2)^2-{1\over
4}\e_{10}B(trR^2)^2\;\;\;,\;\;\;I_5=t_8(trR^2)(trF^2)-{1\over
4}\e_{10}B(trR^2)(trF^2) \pe
\label{3}\ee
\label{inv} \es
As is obvious from the above formulae, apart from the $J_0$
combination, the
other
four-derivative terms are related to the Green--Schwarz anomaly by
supersymmetry.
Thus, in ten dimensions, they are expected to receive corrections only
at one
loop if their perturbative calculation is set up properly (in an
Adler--Bardeen-like scheme).
The $J_0$ invariant is not protected by $N=4$ supersymmetry.
Heterotic/type-II duality in six dimensions implies that it receives
perturbative corrections beyond one loop.
It is however protected in the presence of $N=8$ supersymmetry \cite{kp}.

Here we would like to remind the reader of a few facts about heterotic
perturbation theory.
There are many subtleties in calculating higher-loop contributions
that arise from the presence of supermoduli.
There is no rigorous general setup so far, but several facts are known.
As discussed in \cite{pert} there are several prescriptions for
handling the
supermoduli. They differ by total derivatives on moduli space.
Such total derivatives can sometimes obtain contributions from the
boundaries of moduli space where the Riemann surface degenerates or
vertex operator
insertions collide.
Thus, different prescriptions differ by contact terms.
In \cite{cato} it was shown that such ambiguities eventually reduce to
tadpoles of massless fields at lower orders in perturbation theory.
The issue of supersymmetry is also the subject of such ambiguities.
It is claimed \cite{pert,cato} that in a class of prescriptions
$N\geq 1$ supersymmetry is respected genus by genus provided
there are no disturbing tadpoles at tree level and one loop.
The only exception to this is the case of an anomalous $U(1)$ in $N=1$ 
supersymmetric ground states. In this case there  is a
non-zero D-term at one loop, which naively breaks
supersymmetry. Restoration of supersymmetry implies the presence of
a two-loop contact term that was found by explicit calculation
\cite{D}.
To conclude, if all (multi) tadpoles vanish at one loop and we use the
appropriate prescription for higher loops, we expect supersymmetry to
be valid order by order in perturbation theory.
It is to be remembered, however, that the above statements apply
on-shell.
Sometimes there can be terms in the effective action that vanish
on-shell,
violate the standard lore above, but are required by non-perturbative
dualities. An example was given in \cite{6}.

We now turn again to the terms on which we focus in this paper, which
occur in the presence of $N=4$ supersymmetry.
The CP-odd terms in (\ref{inv}) were explicitly evaluated
at arbitrary order of perturbation theory in \cite{ya}.
There, by carefully computing the surface terms, it was shown
that such contributions vanish for $g>1$.
The CP-even terms are related to the CP-odd ones by supersymmetry
(except
for $J_0$).
If there are no subtleties with supersymmetry at higher loops, 
then these terms also satisfy the non-renormalization theorem.
This was in fact conjectured in \cite{ya}.
In view of our previous discussion on the structure of supersymmetry,
we would expect that once supersymmetry is working well at $g\leq 1$,
it continues to work for $g>1$ for a suitable definition of the 
higher-genus amplitudes.
In view of the above, we will assume that the CP-even terms do not get
contributions beyond one loop.
On the other hand, the $J_0$ term (which is non-zero at tree level) is
not protected by the anomaly.
Thus, it can appear at various orders in the perturbative expansion.
It can be verified by direct calculation that it does not appear at
one loop on the heterotic side.
However, heterotic/type-IIA duality in six dimensions seems to imply
that there is a two-loop contribution to this term on
the heterotic side.
In all of the subsequent discussion, when we refer to $R^4$ terms we
mean the
anomaly-related tensor structures, $I_3$, $I_4$, which can always be
distinguished from $J_0$.

If we now compactify on a torus, although it seems that there might be
no standard anomalies in the lower-dimensional theory, this is
misleading.
Consider for example a compactification on a circle to nine dimensions.
There are no anomalies in nine dimensions, as can be seen by a standard
analysis of massless diagrams. In field theory, that would be the end
of the story.
In string theory however things are a bit different.
Consider the original ten-dimensional gauge symmetry. From a
nine-dimensional
point of view, we still have massless gauge bosons, but also an
infinite tower
of massive gauge bosons (Kaluza--Klein modes and winding modes of the
original
gauge bosons).
If we consider how ten-dimensional gauge transformations act on the
nine-dimensional gauge bosons, we find that they are still the standard
gauge transformations for the massless nine-dimensional bosons, but
they act as transformations of a broken gauge symmetry on the massive
gauge bosons.
Thus, the correct interpretation is that we are in a spontaneously
broken phase
of (part of) the ten-dimensional gauge symmetry.
We know, on the other hand, that a spontaneously broken gauge
symmetry remembers very well potential anomalies visible in the
unbroken
phase.
However, such anomalies would not come from massless nine-dimensional
diagrams.
They would be visible when an infinite series of nine-dimensional
diagrams
are included.
The conclusion is that the anomaly-related terms in ten dimensions are
again
anomaly-related in a lower dimension upon toroidal compactification.
  The important  question is: Are they still expected to get only
one-loop contributions in the lower-dimensional theory?
This question cannot have a unique answer, unless we specify some
properties of the theory in question.
In fact, as shown in \cite{bk2}, the answer to this question is
different for the two dual
theories under consideration, the heterotic and the type-I string.

In the heterotic theory, the  answer is simpler.
Following our discussion, the anomaly CP-odd terms obtain
perturbative contributions only at
one loop, for any toroidal compactification of the heterotic string.
This can be calculated directly, since it requires minor modifications
of the calculation in \cite{ya}.
For the CP-even supersymmetry-related terms the answer is again
expected
to be the same and this is what we assume.
Thus, all perturbative corrections to the CP-even terms in $I_i$ are
expected to come
only from one loop for any $D\leq 10$.
As shown in \cite{bk2}, this is not the case in the type-I dual.
We have already observed that there, supersymmetry does not ``commute"
with the genus expansion. The net result of this upon compactification
is that
there will be ``contact" contributions from higher genera.
In particular, among the terms we are investigating in this paper,
there are the factorizable ones $(trR^2)^2$, $tr R^2tr F^2$,
$(tr F^2)^2$ for which  there are
extra contributions from surfaces with Euler number $\chi=-1,-2$.
The appearance of such extra contributions is controlled on the
heterotic side by world-sheet contact terms at one loop.
Although we do not know the detailed supersymmetry constraints for
the terms in question for $D<10$ we can guess, by analogy with the $N=2$ 
case,
certain recursion relations between different thresholds.
Such recursion relation imply, in the type-I context, the presence of
higher-genus contact terms \cite{bk2}.
This situation is highly reminiscent of the anomalous $U(1)$ case in
the
heterotic string.
This state of affairs also affects the type-I non-perturbative
contributions \cite{bk2}.

We will now consider potential non-perturbative contributions.
The type of instantons that could contribute is governed by
supersymmetry
and the fermionic structure of super-invariants, which can be inferred
from supergravity analysis.
Two derivative terms in the lowest-order effective action contain terms
with up
to four fermions.
The $R^2$ invariant must contain terms with up to eight fermions.
For the rest of the terms of interest, we have:
the super-invariants $I_i$, $i=1,2,\cdots , 5$, must contain terms with
up to eight fermions, 
while $J_0$ must contain terms with up to sixteen fermions.
We are considering a class of theories that are invariant under a
supersymmetry generated by sixteen supercharges.
In general, an instanton configuration will break part or all
of the supersymmetry.
If it breaks all of the supersymmetry, there will be at least sixteen  
fermionic
zero modes in the fluctuation spectrum around the instanton
configuration.
In general the number of zero modes is determined by some appropriate
index theorem.
However, the set will always contain at least a number equal to the
number of
supersymmetries broken by the instanton.
In multi-instanton solutions, there are in general more bosonic moduli
describing
relative positions and orientation.
If the multi-instanton leaves some supersymmetry unbroken, there will
be more fermionic zero modes, supersymmetric partners of the bosonic
moduli
related by the unbroken supersymmetry.
This is the reason why for the terms we will be considering
in this paper, instanton contributions will come from configurations
with a minimal number of instanton moduli.

The next question to be answered is: What part of the supersymmetry can
an instanton configuration  break?
 The answer to this  depends on the number of non-compact
dimensions.
For $D>4$ an instanton can break all or half of the supersymmetries.
In $D=4$ breaking of 1/4 of the supersymmetries is also
allowed.

Now, let us first consider multi-instanton configurations that break all
supersymmetries.
Then we have at least sixteen fermionic zero modes.
Such configurations can give non-zero contributions to terms in the
effective
action that contain terms with at least sixteen fermions.
{}From our last analysis, only $J_0$  is in that class.
Let us now consider instantons that break half of the spacetime
supersymmetries.
In that case we have at least eight zero modes and they can give
non-trivial
corrections to $R^2$, as well as the terms $I_i$.
If we restrict ourselves to $D>4$, we can ask the question whether
there are such
instantons
in the heterotic theory. The answer was already given in \cite{chs},
and the relevant
instanton configuration is the heterotic five-brane.
In order to interpret it as an instanton, on the other hand, we would
have
to wrap its six-dimensional world-volume around a compact
six-dimensional
manifold (so that the instanton action is finite).
This is obviously not possible for $D>4$.
The conclusion is that for $D>4$, in the heterotic theory, there are no
non-perturbative corrections to the terms $R^2$, $I_i$ and of course
to the two-derivative terms.
In $D\leq 4$ we do expect non-perturbative corrections due to the
five-brane. In \cite{ds} it was argued that the instanton corrections
to the $F^4$ terms are absent in the globally supersymmetric case when
$D=4$ but are non-vanishing when $D=3$.
This implies that in $D=4$, the full stringy instanton result is zero or
that
it vanishes in the limit that gravity is decoupled.
The five-brane instanton calculation of $F^4$ terms in $D=4$ remains to
be done.

In the type-I theory the situation is slightly different.
The configurations that break half of the supersymmetries are the
D1-brane
and the D5-brane.
As in the heterotic case, the D5-brane can only give instanton
corrections
when $D<5$.
The D1-brane has an effective description as a soliton of the type-I
effective
theory \cite{dab} and also as a standard D-brane \cite{PW}.
In both descriptions, the spectrum of its zero modes reproduces the
world-sheet
structure  of the heterotic string.
The D1-brane can produce instanton corrections when $D<9$.
In that case, it can wrap around a two-cycle of $T^{10-D}$ producing
at least eight fermionic zero modes.
Multi-D1-brane instantons, if they are some distance apart in target
space,
cannot contribute to the amplitudes in question since, according to our
previous discussion, they have more fermion zero modes and thus, do not
contribute. This is in agreement with heterotic/type-I duality
\cite{bk2}.
Thus, D1-branes will be responsible for non-trivial
instanton corrections to the higher-derivative terms, on the type-I
side.

According to the above discussion, we do not expect instanton
corrections
on the type-I side for $D=9$.
For $4<D<8$ there will be instanton corrections due to the D1-brane.
These were computed for $D=8$ in \cite{bk2} for vanishing Wilson lines.
In this paper we will concern ourselves with $D=8$ and arbitrary Wilson
lines
as well as with $4<D<8$.

One final comment concerns a comparison between the instantons we are
using
here and standard field-theory instantons.
In field theory, we are usually considering two types of instantons.
The first are instantons with finite action, and a typical example
is the BPST instanton \cite{bpst}, present in non-Abelian
four-dimensional gauge theories.
Examples of the other type are provided by the Euclidean Dirac monopole
in three dimensions, which is relevant, as shown in \cite{poly}, to  the
understanding of the non-perturbative behaviour of three-dimensional
gauge theories in the Coulomb phase.
This type of instanton has an ultra-violet (short-distance)-divergent
action, since it is a singular solution to the Euclidean
equations
of motion.
However, by cutting off this divergence and subsequent renormalization,
it can contribute to non-perturbative effects.
Another famous case in the same class is the two-dimensional vortex
of the XY model, responsible for the KT phase transition \cite{KT}.
In four dimensions we also have the BCD merons \cite{mer}, with similar
characteristics, although their role in the non-perturbative
four-dimensional dynamics is not very well understood.

Also in the context of string theory, we  have these two types of
instantons.
Here, how\-ever, the behaviour seems to be somewhat different.
Let us consider first the heterotic five-brane \cite{chs}.
This solution is intimately connected to BPST instantons in the
transverse
space and is smooth provided the instanton size is non-zero.
At zero size the solution has an exact CFT description but the string
coupling is strong. Non-perturbative effects are important and a
conjecture
has been put forth to explain their nature \cite{zerosize}.
Another type of instanton whose effective field-theory description is
regular
is the D3-brane of type-IIB theory.
On the other hand, the other D-brane instantons have an effective
description that is of the singular type. However, their ultra-violet divergence
is cured in their
stringy description.
This is already clear in the case of the type-I D1-brane relevant for
this paper, where the effective description is singular \cite{dab} while the 
stringy description
turns out to be regular and in particular, as we will see later, 
their classical action is finite.

There seems to be a correspondence of the various field-theory
instantons
to stringy ones. We have already mentioned the example of the heterotic
five-brane, but the list does not stop there.
In \cite{bk} it was shown that the three-dimensional Polyakov QED
instanton
as well as various non-Abelian merons have an exact CFT description
and thus correspond to exact classical solutions of string theory.
Moreover, the three-dimensional instanton can be interpreted as an avatar of 
the five-brane zero-size instanton when the theory is compactified to three
dimensions.
Similar remarks apply to the stringy merons, which require the presence
of five-branes with fractional charge \cite{bk}. In that respect
they are solutions of the
singular
type in the effective field theory.
In the context of the string theory, the spectrum of instanton
configurations
is of course richer, since the theory includes gravity.
However, the correspondence of field-theory and some string-theory
instantons
implies that the field-theory non-perturbative phenomena associated
with them are
already included in a suitable stringy description.

\section{One-Loop Heterotic Thresholds}
\setcounter{equation}{0}
\vskip 0.2cm

 In this section we review the calculation of BPS-saturated one-loop
effective couplings
in heterotic  string theory. These have the  form
\cite{Schellekens,Lerche}
\be
{\cal I}^{\rm het}_D  =
-{\cal N}(2\pi)^d 
  \int_{ { F}}{d^2\tau \over
\tau_2^2}\; (\tau_2)^{d/2}   \Gamma_{d,d}\
{\cal A}({\cal F},{\cal R}, \tau)
\co \label{5}\ee
where $d=10-D$ is the number of compact dimensions, ${\cal A}$ is an
(almost) holomorphic modular form of weight zero
related to the elliptic genus \cite{Windey,ellwit} and ${\cal F}$ and
${\cal R}$ stand for the gauge-field
strength and curvature two-forms respectively. $\Gamma_{d,d}$ is the
lattice sum over  momentum and winding modes
for  $d$ toroidally compactified dimensions, ${ F}$ is the usual
fundamental domain,
 and
\be
{\cal N} = {V^{(D)}\over 2^{10} \pi^6}
\ee
 is a
normalization that includes the volume of the uncompactified
dimensions \cite{BaKi}. For simplicity, we first discuss here
the case of vanishing  Wilson lines on the
$d$-hypertorus, reinstating the Wilson line dependence further below.
Then,  the sum over momenta ($p$) and windings ($w$) is given by
\be
\Gamma_{d,d}  =
  \sum_{p,w}
 e^{-{\pi\tau_2} ( p^2 + w^2/\pi^2)
+ i \tau_1 p\cdot w} \co 
\ee
and factorizes inside the integrand.  Our conventions are
\be
 \alpha' = 1 \ ,\ \  q= e^{2\pi i\tau}\ ,
\ \  d^2\tau = d\tau_1 d\tau_2  \co 
\ee
while   winding  and   momentum  are normalized so that
$p\in {1\over R} {\Bbb Z}$ and
$w\in 2\pi R\ {\Bbb Z}$
for a circle of radius $R$.  The
Lagrangian form of the above lattice sum,
obtained by a Poisson resummation, reads
\be
\Gamma_{d,d}
 = {1\over \tau_2^{d/2} }
\sqrt{\det G} \sum_{m^I,n^I \in {\Bbb Z}}
e^{-{\pi\over\tau_2}\sum_{I,J}
 (G+B)_{IJ} ( m^I + n^I \tau) ( m^I+ n^I  \bar\tau)}
\ee
with $G_{IJ}$ the metric and $B_{IJ}$ the (constant)
antisymmetric-tensor background
on the compactification torus. For a circle of radius $R$ the metric
is $G= R^2$.

The  modular function   ${\cal A}$ inside the integrand
  depends on the
 vacuum. It is {\it quartic, quadratic} or {\it linear} in ${\cal F}$
and
 ${\cal R}$,
for vacua with {\it maximal, half} or a {\it quarter}
 of unbroken supersymmetries.
The corresponding amplitudes have the property of saturating
exactly the fermionic zero modes in a Green--Schwarz light-cone
formalism, so that the contribution from left-moving oscillators
cancels out \cite{Lerche}.
 In the covariant NSR formulation this same
fact follows from $\th$-function
 identities. As a result ${\cal A}$ should
have been
holomorphic in   $q$, but the use of a   modular-invariant
regulator  introduces some
extra  $\tau_2$-dependence \cite{Lerche}.
As  described in more detail
in Appendix \ref{elg4},   ${\cal A}$  takes  the generic form
of a finite polynomial in $1/\tau_2$,  with coefficients that have
Laurent expansions with at most simple poles in $q$,
\be
{\cal A}({\cal F},{\cal R},\tau) =  \sum_{\n=0}^{\n_{\rm max}}\
 \sum_{n=-1}^\infty  {1\over\tau_2^\n}\; q^n \
{\cal A}^{(\n)}_n({\cal F},{\cal R})
\pe  \label{exp}
\ee
The poles in $q$ come from the would-be tachyon. Since this is not
charged under the gauge group, the poles are only present in the purely
gravitational terms of the effective action. This can be verified
explicitly in eq. (\ref{genus}) below.
The $1/\tau_2^\n$ terms  play an important role in what follows.
They come from corners of the moduli space where vertex operators,
whose fusion can produce a massless state, collide. Each pair of
colliding operators  contributes  one factor of $1/\tau_2$.
For maximally supersymmetric vacua, the effective action of interest
starts with terms having four external legs, so that $\n_{\rm max} = 2$.
For vacua respecting  half the supersymmetries ($N=1$ in six dimensions
or $N=2$ in four) the one-loop effective action starts with terms
having two external legs and thus $\n_{\rm max} = 1$.

Much of what we will say in the sequel depends only on the above
generic properties of ${\cal A}$.  It will apply  in particular
to the most often studied case of  four-dimensional
vacua with $N=2$.  For definiteness we will, however, focus
our attention on
the toroidally compactified $SO(32)$ theory, for which
 \cite{Schellekens,Lerche}
\be
\eqalign{
\ \ \  \cA( {\cal F},{\cal R},\tau )= &\  t_8\;  tr{\cal F}^4
\;+\;\frac{1}{2^7\cdot 3^2\cdot 5} \   {E_4^3\over \eta^{24}}\  t_8\;
tr{\cal R}^4
\;+\; {1\over 2^9\cdot 3^2} {\hat E^2_2 E_4^2\over \eta^{24}}
t_8\; ( tr{\cal R}^2)^2
 \cr &
+{1\over 2^9\cdot 3^2}\; \Bigl[  {E_4^3\over \eta^{24}} +
 {\hat E^2_2 E_4^2\over \eta^{24}}
-2 {\hat E_2E_4E_6\over \eta^{24}} -2^7\cdot
3^2\Bigr]\   t_8\; ( tr {\cal F}^2)^2 \cr \ \ \ \ \ &
+ {1\over 2^8\cdot 3^2}\; \Bigl[   {\hat E_2E_4E_6\over \eta^{24}}
 - {\hat E^2_2 E_4^2\over \eta^{24} }
\Bigr]\   t_8\;   tr {\cal F}^2   tr {\cal R}^2  \pe \cr} \
\label{genus}
\ee
Here $t_8$ is the well-known tensor appearing in four-point amplitudes
of the heterotic string \cite{GSW},
and  $E_{2k}$ are  the
Eisenstein series, which are (holomorphic for $k>1$)
 modular forms of weight $2k$. Their explicit expressions are
 collected for convenience in Appendix \ref{modf}.
The  second Eisenstein series $\hat E_2$ is special, in that it
 requires non-holomorphic regularization.
 The entire non-holomorphicity of ${\cal A}$ in  eq. (\ref{genus}),
arises through  this modified Eisenstein series.

We will also give here the gravitational thresholds in the case of
non-trivial Wilson lines: 
\be
{\cal I}^{\rm het}_D  =
-{\cal N}(2\pi)^d 
  \int_{ { F}}{d^2\tau \over
\tau_2^2}\; (\tau_2)^{d/2}   \Gamma_{d,d+16}\
\hat {\cal A}({\cal R}, \tau)
\label{5ex}\ee
where
\be
\hat \cA( {\cal R},\tau )= t_8\; \frac{1}{2^7\cdot 3^2\cdot 5}
\   {E_4\over \eta^{24}}\  t_8\;
tr{\cal R}^4
\;+\; {1\over 2^9\cdot 3^2} {\hat E^2_2\over \eta^{24}}
t_8\; ( tr{\cal R}^2)^2
\pe \label{gravell}\ee
An explicit form of the lattice sum in the Lagrangian representation is
given by
\bs
\be
\Ga_{d+16,d}(G,B,Y)={\sqrt{{\rm det~G}}\over
\tau_2^{d/2}}
\sum_{m^{I},n^{I}\in \Bbb{Z}}\exp\left[-{\pi\over \tau_2}(G+B)_{IJ }
(m^{I}+ n^{I} \t )(m^{J}+ n^{J} \bar\t )\right]
\;\;\;\;\;\; \;\;\;\;\;\;
\label{B2222}\ee
$$\times {1\over
2}\sum_{a,b=0}^1~\prod_{i=1}^{16}~e^{-i\pi(m^{I}Y^i_{I}
Y^i_{J}n^{J}+ b~n^{I}Y^i_{I})}~\th\left[^{a+2n^{K}Y^i_{K}}_
{b+2m^{K}Y^i_{K}}\right] (0 |\t)
\;\;\;\;\;
$$
\be
\;\;\;\;\;\;\;\;\;
\;\;\;\;\;\;\;\;\;
\;\;\;\;\;\;\;\;\;
={\sqrt{{\rm det~G}}\over
\tau_2^{d/2}}
\sum_{m^{I},n^{I}\in \Bbb{Z}}\exp\left[-{\pi\over \t_2}(G+B)_{IJ}
(m^{I}+ n^{I} \t )(m^{J}+ n^{J} \bar\t )\right]
\label{B222}\ee
$$
\;\;\;\;\;\;\;\;\; \;\;\;\;\;\;
\;\;\;\;\;
\times \exp\left[i\pi\sum_i n^{I}\left(m^{J}+n^{J} \tau
\right)Y^i_{I}Y^i_{J}\right]~{1\over
2}\sum_{a,b=0}^1\prod_{i=1}^{16}
\th[^a_b](Y^i_{K}(m^{K}+\t n^{K})|\tau) 
$$
\es
where $G,B$ are the constant metric and antisymmetric tensor and $Y$
are the constant Wilson lines.

In the toroidally compactified  heterotic string, 
all one-loop  on-shell amplitudes with fewer
than four external legs vanish identically \cite{AS}.
This is not true for off-shell amplitudes.
In \cite{6} it was shown that heterotic/type-II duality implies
an antisymmetric tensor-gravitational Chern--Simons CP-even coupling, 
which vanishes on shell.
Consequently  eq. (\ref{5}) directly gives the effective action,
without
having to subtract  one-particle-reducible diagrams,  as is the case
at tree level \cite{slo}. Notice also that this four-derivative
effective action has infrared divergences  when more than one
dimensions are
compactified.  Such IR divergences can be regularized in
a modular-invariant way with
a curved background \cite{KK,chem}. This should be kept in mind,
 even though for the sake of simplicity we will be working
in this paper  with a simpler cutoff procedure to be specified later.

\section{ Two-Torus Compactification}
\setcounter{equation}{0}
\vskip 0.2cm

The comparison of the two theories in perturbation theory for $D=9$ was
discussed in detail in \cite{bk2}.
They agree at one loop. Moreover duality implies higher contact
contributions on the type-I side. It was argued in \cite{bk2} that such
contributions
are required by supersymmetry.
Here, we will review the next simplest situation, corresponding to
compactification on a
two-dimensional torus with zero Wilson lines, which was treated in
\cite{bk2}.
In this case, there are world-sheet
instanton contributions  on the heterotic side, and our aim in
this and the following sections will be to understand them as
(Euclidean)
D1-brane  contributions on the type-I side.

 The target-space torus
 is  characterized by  two complex moduli,
the  K{\"a}hler-class
\be
 T = T_1 + iT_2 = \frac{1}{\alpha' } (B_{89} + i\sqrt{G})
\ee
and the complex structure
\be
U = U_1 + i U_2 = ( G_{89} + i\sqrt{G} )/ G_{88} \ ,
\ee
where $G_{IJ}$ and $B_{IJ}$ are the
 $\sigma$-model metric and antisymmetric
tensor  on the heterotic side.
The one-loop  thresholds now read
\be
{\cal I}^{\rm het}= -
{V^{(8)}\over 2^8 \pi^4} \; \int_{ {\cal{F}} }{d^2\tau\over
  \tau_2}\Gamma_{2,2}(T,U)\;  {\cal A}({\cal F},{\cal R},\tau) \ ,
\label{dd8}\ee
where the lattice sum takes the form  \cite{DKL}
\be
    {\Gamma_{2,2}} (T,U) =
{ T_2\over \tau_2 }
 \sum_{ A \in {\rm Mat}(2 \times 2, {\Bbb{Z}}) }
 e^{ 2\pi i  T {\rm det} A}
e^{- \frac{\pi T_2 }{ \tau_2 U_2 }
\big| (1\; U)A  \big( {\tau \atop 1} \big) \big| ^2 } \pe
\label{DKL}
\ee
Following Dixon, Kaplunovsky and Louis \cite{DKL}, we
decompose the set of all matrices $A$  into orbits of $PSL(2,Z)$,
which is the group of the above transformations up to an overall sign.
 There are three types of orbits,
$$
\eqalign{& {\rm invariant}:  \ A=0 \cr
&  {\rm degenerate}: \  {\rm det} A = 0 ,\;  A\not= 0   \cr
&  \mbox{non-degenerate}: \  {\rm det} A \not= 0  \cr}
$$
A  canonical choice of representatives for the  degenerate orbits
is
\be
A  = \left( \matrix{0  & j\cr 0  & p  \cr}\right)
\co \ee
where the
 integers $j,p$ should not both vanish,
but are  otherwise arbitrary.
Distinct elements of  a degenerate orbit are in one-to-one
correspondence
with the set of modular transformations that map the fundamental
domain on
the strip.
In what concerns the  non-degenerate
orbits, a canonical choice of representatives is
\be
\pm A = \left(\matrix{ k& j\cr 0&p\cr}\right) \ \
{\rm with}\ \   0\le j <k \sp \ p\not= 0\ .
\ee
Distinct elements of a non-degenerate orbit
 are  in one-to-one correspondence with the fundamental
domains of $\tau$ in the double cover of the upper-half  complex plane.

Trading  the sum
over orbit elements for  an extension of the integration region
of $\tau$, we can thus express eqs. (\ref{dd8}), (\ref{DKL}) as
follows: 
\be
\eqalign{
{\cal I}^{\rm het} =
-{V^{(8)}  T_2 \over 2^8\pi^4}  \times &
\Biggl\{  \int_{{\cal{F}}}{d^2\tau \over
\tau_2^2}  {\cal A}
\ + \
\int_{\rm strip} {d^2\tau\over\tau_2^{\ 2}}
\sum_{(j,p) \neq (0,0)}
 e^{- \frac{\pi T_2 }{ \tau_2 U_2 }
\big| j +pU \big| ^2 }
\;{\cal A} \cr
&+
2\; \int_{\Bbb C^+} {d^2\tau\over\tau_2^{\ 2}}
 \sum_{{0 \leq j<k} \atop { p\neq 0}}
        e^{2\pi i Tpk}
        \; e^{- \frac{\pi T_2 }{ \tau_2 U_2 }
        \big|k\tau + j+ pU \big| ^2 }\; {\cal A}
\Biggr\} \ \equiv {\cal I}_{\rm pert} +{\cal I}_{\rm inst} .
\cr}
\label{3terms}
\ee
The three terms inside the curly brackets are constant,
power-suppressed
and exponentially suppressed in the large compactification-volume
limit.
They correspond respectively to tree-level, higher-perturbative and
 non-perturbative contributions on the type-I side.
Substituting the form (\ref{exp}) of the elliptic genus in
(\ref{3terms}),
we may write, for each of the three contributions: 
\be
 {\cal I}=- {V^{(8)} T_2 \over 2^{8}\pi^4}
 \sum_{\n=0}^{\n_{\rm max}} \sum_{n=-1}^\infty
I_{\n,n} {\cal A}^{(\n)}_n({\cal F},{\cal R})
\co \ee
where the corresponding integrals $I_{\n,n}$ are computed 
in Appendix \ref{1li} and
further rewritten in Appendix \ref{lte} to exhibit the instanton
expansion.

In particular, for the higher perturbative contributions we need
\be
I^{\rm pert}_{\n} =
\int_0^\infty {d\tau_2\over\tau_2^{2+\n}}\;\sum_{(j,p) \neq (0,0)}
 e^{- \frac{\pi T_2 }{ \tau_2 U_2 }
\big| j+p U \big| ^2 }
=
\n !  \left({U_2\over\pi T_2}\right)^{\n +1}  \;
\sum_{(j,p) \neq (0,0)} \vert j + p  U\vert^{-2(\n+1) }\pe 
\ee
In the open-string channel of the type-I side, this properly takes into
account the (double) sum over Kaluza--Klein momenta
\cite{BaKi}. Notice that the holomorphic anomalies  in ${\cal A}$
lead again to higher powers of the inverse volume, which translate
to higher-genus contributions on the type-I side. Notice also that
the $\n=0$ term has a logarithmic infrared divergence, which must be
appropriately regularized.
In all $D=8$ calculations we regularize the thresholds by removing the
contribution from the massless states.

 We now turn to the contributions  of the world-sheet instantons, in
which case we are led to consider the integrals
\be
I_{\n,n}^{\rm inst} = 2 \sum_{0 \leq j < k \atop p \neq  0}
\int_{\Bbb C^+} {d^2\tau\over\tau_2^{\ 2}} \; e^{2\pi i Tpk}
         \; e^{- \frac{\pi T_2 }{ \tau_2 U_2 }
        \big|k\tau + j + pU \big| ^2 }\; {1\over \tau_2^{\n}}
e^{2i\pi\tau n}
\pe \ee
To write the final result, we expand the elliptic genus as
\be
{\cal A}\equiv {\cal A}_0=\sum_{\nu=0}^{\nu_{\rm max}}\hat
E_2^{\nu}\Phi_{\nu}(\tau)
\label{ell1}\ee
and define the following relatives of the elliptic genus
\be
{\cal A}_s=D_{\tau}^s\sum_{\nu=s}^{\nu_{\rm max}}\left(\nu\atop
s\right)\hat E_2^{\nu-s}\Phi_{\nu}(\tau)
\co \label{ell2}\ee
where $D_{\tau}$ are the appropriate (non-holomorphic) covariant
derivatives defined in Appendix \ref{modf1}.
In the next section we show that $\cA_s$ is also an elliptic genus
relevant to thresholds involving the moduli.
Then, we find the following expression for the instantonic
contributions
\be
 {\cal I}_{\rm inst}=-{V^{(8)}\over 2^{6}\pi^4}
  {\rm Re}\sum_{s=0}^{\nu_{\rm max}}
\left( {3 \over 2 \p} \right)^s
\sum_{{0 \leq j<k} \atop { p > 0}}
      {1\over (k p)^{s+1} T_2^s}\;  e^{2\pi i Tpk}
        \; {\cal A}_s\left({ pU + j \over k}\right)
\co \label{inst}
\ee
which is one of the main results of Ref. \cite{bk2}. In particular, it
was
shown there that this form
 reproduces the sum of D1-instantons on the type-I side.

Expression (\ref{inst}) has an elegant rewriting in terms of
Hecke operators  $H_N$.
 On  any modular form $F_d(z)$ of weight $d$,
the action of a Hecke operator,  defined by  \cite{Serre}
\be
H_N[F_d](z) = {1\over N}
\sum_{k,p>0\atop kp=N} \sum_{0\le j <k } { p^{d}}\;
  F_d\left(pz+j\over
k\right) \ ,
\label{hec}\ee
gives another modular form of the same  weight.
The Hecke operator is self-adjoint
with respect to the inner product defined by integration of modular
forms on a fundamental domain.
Using the definition (\ref{hec})
 one finds
\be
{\cal I}_{\rm inst} = -{V^{(8)}\over 2^{6}\pi^4}
{\rm Re} \sum_{s=0}^{\nu_{\rm max}}
\left( {3 \over 2 \p} \right)^s
\sum_{N=1}^\infty
      {1\over  (N T_2)^s}\;  e^{2\pi i NT}
        \; H_N[ {\cal A}_s](U)
\pe \label{inst1}
\ee
As we will argue in the next section, this form of the instanton sum
should be related to a matrix-model interpretation of the D-instantons.

\boldmath
\section{Further $D=8$ Thresholds, Supersymmetric Recursion Relations
and Generalized Prepotentials \label{fth} }
\setcounter{equation}{0}
\unboldmath
In this section we will further analyze one-loop threshold corrections
to low-energy couplings beyond the ones described up to now.
We will show that elliptic genera ${\cal A}_s$, with $s=1,2$, that are
defined
in  (\ref{ell2}) and control the higher-genus corrections in
(\ref{inst})
are appearing in threshold corrections of other terms in the effective
action.
Such thresholds are related via recursion relations to those of the
$F^4$ and $R^4$ terms. We will argue in analogy with $N=2$ supersymmetry
in four dimensions
that such relations are dictated by supersymmetry.

We start by reminding the reader of an analogous situation in
heterotic ground states with four-dimensional $N=2$ supersymmetry, which can
be obtained
from six-dimensional ground states upon compactification on a two-torus.
It was shown in \cite{KK1} that the one-loop Wilsonian threshold
correction to the four-dimensional gauge couplings (for zero Wilson
lines) is almost universal and has the form
\be
\Delta^{F^2}_i=\int_{\cal F}{d^2\tau\over
\tau_2}\left[\Gamma_{2,2}{\cal A}_0-b_i\right]=\int_{\cal
F}{d^2\tau\over
\tau_2}\left[\Gamma_{2,2}\left(\Phi_1(\tau)\hat
E_2+\Phi_0(\tau)\right)-b_i
\right]
\co \label{add2}\ee
where
\be
\Phi_1=-{k_i\over 12}{E_4E_6\over \eta^{24}}\;\;\;,\;\;\;
\Phi_0={k_i\over 12}(j-1008)+b_i
\label{add3}\ee
and $i$ labels a non-Abelian factor of the gauge group. In particular,
$k_i$ is the level
of the associated current algebra that determines the tree-level gauge
coupling,
$b_i$ is the $\beta$-function of massless states and
$j$ is the modular-invariant.
The expression (\ref{add2}) parallels the threshold expressions studied
in this paper.

On the other hand, there is a one-loop correction to the K\"ahler
potential
that governs the kinetic terms of the two-torus moduli $T,U$.
We will focus for simplicity on the kinetic terms of $T$.
The K\"ahler metric has been calculated in \cite{agnt,KK1,stie}, with the
result
\be
K^{(1)}_{T\overline{T}}=-{1\over T_2^2}\int_{\cal F}{d^2\tau\over
\tau_2^2}
{i\over \pi}\partial_{\tau}\left(\tau_2\Gamma_{2,2}\right){E_4E_6\over
\eta^{24}}
={1\over T_2^2}\int_{\cal F}{d^2\tau\over \tau_2}
\Gamma_{2,2}D_{\tau}\Phi_1={1\over T_2^2}\int_{\cal F}{d^2\tau\over
\tau_2}
\Gamma_{2,2}{\cal A}_1 \co 
\label{add5}\ee
where ${\cal A}_1$ is the descendant of the $F^2$ elliptic genus.
The two threshold corrections are related as a consequence of  supersymmetry
\cite{agnt}: 
\be
\partial_{T}\partial_{\overline T}~{\Delta^{F^2}_i\over k_i}={3\over 2}
K^{(1)}_{T\overline{T}}+{b_i\over k_i~T_2^2}
\co \label{add6}\ee
which is valid away from enhanced symmetry points\footnote{There are extra
corrections there, see \cite{sing,KK1}.}.
That (\ref{add2}) and (\ref{add5}) satisfy (\ref{add6}) can be shown as
follows.
The lattice sum satisfies the following identity: 
\be
T_2^2\partial_{T}\partial_{\overline T}(\tau_2\Gamma_{2,2})=\tau_2^2
\partial_{\tau}\partial_{\bar \tau}(\tau_2\Gamma_{2,2})
\pe \label{add7}\ee
Act on (\ref{add2}) using (\ref{add7}) and integrate twice by parts,
then
eq. (\ref{add6}) follows, where the last constant terms come from the
boundary
and where one has to use the relation
\def\aa{{\cal A}}
\be
\aa_1={2\over 3}\tau_2^2\partial_{\tau}\partial_{\bar\tau}~\aa_0
\pe \label{add8}\ee
In this relation, $\aa_1$ would have been zero were it not for the
non-holomorphicity of the elliptic genus $\aa_0$.
To put it otherwise, the world-sheet contact terms responsible for the
non-holomorphicity of the elliptic genus are crucial for spacetime
supersymmetry.

Similar arguments should be applicable to $N=4$ supersymmetry in $D=8$.
Unfortunately in this case the detailed structure of supersymmetry
relevant for higher-derivative terms is not known in detail.
Our results for the thresholds on the heterotic side, presented in 
Appendix \ref{rrp}, strongly suggest that there is a structure
similar to $N=2$ supersymmetry in four dimensions, and that several
couplings can be written in terms of holomorphic prepotentials.
Despite this lack of knowledge, there is, as we will now show, a
generalization
of the structure we presented above for $D=4$, $N=2$ ground states, and
similar recursion relations exist as well. We conjecture that such
recursion
relations are due to supersymmetry.

{}From now on we will specialize to the $O(32)$ string compactified on
a torus.
  Let us consider first the one-loop correction of a four-derivative term
involving the toroidal moduli only.
At tree level such a term is obtained from a dimensional reduction
of the $tr R^2$ term, which does not receive loop corrections.
As we shall see, the one-loop correction is entirely due to world-sheet
instantons.
The torus moduli are $G_{IJ}, B_{IJ}$. We will use some arbitrary
basis $\phi_i$ for the moduli.
The appropriate vertex operators for $\phi_i$ are
\be
V_{\phi_i}=v^i_{IJ}
\partial X^I (\bar\partial
X^J-ip_{\mu}\bar\psi^{\m}\bar\psi^J)e^{ip\cdot X}
\co \label{b40}\ee
where
\be
v^i_{IJ}={\partial\over \partial\phi_i}(G_{IJ}+B_{IJ})
\pe \label{b41}\ee
Doing the direct calculation of the torus amplitude, we obtain the
following term in the effective action\footnote{We will not worry about
overall,
moduli-independent normalization of the thresholds.}
\be
Z^{ijkl}(g^{\m\n}g^{\r\s}-g^{\m\r}g^{\n\s}+g^{\m\s}g^{\n\r})
\partial_{\m}
\phi_i\partial_{\n}\phi_j\partial_{\r}\phi_k\partial_{\s}\phi_l
\label{b42}\ee
where
\be
Z^{ijkl}=v^i_{I_1J_1}v^j_{I_2J_2}v^k_{I_3J_3}v^l_{I_4J_4}
(G^{I_1I_2}G^{I_3I_4}
-G^{I_1I_3}G^{I_2I_4}+G^{I_1I_4}G^{I_2I_3}){\cal I}^{J_1J_2J_3J_4}
\label{b43}\ee
and
\be
{\cal I}^{J_1J_2J_3J_4}=\sqrt{G}\int_{\cal F}{d^2\tau\over \tau_2^2}
\sum_{m^J,n^J }\left[\prod_{i=1}^4{m^{J_i}+n^{J_i}\bar \tau\over
\tau_2}
\right]
e^{-{\pi\over \tau_2}(G+B)_{KL}
(m^K+n^K\tau)(m^L+n^L\bar \tau) } 
\, \left( {E_4^2\over \eta^{24}} \right) 
\pe \label{b44}\ee

Let us now focus on $D=8$ where the lattice is two-dimensional and the
relevant
moduli\footnote{We set the Wilson lines to zero.}
 are $T,U$.
Then, for the $(\partial T\partial\bar T)^2$ we obtain, using
(\ref{b44}), the relevant integral:  
\be
Z^{T^2\bar T^2}={1\over T_2^4}\int_{\cal F}{d^2\tau\over \tau_2^2}
D^2(\tau_2\Gamma_{2,2}){E_4^2\over \eta^{24}}
={1\over T_2^4}\int_{\cal F}{d^2\tau\over \tau_2}
\Gamma_{2,2}~D^2\left({E_4^2\over \eta^{24}}\right)={1\over
T_2^4}\int_{\cal F}{d^2\tau\over \tau_2}
\Gamma_{2,2}~{\cal A}_2
 \label{b45}\ee
where, in the second step, we have integrated by parts twice.
The boundary terms
\be
\int_{\cal F}d^2\tau ~\partial_{\tau}\left[{1\over \tau_2^2}
{E^2_4\over \eta^{24}}~
\partial_{\tau}(\tau_2\Gamma_{2,2})-\tau_2\Gamma_{2,2}
\left(\partial_{\tau}
+{i\over \tau_2}\right)\left({1\over \tau_2^2}{E_4^2\over
\eta^{24}}\right)
\right]
\ee
can be verified to vanish and ${\cal A}_2$ is given in (\ref{ell2}).

We also have terms of the form $(\partial\phi)^2 trF^2$ and
$(\partial\phi)^2 tr R^2$.
By direct calculation we obtain the one-loop term of the form\footnote{
Such threshold integrals were calculated in \cite{stie}.}
\be
Z^{ij}\partial_{\m}\phi_i\partial_{\n}\phi_j
tr\left(F^2_{\m\n}-{1\over 4}g_{\m\n}F^2\right)
\co \label{b46}\ee
where
\bs
\be
Z^{ij}=v^i_{I_1J_1}v^j_{I_2J_2}G^{I_1I_2}{\cal I}_F^{J_1J_2}
\label{b47}\ee
\be
{\cal I}_F^{J_1J_2}=-
\sqrt{G}\int_{\cal F}{d^2\tau\over \tau_2^2}
\sum_{m^J,n^J}\left[\prod_{i=1}^2{m^{J_i}+n^{J_i}\bar
\tau\over \tau_2}\right]
e^{-{\pi\over \tau_2}(G+B)_{KL}
(m^K+n^K\tau)(m^L+n^L\bar \tau) } 
 tr\left[Q^2 -{k\over 4\pi\tau_2}\right]
\pe 
\label{b48}\ee
\es
In the $O(32)$ case at hand 
\be
tr\left[Q^2 -{k\over 4\pi\tau_2}\right]={1\over 12}
{\hat E_2E_4^2-E_4E_6\over \eta^{24}}
\pe \label{b49}\ee
A similar computation gives a term as in (\ref{b46}), with
$F\to R$ and
\be
{\cal I}_R^{J_1J_2}=-
{\sqrt{G}\over 12}\int_{\cal F}{d^2\tau\over \tau_2^2}
\sum_{m^J,n^J}\left[\prod_{i=1}^2{m^{J_i}+n^{J_i}\bar
\tau\over \tau_2}\right]
e^{-{\pi\over \tau_2}(G+B)_{kl}
(m^k+n^k\tau)(m^l+n^l\bar \tau)} 
\, \left( {\hat E_2E_4^2\over \eta^{24}} \right) 
\pe \label{b500}\ee

Specializing to $D=8$ we find that for the terms $\partial
T\partial\bar T
tr R^2$ and  $\partial T\partial\bar T
tr F^2$ the threshold correction is given by
\bs
\be
Z^{T\bar TF^2}=-{1\over T_2^2}
\int_{\cal F}{d^2\tau\over \tau_2^2}
\partial_{\tau}(\tau_2\Gamma_{2,2}){\hat E_2E_4^2-E_4E_6\over
12~\eta^{24}}
={1\over T_2^2}\int_{\cal F}{d^2\tau\over \tau_2}
\Gamma_{2,2}~D\left({{\hat E_2E_4^2-E_4E_6\over 12~\eta^{24}}}\right)
\label{b51}\ee
\be
Z^{T\bar TR^2}=-{1\over T_2^2}
\int_{\cal F}{d^2\tau\over \tau_2^2}
\partial_{\tau}(\tau_2\Gamma_{2,2}){\hat E_2E_4^2\over 12~\eta^{24}}
={1\over T_2^2}\int_{\cal F}{d^2\tau\over \tau_2}
\Gamma_{2,2}~D\left({{\hat E_2E_4^2\over 12~\eta^{24}}}\right)
\pe \label{b52}\ee
\label{frt} \es
The elliptic genera appearing in eqs. (\ref{frt}) are
essentially
${\cal A}_1$ in (\ref{ell2}) for the appropriate terms.

We can now discuss recursion relations, which are supposed to hold 
because of supersymmetry.
We consider as a starting point the $(trF^2)^2$ threshold
\be
Z^{(F^2)^2}=\int_{\cal F}{d^2\tau\over
\tau_2}\Gamma_{2,2}\aa_0^{(F^2)^2}
\pe \label{b53}\ee
It can be verified that the  elliptic genus (\ref{ell1})
and its relatives defined in (\ref{ell2}) satisfy the following
recursion
relation
\be
\aa_s={1\over s!}\left({2\over 3}\right)^s~D^s(-i\pi\tau_2^2
\partial_{\bar\tau})^s~\aa_0
\pe \ee
By straightforward algebra, using the form of the covariant derivatives
from Appendix \ref{modf} ($D=D_{-2}$, $D^2=D_{-2}D_{-4}$ etc.), we find
\bs
\be
\aa_1={2\over 3}\tau_2^2\partial_{\tau}\partial_{\bar\tau}~\aa_0
\;\;\;\;\;\;\;\;\;\;\;\;\;\;\;\;\;\; 
\;\;\;\;\;\;\;\;\;\;\;\;\;\;\;\;\;\; 
\label{b55}\ee
\be
\aa_2={1\over 2}\left({2\over
3}\right)^2\left[(\tau_2^2\partial_{\tau}\partial_{\bar\tau})^2
-{1\over 2}\tau_2^2\partial_{\tau}\partial_{\bar\tau}\right]\aa_0
\pe \label{b56}\ee
\es
These are special cases of the relations (\ref{z12}) and (\ref{z11}).  

Again we emphasize that these recursion relations are due to the
non-holomorphicity of the elliptic genus.
Following the same procedure as in the $N=2$ case we can derive the
following recursion relations
\bs
\be
\partial_T\partial_{\overline T}Z^{(F^2)^2}={3\over
2}Z^{T\overline{T}}+{{\rm constant}\over T_2^2}
\;\;\;\;\;\;\;\;\;\;\;\;\;\;\;\;\;\; 
\;\;\;\;\;\;\;\;\;\;\;\;\;\;\;\;\;\; 
\label{b57}\ee
\be
\left(T_2^2\partial_T\partial_{\overline T}-{1\over 2}\right)(T_2^2
Z^{T\overline{T}F^2})=3T_2^4~Z^{T^2\overline{T}^2}+{\rm constant '}
\pe \label{b58}\ee
\es
The constants come from boundary terms.
Similar recursion relations can be written down for all the
factorizable
terms we are considering in the paper.

We believe that these relations are a consequence of supersymmetry, 
as in the $N=2$ case.
They only exist owing to the world-sheet contact terms in the heterotic
result.
These contact terms imply a higher-genus contribution in the type-I
side.
It is natural to conjecture that their presence in the type-I theory is
due
to the different realization of supersymmetry.

Such recursion relations between elliptic genera and
differential equations satisfied by the (2,2) torus lattice sum imply
the existence of prepotentials, generalizing the $N=2$ situation in four
dimensions.

We  consider the following integrals
\be
\Psi_s=\int_{\cal F}{d^2\t\over
\tau_2}\left[\Gamma_{2,2}(T,U){\cal A}_s -C\delta_{s,0}\right]
\co \label{z100}\ee
where $s=0,1,2,\cdots ,\nu_{\rm max}$ and ${\cal A}_s$ are the
relative elliptic genera.
$C$ is the coefficient of the $q^0$ term in $\cA_0$.
The IR is regulated by subtracting the contribution of the massless
states;  $\Psi_s$ is real.

The (2,2) lattice sum satisfies various differential identities
summarized in Appendix \ref{lpr}.
It is shown in Appendix \ref{rrp} that, using such equations, the
thresholds (\ref{z100}) can in general be written as
\be
\Psi_s=-C\delta_{s,0}\log(T_2U_2)+\sum_{\nu=s}^{\nu_{\rm max}}
{(\nu+s)!\over 6^s(\nu-s)!s!}\left[D_T^{\nu}D_U^{\nu}
f_{\nu}(T,U)+cc\right]
\co \label{z102}\ee
where $D_T$, $D_U$ are the appropriate covariant derivatives defined
in Appendix \ref{modf}.
The functions $f_{\nu}$ depend holomorphically on the moduli $T,U$.
They are prepotentials generalizing the usual case of $N=2$ 
four-dimensional supersymmetry, which corresponds to $\n_{\rm max}=1$  
\cite{hm2,stie}.
They transform as modular forms of weight $-2\nu$ in $T$ and $U$, up to
additive
pieces that are annihilated by the covariant derivatives.
The full threshold is duality-invariant.
Explicit expressions of the prepotentials can be found in Appendix
\ref{gpp}.
\section{D1-Instanton Interpretation}
\setcounter{equation}{0}

\begin{figure}
\begin{center}
\leavevmode
\epsfbox{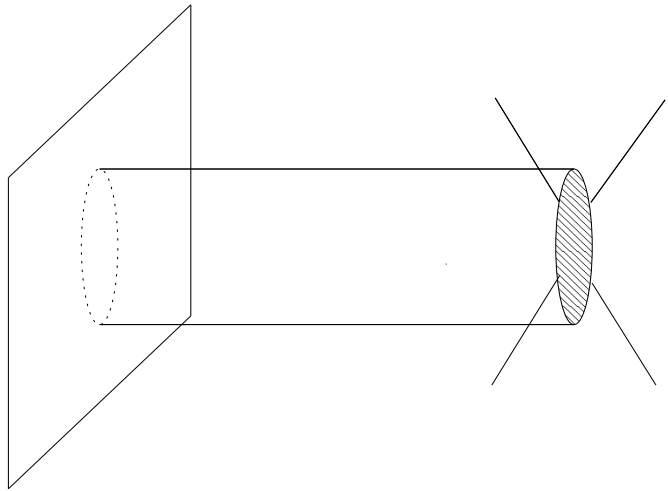}
\end{center}
\caption[]{A D1-brane instanton correction to $tr F^4$.}
\label{f5}\end{figure}

In the type-I theory a flat Euclidean D1-brane, wrapped around the
target space
two-torus, provides us with a supersymmetric instanton that has maximal
supersymmetry. Since the Dirichlet boundary conditions are imposed in
the eight spacetime dimensions, this is a defect localized in spacetime and
thus an
instanton.
Maximal supersymmetry implies that the number of zero modes is minimal
and we expect that it is the only such instanton that would contribute
corrections to the effective terms under consideration.
For example, an instanton contribution to $tr F^4$ at $\chi=0$ should be
generated by the diagram depicted in Fig. \ref{f5}.
We will be guided in our computation  of the instanton corrections by
the heterotic result (\ref{inst}).

The Nambu--Goto world-sheet Euclidean action of the D1-brane is known
\cite{tasi} to be
\be
S_{D1}={1\over 2\pi \alpha'}\int d^2\sigma e^{-\Phi/2}\sqrt{|{\rm det}
\hat G|}-{i\over 2 \pi \alpha'}\int B
\co \label{30}\ee
where $\hat G$ is the induced metric on the world-sheet
\be
\hat G_{ij}=G_{\m\n}\partial_iX^{\mu}\partial_j X^{\nu}
\co \label{31}\ee
$G_{\m\nu}$ is the type-I spacetime metric ($\s$-model frame),
$B$ is the type-I (RR) antisymmetric tensor and the factor
$e^{-\Phi/2}$ is due to the fact that the action comes from the disk.
The tension $1/2\pi \alpha'$ has been computed directly in
\cite{Polch}.

We will now evaluate the classical action of the D1-brane wrapped
around the target space torus.
Using Cartesian coordinates $X^1,X^2\in [0,2\pi]$ for the target space
torus and
$\sigma_{1,2}\in [0,2\pi]$ for the D1-brane,
the $\sigma$-model type-I torus metric is
\be
G={\sqrt{{\rm det} G}\over U_2}\left(\matrix {1&U_1\cr
U_1&|U|^2\cr}\right) \pe
\label{32}\ee
The complex structure $U$ defines complex coordinates as usual
\be
Z=X^1+UX^2\;\;\;,\;\;\;\bar Z=X^1+\overline{U}X^2 \pe
\label{33}\ee
The map that wraps the D1-brane world-sheet around the two-torus
is
\be
\left(\matrix{X^1\cr X^2\cr}\right)=\left(\matrix{n_1&m_1\cr
n_2&m_2\cr}\right)\left(\matrix{\s_1\cr \s_2\cr}\right) \pe
\label{34}\ee
To have a non-trivial wrapped configuration with the same orientation,
$n_1m_2-n_2m_1>0$.
For $n_1m_2-n_2m_1 <0$, the orientation is reversed and the induced
complex structure is complex-conjugated.
As we will see below, the first case corresponds to instantons, 
the second to anti-instantons.

The complex structure of the original torus (\ref{33}) induces a
complex structure of the D1-brane.  Defining
\be
z=\s_1+{\cal U}\s_2\;\;\;,\;\;\;\bar z=\s_1+\overline{{\cal U}}\s_2
\label{38}\ee
the map from $Z$ to $z$ is holomorphic, $Z=f(z)$.
If the map changes the orientation, this acts as complex conjugation
on the complex structure.
Using (\ref{33}) and (\ref{34}) we find that
\be
Z=({n_1+Un_2})\left[\s_1+{m_1+Um_2\over n_1+Un_2}\s_2\right]
\co \label{36}\ee
which implies that the induced complex modulus is
\be
{\cal U}={m_1+Um_2\over n_1+Un_2}\;\;\;,\;\;\;n_1m_2-n_2m_1 >0
\;\;\;,\;\;\;{\rm Im}\,{\cal U}>0
\label{37}\ee
and
\be
\;\;\;\; {\cal U}={m_1+\overline{U}m_2\over
n_1+\overline{U}n_2}\;\;\;,\;\;\;n_1m_2-n_2m_1 <0
\;\;\;,\;\;\;{\rm Im}\,{\cal U}>0 \pe
\label{377}\ee

Modular transformations of the target-space torus act on $X^1,X^2$
by $SL(2,Z)$ transformations. From (\ref{34}) we deduce that they
also act on the  matrix of ``winding numbers" by left $SL(2,Z)$
transformations.
Modular transformations on the D1-brane coordinates $\s_1,\s_2$,
act on the winding number matrix by right modular transformations.
Configurations are equivalent if they are related by $SL(2,Z)$
transformations
of the D1-brane coordinates $\s_i$.
The reason is that, since we are using the Nambu--Goto type action, we
have already
``integrated out" the world-sheet metric.
Thus, we can use the right $SL(2,Z)$ action to pick representative
configurations with
\be
\left(\matrix{n_1&m_1\cr n_2&m_2\cr}\right)=\left(\matrix{k&j\cr
0&p\cr}\right)\;\;\;,\;\;\;p>0\;\; \sp 0\leq j <|k| \pe
\label{35}\ee
For such configurations ${\cal U}=(pU+j)/k$.

We can now evaluate the D1-brane classical action. Using
(\ref{31}), (\ref{34}), (\ref{35}) we find
\bs
\be
\sqrt{|{\rm det}\hat G|}=\sqrt{{\rm det}G}|pk|
\label{39}\ee
\be
\int B_{ij}dX^i\wedge dX^j=pk~B_{12}
\pe \label{40}\ee
\es
Denoting also $\l_{\rm I}=e^{\Phi/2}$ we obtain
\be
S_{\rm class}={2\pi\over \a'}\left[|pk|{\sqrt{{\rm det} G}\over
\l_{\rm I}}-ipkB_{12}\right]
\pe \label{41}\ee

As described in Appendix \ref{htd} the mapping between heterotic and
type-I
variables
is $\left.T_1\right|_{\rm
  het}=\left.T_1\right|_{\rm I}$ , $U_{\rm het}=U_{\rm I}$ and
  $\left.T_2\right|_{\rm het}=\left.T_2\over\lambda\right|_{\rm I}$. We
can express this in terms of heterotic variables
\be
\left.T_2\right|_{\rm het}={\sqrt{{\rm det} G}\over
\a'\l_{\rm I}}\;\;\;,\;\;\;\left.T_1\right|_{\rm het}={B_{12}\over \a'}
\label{42}\ee
to obtain
\be
e^{-S_{\rm class}}=\exp\left[-2\pi\left(|pk|T_2-ipk T_1\right)\right]
\pe
\label{43}\ee
When $k>0$, we have instantons and
\be
e^{-S_{\rm class}}=e^{2\pi i pk T}
\co \label{44}\ee
which is to be summed over $k,p>0$, $0\leq j <k $.
For $k<0$, we have anti-instantons and
\be
e^{-S_{\rm class}}=e^{-2\pi i pk \overline{T}}
\co \label{45}\ee
which is again to be summed over $k,p>0$, $0\leq j < k$.
This precisely matches the instanton expansion on the heterotic side
in (\ref{inst}).

We now come to the issue of determinants.
Since the D1-brane has the same world-sheet structure as the heterotic
string
\cite{PW}, we would expect that, up to volume factors, the $\chi=0$
(one-loop)
contribution to the determinant should be the heterotic elliptic genus
evaluated at the modulus of the wrapped D1-brane, $\tau\to {\cal U}$.
This is suggested by the heterotic expansion (\ref{inst}) and is also
natural
on the type-I side.
For anti-instantons, $\tau\to \overline{\cal U}$.
Finally, there is an overall factor of $\sqrt{{\rm det}G}/\sqrt{{\rm
det}\hat G}$, the ratio of volumes of the target space torus to the
D1-brane torus.
This can be understood as follows.
The inverse of $\sqrt{{\rm det}\hat G}$ is coming from the
normalization of zero modes, while the $\sqrt{{\rm det} G}$
factor is the standard volume factor
of the target space torus.

This concludes the discussion for the $tr F^4$ and $tr R^4$ terms.
For the rest, there are extra contributions coming from an instanton
calculation
for $\chi=-1,-2$.
The holomorphic determinants here are related to the heterotic elliptic
genus
via (\ref{inst}) and are the relevant quantities that appear in the
calculation of the generalization of the K{\"a}hler potential in the
$N=4$ case (see Section \ref{fth}).
Moreover, there is an extra overall factor of $({\rm det }\hat
G)^{\chi/2}$ related to zero modes.
It would be interesting to directly understand the type-I calculation
of these terms.

One final comment is in order here.
The world-sheet theory of $N$ D1-branes is a gauge theory
with (8,0) supersymmetry in two dimensions.
It has an $SO(N)$ gauge group, eight scalars that transform in the symmetric
tensor of $SO(N)$ and parametrize the relative distance moduli as well as
another eight, which are $SO(N)$ singlets and parametrize the centre-of-mass
position
in transverse space.
These are accompanied by left-moving  fermions coming from the DD Ramond
sector. There are also DN fermions transforming
 in the $(N,32)$ under $SO(N)\times SO(32)$.
Thus, an $SO(N)$ matrix theory describes the dynamics of $N$ D1-branes,  
\cite{rey}.
As was observed in \cite{talk}, in analogy with the type-II case, 
we would expect that the IR limit
is parametrized by separate coordinates of the $N$ D1-strings, with an
orbifold identification when two of them coincide \cite{matrix}.
On the other hand, it was shown in \cite{DMVV} that for symmetric CFTs
the elliptic genus of an $S_N$ orbifold is given by the action of a
Hecke operator of order $N $ on the original elliptic genus.
Although this was shown in the type-II context, it is also valid for
heterotic orbifolds.

The above discussion provides an interpretation of eq. (\ref{inst1}), 
which expresses the instanton sum as a sum over Hecke operators acting
on the elliptic genus.
The $N$-th term in the sum should come from $N$ D1-brane instanton
configurations.
This interpretation should be directly derivable from the appropriate
matrix model \cite{talk}.

\boldmath
\section{$D=8$ Heterotic Thresholds with \\ Non-Zero Wilson Lines}
\setcounter{equation}{0}
\unboldmath

We will now include generic Wilson lines $Y^i_{I}$, $i=1 \ldots 16$,
$I=1,2$ which generically break the gauge group to the Cartan, 
$O(32)\to U(1)^{16}$.
We define the following complex moduli
\bs
\be
G =  {(2T_2U_2-\bar y_2\cdot\bar y_2)\over 2 U_2^2 } \pmatrix{ 1 & U_1
\cr U_1 & |U|^2 \cr}
\sp B_{12} = T_1 - { \by_1 \cdot \by_2 \over 2 U_2}
\ee
\be
y^i= (y_1 + i y_2)^i   = - Y_2^i + U Y_1^i
\ee
\label{cc12} \es
where we denote with $\bar y$ the sixteen-dimensional complex vector of
Wilson lines.
Note that the volume of the two-torus in this parametrization is
\be
\cV \equiv \sqrt{\det G}
=T_2-{\bar y_2\cdot\bar y_2\over 2 U_2}
\pe \label{volume}\ee

We focus for simplicity on the gravitational one-loop thresholds
given in
(\ref{5ex}) and (\ref{gravell}): 
\be
{\cal I}^{\rm het}_{D=8}  =
-{\cal N}_8
  \int_{ {\cal  F}}{d^2\tau \over
\tau_2^2}\; (\tau_2   \Gamma_{2,18}(T,U,\bar y))\
\hat {\cal A}({\cal R}, \tau)
\pe \label{55ex}\ee
The appropriate elliptic genus for a given term can be written
as
\be
\hat {\cA}=\sum_{\nu=0}^{\nu_{\rm max}}\hat E_2^{\nu}\Phi_{\nu}(\tau)
\co \label{771}\ee
where $\nu_{\rm max}=0$ for $tr R^4$ and $\nu_{\rm max}=2$ for $(tr R^2)^2$.
Here $\Phi_{\nu}$ is a modular form of weight $-8-2\nu$ and the explicit
form of the relevant $\Phi_\n$'s can be read from (\ref{gravell}).
The integral can be done explicitly and the result can be expressed in
terms of polylogarithms. This is described in Appendix \ref{1li}.
The trivial and degenerate orbits produce a result that is perturbative
on the type-I side. The non-degenerate orbits give a result that is
non-perturbative on the type-I side.
We are interested in the large volume expansion of the non-degenerate
orbit
contribution.
This is derived in Appendix \ref{lte}, and we will reproduce it here.
We introduce the $O(32)_1$ affine lattice sum
\be
\chi(\bar y|\t) =\sum_{\bar r}e^{i\pi\tau \bar r\cdot\bar r}e^{2i \pi
\bar r\cdot \bar y}={1\over
2}\sum_{a,b=0}^1\prod_{i=1}^{16}\th[^a_b](y^i|\tau)
\label{772}\ee
where $\bar r$ is a vector in the Spin(32)/Z$_2$ lattice.
The full affine character is given by $\chi$ divided by $\eta^{16}$.
Under modular transformations
\bs
\be
\chi(\bar y+\bar \e_1+\tau\bar\epsilon_2| \tau )
= e^{-i\pi(\tau\bar
\e_2\cdot\bar \e_2+2\bar \e_2\cdot \bar y)}\chi(\tau,\bar
y)\;\;\;,\;\;\;
\chi(\bar y| \tau+1)=\chi(\bar y| \tau)
\label{773}\ee
\be
\chi\left({\bar y\over \tau}| -{1\over \tau}\right)= \tau^8
e^{i\pi{\bar
y\cdot \bar y\over \tau}}\chi(\tau,\bar y)
\label{774}\ee
\es
where $\bar\e_{1,2}$ are lattice vectors.
These transformations define a generalized Jacobi form of type
$(d,m)=(8,1)$.
Properties of such forms are reviewed in Appendix \ref{modf2}.
We will introduce also the covariant derivative on generalized Jacobi
forms
\be
\tilde{D}
= D_\t   + {i \over \p \t_2 } \by_2 \cdot \pa_{\bar y}
- m {\by_2 \cdot \by_2 \over \t_2^2}
\co \label{7jde}
\ee
which is such that $\tilde{D}F_{d,m}$ is a Jacobi form of type
$(d+2,m)$.
$D_{\t}$ is the usual covariant derivative on weight $d$ modular forms
defined
in Appendix \ref{modf1}.
We will now define the relatives of the character-valued elliptic genus
as
\be
\hat\aa_s(\bar y,\tau)=\tilde D^s\sum_{\nu=s}^{\nu_{\rm max}}\left(\nu\atop
s\right)\chi(\bar y| \tau )\hat
E^{\nu-s}_2\Phi_{\nu}(\tau)
\co \label{bb54}\ee
with $\hat\aa_0=\chi(\bar y|\tau)\hat \aa(\tau)$.
Note that by setting the Wilson lines $\bar y$ to zero (\ref{bb54})
reduces to (\ref{ell2}).
The non-degenerate orbit part of the threshold can be written as
\be
 {\cal I}_{\rm inst}=-4{\cal N}_8
  {\rm Re}\sum_{s=0}^{\nu_{\rm max}}
\left( {3 \over 2 \p} \right)^s
\sum_{{0 \leq j<k} \atop { p > 0}}
      {1\over (k p)^{s+1} \cV^s}\;  e^{2\pi i Tpk}
        \; \hat {\cal A}_s\left( p \bar y ,  { pU + j \over k}\right)
\pe \label{7inst}
\ee
which generalizes the zero Wilson-line result (\ref{inst}).
It is also written as an expansion in inverse powers of $\cV$, 
which is the volume of the two-torus (see eq. (\ref{volume})).
Using the generalized Hecke operators $V_N$ \cite{ez} of Appendix
\ref{modf3}, the result
can also be recast in the following form: 
\be
 {\cal I}_{\rm inst}=-4{\cal N}_8
  {\rm Re}\sum_{s=0}^{\nu_{\rm max}}
\left( {3 \over 2 \p} \right)^s
\sum_{N=1}^{\infty} {1\over (N \cV)^s}\;  e^{2\pi i N T}
V_N[ \hat {\cal A}_s] (\by |U) \co \label{ghecke}
\ee
which is the analogue of (\ref{inst1}) obtained with zero Wilson lines.

Before we proceed with the D1-instanton interpretation of the result, we
should mention that the thresholds in the presence of Wilson lines can
also be written
in terms of generalized prepotentials.
As shown in Appendix \ref{gpp}, the generalization of (\ref{z102}) is
\be
\Psi_s=-C\delta_{s,0}\log(T_2U_2-\bar y\cdot\bar
y/2)+\sum_{\nu=s}^{\nu_{\rm max}}
{(\nu+s)!\over 6^s(\nu-s)!s!}\left[\square^{\nu}
f_{\nu}(T,U,\bar y)+cc\right]
\co \label{z101}\ee
where $\square$ acting on a $(d,m)$ Jacobi form is
\be
\square=
\frac{1}{2\pi^2} \left(\pa_{\bar y}\cdot \pa_{\bar y}-2\pa_T\pa_U
+{16 -2 d   \over (\bar y_2\cdot \bar y_2-2T_2U_2) } \left[\frac{d}{2} +i(
\bar y_2
\cdot \pa_{\bar y}+T_2\pa_T+U_2\pa_U) \right]
\right) \pe 
\label{7box} \ee
It reduces to $D_TD_U$ in the absence of Wilson lines.
More explicit forms for the generalized prepotentials in this case
are given in Appendix \ref{gpp}.

We will now interpret the result in terms of the D1-brane.
The coupling of the D1-brane to bulk gauge fields  is a one-loop effect
given by the diagram in Fig. \ref{f5}. Thus, the coupling to Wilson
lines is
also a
one-loop effect; consequently, it is independent of the type-I dilaton.
We can evaluate the induced Wilson lines on the D1-brane world-sheet as
\be
\bar w=-\bar W_2+{\cal U}\bar W_1=p\bar y
\ee
where we have used $\bar W_i=\bar Y_{\a}\partial_{i}X^{\a}$,
${\cal U}=(pU+j)/k$ and the map (\ref{34}), (\ref{35}).
This explains the dependence of the generalized elliptic genus in
(\ref{7inst}).
Thus, part of the one-loop determinant is the heterotic genus evaluated
at the induced world-sheet modulus ${\cal U}$ and the induced Wilson
lines $\bar w$.

The exponential factor $\exp[2\pi i kp T]$ is composed of two parts.
Using (\ref{cc12}) we find that the first part is the same as was discussed
in Section 6 and that it is generated by the D1-brane classical action.
There is a left-over piece depending on the Wilson lines, which after
some algebra can be written in terms of induced data as $\exp[i\pi
{\bar w\cdot \bar w_2\over {\cal U}_2}]$.
This is the Quillen anomaly of the one-loop determinant of the 32
world-sheet
fermion fluctuations of the D1-brane coupled to the induced Wilson
lines $\bar w$.
There are also the usual factors of volume, as in the case with zero
Wilson
lines.
The terms in (\ref{7inst}) with $s>0$ correspond to higher-loop
contributions around the instanton, on the type-I side.
We conclude that the one-loop determinants around the D1-instanton
are composed of the heterotic elliptic genus evaluated at $\tau={\cal
U}$ multiplied on the one hand by the $O(32)$ affine character evaluated at 
$\tau={\cal U}$ and at the induced Wilson lines $\bar y\to \bar w$ and also
multiplied by  the
anomaly factor of the world-sheet fermions.
Again we sum over all possible wrappings of the D1-brane, modded out by
the world-sheet diffeomorphisms.

Since, here, we can also write the result in terms of the generalized
Hecke
operators $V_N$ as in (\ref{ghecke}), it is this form that should
correspond to the D1
matrix model with non-trivial Wilson lines.

\newpage
\boldmath
\section{Heterotic Thresholds in $D<8$}
\setcounter{equation}{0}
\unboldmath

We will now discuss heterotic thresholds in toroidal compactifications
to $D<8$. As we argued earlier, if $D>4$ then the heterotic result is
still one-loop only and can be evaluated.
Using heterotic/type-I duality we find again the non-perturbative
type-I corrections and we show that their corresponding D1-brane
interpretation
is in agreement with the D1-brane rules given in Section 6.

Our starting point is the general form of the one-loop thresholds
\be
{\cal I}^{\rm het}_{D}  =
-{\cal N}_D
  \int_{ {\cal  F}}{d^2\tau \over
\tau_2^2}\; (\tau_2^{d/2}   \Gamma_{d,d}(G,B ))\
{\cal A}(\tau)
\co \label{ddh}
\ee
where the $D + d =10$ and the $d$-dimensional lattice sum $\Ga_{d,d}$
is given by
\be
\Ga_{d,d}(G,B)  = {\sqrt{G} \over \t_2^{d/2} }
\sum_{m^i,n^i \in \Bbb Z }
\exp\left[-{\pi\over \tau_2}(G+B)_{ij }
(m^{i}+ n^{i} \t )(m^{j}+ n^{j} \bar\t )\right]
\co \label{ddt} \ee
where $G$ and $B$ are the $d$-dimensional metric and antisymmetric
tensor
respectively. Alternate forms of the lattice sum can be found in
Appendix \ref{htt}.

The corresponding integral (\ref{ddh}) can be evaluated again, using the
method
of orbits. We refer to Appendix \ref{htt} for the main steps, and quote
here only the result of the non-degenerate orbit, which comprises the
type-I instantonic contributions: 
\be
 {\cal I}_{\rm inst}=-2 {\cal N}_D
\sum_{s=0}^{\n_{\rm max}}
 \left(  { 3 \over 2 \p} \right)^s
 \sum_{m,n}' { \sqrt{G} \over (T^{m,n}_2)^{s+1} }
 e^{2 \pi i T^{m,n} }
\cA_s( U^{m,n})
\label{gic} \ee
where we have used the definition (\ref{ell2}) of the elliptic genera.
Here, the induced K\"ahler and complex structure moduli are given by
\bs
\be
T^{m,n} =  m B n + i \sqrt{ (m G m) (n G n  ) - (m Gn)^2 }
\;\;\;\;\;\;\;\;\;\;\;\;\;\;\;\;\; \ee
\be
U^{m,n} = \left( - m G n + i \sqrt{ (m G m) (n G n  ) - (m Gn)^2 } \;
\right)
/ n G n
\ee
\label{utm} \es
and the $\sum_{m,n}'$ is over the non-degenerate orbits, which are
parametrized
by the following integer-valued $2\ti d$ matrices
\bs
\be
\mbox{non-degenerate orbit} : \;\;\;\; A^T=
 \left( \matrix{n_1 &  \ldots & n_k & 0 &  \ldots &   0  \cr
 m_1 &  \ldots & m_k & m_{k+1} &  \ldots  & m_d   \cr } \right)
\;\;\;\;\;\;\;\;\;\;\;\;\;\;\;\;\;\; 
\;\;\;\;\;\;\;\;\;\;\;\;\;\;
\ee
\be
\;\;\;\;\;\;\;\;\;\;\;\;\;\;\;\;\;
\;\;\;\;\;\;\;\;\;\;\;\;\;\;\;\;\;
 1 \leq k < d   \sp n_k > m_k \geq 0 \sp (m_{k+1},\ldots,m_d)  \neq
(0,\ldots,0)
\pe  \ee
\label{ndm}
\es
Note that for $d=2$ the general result (\ref{gic}) reduces to the one
given in
(\ref{inst}).

Turning to the D1-brane interpretation of the result, we first wish
to establish that the exponential factor $ e^{2 \pi i T^{m,n} } $
agrees
with the classical action of a D1-brane. The map that describes the
wrapping of the D1-brane world-sheet around a 2-cycle in the $d$-torus
is
\be
X^i = n_i \s_1 + m_i \s_2  \sp i = 1 \ldots d
\co \label{dtm}
\ee
where $X^i$ are the coordinates on $T^d$ and $\s_{1,2}$ the D1-brane
coordinates. We observe that modular transformations on the D1-brane
coordinates act on the matrix $A$ that enters (\ref{dtm})
\be
 A =
 \left( \matrix{n_1 &  m_1  \cr
\vdots & \vdots \cr
 n_d  &  m_d \cr} \right)
\ee
by right $SL(2,Z)$ transformations, which forces us to pick the
representative
configurations described by the matrices in (\ref{ndm}).

In terms of the matrix $M_I^i =(A_i^I)^T =  (n^i,m^i)$, $I=1,2$, we
see that
the induced metric and antisymmetric tensor fields are
\be
\hat{G}_{IJ} = M_I^i G_{ij} M_J^j
\sp \hat{B}_{IJ} = M_I^i B_{ij} M_J^j
\pe \label{inf}
\ee
In particular, going through the same steps as in Section 6, we find
from the D1-brane classical action (\ref{30}) and (\ref{utm}),
(\ref{inf})
that $e^{-S_{\rm class}}$ precisely reduces to the
exponential factor $ e^{2 \pi i T^{m,n} }   $,
which is to be summed over the ranges indicated in (\ref{ndm}).
We also note that we correctly observe the overall factor
$\sqrt{G}/\sqrt{\hat{G}} = \sqrt{G}/T^{m,n}_2 $. Moreover, the
fluctuation
determinant is evaluated at the induced modulus $U^{m,n}$ of the 
wrapped D1-brane.

This establishes the claim that the D1-brane rules in $D<8$ are
consistent
with those obtained for $D=8$. In summary, we have found the
intuitively
expected result that
the instantonic contributions consist of all possible inequivalent
wrappings of the D1-brane around two-tori that are embedded in the
$d$-dimensional  target space torus modulo reparametrizations of the
D1-brane world-sheet.

In the eight-dimensional case we have shown that differential
equations satisfied by the (2,2) toroidal lattice sum translate into  
recursion relations for the thresholds, which can be solved in terms of  
holomorphic prepotentials.
There is a generalization of such equations for the $(d,d)$ toroidal lattice
sum.

It was noted in Refs. \cite{KK,KK5} that the toroidal partition function 
$\Ga_{d,d}(G,B;\t) $ satisfies the following differential
equation: 
\be
\left[
\left( \sum_{i \leq j} G_{ij} { \pa \over \pa G_{ij} } +
{ 1- d \over 2} \right)^2
+ \frac{1}{2} \sum_{ijkl} G_{ik} G_{jl} {\pa^2 \over \pa B_{ij} \pa B_{kl} }
 -\frac{1}{4}
 - 4 \t_2^2 {\pa^2 \over \pa \t \pa \bar{\t} }
 \right] \Ga_{d,d}(G,B;\t)  = 0
\label{de1} \ee
which in the case $d=2$ reduces to
\be
\left[  T_2^2 \pa_T  \pa_{\bar{T}}
 -  \t_2^2  \pa_\t \pa_{\bar{\t}}
 \right] \Ga_{2,2}(T,U;\t)  = 0 \pe 
\label{2de} \ee
However, the general differential equation in (\ref{de1}) is not
invariant under
the full $O(d,d,Z)$ duality group. It  may be verified that
it is invariant under integer $B$ shifts and $SL(d,Z)$ basis changes,
but there is no invariance under the
 remaining generators of the duality group, which
are the inversion and factorized duality. The latter
two transformations act on the matrix $E \equiv G +B$ as follows: 
\be
E \ra E^{-1} \sp E \ra [(1-e_i)E + e_i][ e_i E + (1-e_i)] ^{-1}
\sp (e_i)_{k,l} = \d_{ik} \d_{il}
\pe \ee
For example, in the  $d=2$ case
the factorized dualities correspond to $ T \ra U$ and $T \ra 1/U$ for $i=1$
and 2 respectively, which do not leave 
the differential equation in (\ref{2de}) invariant. 

This implies that there must be further constraints on $\Ga_{d,d}$
generalizing the $d=2$ relation
\be
\left[  T_2^2 \pa_T  \pa_{\bar{T}}
-  U_2^2 \pa_U  \pa_{\bar{U}} \right]  \Ga_{2,2}(T,U;\t)  = 0
\pe \label{utd} \ee
To find the generalization of this relation we note that there
is another $ O(d,d,Z)$ invariant differential equation
on the lattice sum, which
reads
\be
\left[  \sum_{ijkl}
 G_{ik} G_{jl} {\pa^2 \over \pa E_{ij} \pa E_{kl} }
+   \sum_{ij} G_{ij} { \pa \over \pa E_{ij} }
+\frac{1}{4} d (d-2) -4 \t_2^2 \pa_{\t} \pa_{\bar{\t}} \right]
\Ga_{d,d}(E;\t)  = 0
\pe \label{de2} \ee
As a consequence we find that the difference between (\ref{de1}) and
(\ref{de2}) is the differential equation,
\be
\left[ \sum_{ijkl} (G_{ij} G_{kl} - G_{jk} G_{il})
{\pa^2 \over \pa E_{ij}  \pa E_{kl} }
+ (1-d)\sum_{ij} G_{ij}  {\pa \over \pa E_{ij}  }\right]
\Ga_{d,d}(E;\t)  = 0
\co \label{de3} \ee
which, for $d=2$, turns out to precisely reduce to (\ref{utd}).
In fact, there is an entire family of constraints
\be
\left[ \sum_{ijkl} (P_{ij} P_{kl} - P_{kj} P_{il})
{\pa^2 \over \pa E_{ij}  \pa E_{kl} }
+ \sum_{ij} [ (PG^{-1}  - {\rm Tr}(P G^{-1})\one ) P ]_{ij}  {\pa
\over \pa E_{ij} }\right]
\Ga_{d,d}(E;\t)  = 0
\co \label{de4} \ee
which include (\ref{de3}) for $P=G$.
Here $P$ is an arbitrary matrix.

Clearly (\ref{de3}) and its generalization (\ref{de4})
are not invariant under the duality group, since (as (\ref{de1}))
the inversion and factorized duality are broken, but  these transformations
should be used to form a complete irreducible set of differential equations.
For example, under the inversion, we find that (\ref{de4}) with matrix $P$ 
is transformed into the same differential equation with matrix
\be
P' = E \tilde{P} E \sp \tilde{P} = P \ve_{E \ra E^{-1} }
\pe \ee

It is an open problem to find the general solution of such equations  
which will
define the analog of prepotentials in the lower dimensional case.

\section{Conclusions and Remarks}

We have analyzed here heterotic/type-I duality in eight
dimensions with arbitrary Wilson lines as well as in $D<8$ dimensions
with zero Wilson lines.

We focused in particular on $R^4$ terms in the effective
action
that obtain corrections from short multiplets.

In eight dimensions, the heterotic result is one-loop only.
However, non-perturbative instanton corrections are necessary on the
type-I side.
We identified the relevant instanton configurations with a D1-brane
wrapped
around the compact two-torus.
The heterotic result implies a concrete way to count different
instanton configurations.
Multiple overlapping D1-branes have to be included, however, in order
to restore $T$-duality.
Moreover, we have to sum over D1-branes wrapped in any possible way
around ${\cal T}^2$ modulo the modular transformations of the
D1-world-volume.
Most interestingly, the fluctuation determinant around a given
D1-instanton
configuration is given by the heterotic elliptic genus evaluated at the
complex structure modulus induced on the world-sheet of the wrapped
D1-brane.
The instanton result can be written in terms of Hecke operators. In
this form it
provides a
potentially interesting link with a $SO(N)$ matrix model of D1-branes.
Finally, we have shown that the thresholds can be expressed in terms of
generalized
holomorphic prepotentials.

We have also considered the heterotic perturbative thresholds in $D=8$
in the presence of arbitrary Wilson lines. We have calculated exactly
the one-loop perturbative contribution. In this case, 
heterotic/type-I duality predicts that the D1-instanton determinant is
the
affine character-valued genus evaluated at the induced complex
structure
of the D1-brane world-volume and the induced Wilson lines on this
world-sheet.
Moreover, we found the exponential factors to be  in agreement with the
classical D1-brane action as well as the Quillen anomaly of the 32
fermions.

Finally, we have discussed the heterotic perturbative thresholds in
toroidal compactifications to $D< 8$. In this case, again using
heterotic/type-I duality, we find agreement with the D1-brane rules
obtained from $D=8$. In particular, we observe all possible wrappings
of
the D1-brane around the various two-tori that are embedded in the
$d$-torus.
Moreover, the exponential factor corresponding to the classical action
as well as the fluctuation determinants are in agreement with the $D=8$
result
as well.

There are several questions, however, that remain open.
An essential quantitative test of heterotic/type-I duality can be
obtained
by directly calculating relevant higher-genus terms on the type-I side.
Already in ten dimensions, the $\chi=-1$, $(tr F^2-tr R^2)^2$ term
should match
the corresponding tree-level term on the heterotic side.
In $D<10$, further higher-genus contact terms, corresponding to
one-loop world-sheet contact terms on the heterotic side, should be checked.
This state of affairs in duality comparisons is not new.
Similar situations arise in $N=2$ heterotic/type-II dual pairs with $N=2$ 
supersymmetry, and heterotic/type-I dual pairs with $N=2$ supersymmetry.

At the effective supergravity level, knowledge of the holomorphic
($D=8$) or quaternionic ($D=6$) structure of the special derivative
terms is missing.
An analogue of the higher F-terms of $N=2$ supersymmetry should exist for
$N=4$ supersymmetry.
The expressions that we have obtained in Appendix \ref{gpp}
for the heterotic
thresholds
in terms of generalized prepotentials are very suggestive in this
respect.

Since our results on the heterotic side are supposed  to be
non-perturbatively exact for $D>4$, a direct quantitative check could
be made of the conjectured
F-theory/heterotic duality in eight dimensions \cite{F}.
Techniques however are necessary on the F-theory side to calculate the
relevant amplitudes.

The heterotic result can provide a (missing) quantitative test
of string--string duality in six dimensions.
The type-IIA theory compactified on $K3$ down to six dimensions is
conjectured
to be equivalent to the heterotic string compactified on ${\cal T}^4$.
As in the heterotic case, we do not expect non-perturbative corrections 
either on the type-II side for the $F^4,R^4,R^2F^2$ terms.
This can be seen as follows: the relevant D$p$-branes of the
ten-dimensional
IIA theory have $p=0,2,4,6,8$ with world-sheets being
$1,3,5,7,9$-dimensional.
To obtain an instanton contribution we need appropriate supersymmetric
cycles on $K3$ with dimension belonging to the list above.
It is known that there are  no such cycles.
Moreover, we also have the five-brane, which is magnetically coupled to
the NS-NS
antisymmetric tensor. Since its world-sheet is six-dimensional it can
only give
instanton corrections in $D<5$ dimensions.
Thus, in $D=6$, heterotic/type-II duality can be tested for the special
terms
in perturbation theory.
Preliminary investigation suggests that the relevant objects on the
type-II side are the $N=4$ topological amplitudes defined in \cite{BV}.
Preliminary investigation shows that for example the tree-level $F^4$
terms on the type-II side match the one-loop corrections to such terms
on the heterotic side as required by duality.
We can further compactify both theories on a circle to five dimensions.
There are still no non-perturbative corrections on the heterotic side.
In the type-II theory, we expect instanton corrections from the D2- and
D4-branes,  which are electrically (magnetically) charged under the
3-form.
The D2-brane can wrap around ${\cal S}^1$ and a supersymmetric two-cycle
of $K3$.
The D4-brane can wrap on ${\cal S}^1$ and the whole of $K3$.
These non-perturbative type-II corrections are expected to reproduce
the heterotic cross-terms coupling the (4,4) and the (1,1) lattice.
A more thorough investigation is needed, however.

Finally, although we do think that we understand the conceptual rules of
instanton calculations in string theory, there are several issues that
remain to be answered in this respect.
A direct D-brane calculation of the D1-instanton determinant should be
done. Such techniques are also of importance for five-brane 
instanton calculations in
four-dimensional ground states. Knowing how to do this calculation for
the D5-brane will provide, via various dualities, the rules for NS5-brane
instantons in heterotic and type-II string theory.

\vskip 0.8cm
\centerline{\bf Acknowledgements}
\vs .1cm
This research was partially supported by EEC grant
TMR-ERBFMRXCT96-0090.
N. Obers acknowledges the hospitality
of the Niels Bohr institute during part of this work.
We thank  C. Bachas and P. Vanhove for participating in the
earlier stages of this project and for numerous discussions and
insights.
We would also like to thank M. Henningson for contributing to the
threshold evaluation in $D<8$.

\vskip 0.3cm
\newpage 
\appendix
\boldmath
\section{Modular functions \label{modf} }
\subsection{$SL(2,Z)$ modular functions and covariant derivatives
 \label{modf1}}
\renewcommand{\theequation}{A.\arabic{equation}}
\setcounter{equation}{0}
\unboldmath

We list in this appendix the $\th$-function definitions we use, and
those associated with modular forms.
We also discuss modular-covariant derivatives and
a number of identities involving these.

Our conventions for the $\th$-function  are
\be
\th [^a_b] (v |\t )=
 \sum_{p \in \Bbb Z}  e^{\pi i \t \left(p-{a \over 2}\right)^2 +
2\p i \left(v -{b \over 2} \right)\left(p -{a \over 2} \right) }
\label{the} \ee
so that the Jacobi $\th$-functions are given by
\be
\th_1= \th[^1_1]     \sp
\th_2= \th[^1_0] \sp
\th_3= \th[^0_0] \sp
\th_4= \th[^0_1]
\label{tfn} \co \ee
and the Dedekind function is
\be
\eta(\tau)=q^{1/24}\prod_{n=1}^{\infty} (1-q^n)
\ee
where $q=e^{2i\pi\tau}$.

Holomorphic modular forms $F_d(\tau)$ of weight $d$ transform under
the modular group as
\be
F_d (\t+1)= F_d (\t)
\sp
F_d (-1/\t) =  \tau^d~ F_d (\t)
\label{A1}\ee
A set of modular forms, relevant for our purposes, are the Eisenstein
series
\be
E_{2k} = - {(2k)!\over (2\pi i)^{2k} B_{2k}}\ G_{2k}\ ,
\label{A2}\ee
with  $B_{2k}$  the Bernoulli numbers  and
\be
G_{2k}(\tau) = \sum_{(m,n)\not= 0}\ (m\tau + n)^{-2k}
\ee
for $k>1$.
For $k=1$ the Eisenstein series diverges. Its 
modular-invariant regularization,  denoted with a hat and used  in this paper,
is
\be
{\hat G}_2 (\t)  = \limit{\rm lim}{s\to 0} \sum_{(m,n)\not= 0}\ (m\tau
+
n)^{-2}
 \vert m\tau + n\vert ^{-s}
\pe \ee
The (hatted)  Eisenstein series are modular forms of weight $2k$.
The ring of holomorphic modular forms is generated by $E_4$ and $E_6$.
If we include (non-holomorphic) covariant derivatives (to be discussed
below)
then the generators of this ring are $\hat E_2$, $E_4$, $E_6$.

Expressed as power series in $q$, the first few of the Eisenstein
series
 are
\bs
\be
E_{2}(q)=
{12\over i \pi}\partial_{\t}\log \eta
=1-24\sum_{n=1}^{\infty}{n\, q^n\over 1-q^n}
\, \
\label{A3}
\ee
\be
E_{4}(q)=
{1 \over 2}\left(
{\vartheta}_2^8+
{\vartheta}_3^8+
{\vartheta}_4^8
\right)
=1+240\sum_{n=1}^{\infty}{n^3q^n\over 1-q^n}
\, \
\label{A4}
\ee
\be
E_{6}(q)=
\frac{1}{2}
\left({\vartheta}_2^4 + {\vartheta}_3^4\right)
\left({\vartheta}_3^4 + {\vartheta}_4^4\right)
\left({\vartheta}_4^4 - {\vartheta}_2^4\right)
=1-504\sum_{n=1}^{\infty}{n^5q^n\over 1-q^n}
\pe
\label{A5}\ee
\label{e2s} \es
The modified first Eisenstein series is
\be
 {\hat E}_2 =
 E_2-{3\over \pi\tau_2} \pe
\label{a6}\ee
We can write the (weight 12) cusp form $\eta^{24}$ and the 
modular-invariant $j$-function in terms of $E_4$ and $E_6$
\be
\eta^{24}={1\over 2^6\cdot 3^3}\left[E_4^3-E_6^2\right]
\;\;\;,\;\;\;
j={E_4^3\over \eta^{24}}={1\over q}+744+\cO (q)
\pe \label{a7}\ee

There is a (non-holomorphic) covariant derivative that maps modular
forms
of weight $d$ to forms of weight $d+2$, defined as
\be
F_{d+2}=\left({i\over \pi}\partial_{\t}+{d/2\over
\pi\tau_2}\right)F_{d}=
-2\left(q\partial_q-{d\over 4\pi\tau_2}\right)F_d
\equiv D_d\;F_d \pe
\label{a9}\ee
The covariant derivative satisfies the distributive property
\be
D_{d_1+d_2}~(F_{d_1}~F_{d_2})=F_{d_{2}}(D_{d_1}F_{d_1})+
F_{d_{1}}(D_{d_2}F_{d_2} ) \pe
\label{a8}\ee
We will suppress the index $d$ from the covariant derivative and
write multiple derivatives as $D^n$. For example a double
derivative on a weight $d$ form is
\be
D^2 F_d\equiv\left({i\over \pi}\partial_{\t}+{(d+2)/2\over
\pi\tau_2}\right)\left({i\over \pi}\partial_{\t}+{d/2\over
\pi\tau_2}\right)
F_d \pe
\label{a11}\ee
The following formulae allow the computation of the covariant
derivative
of any form:
\bs
\be
D\;\hat E_2={1\over 6}E_4 - {1\over 6}\hat E_2^2
\;\;\;,\;\;\;
D\;E_4={2\over 3}E_6-{2\over 3}\hat E_2\;E_4
\label{a12}\ee
\be
D\;E_6=E_4^2-\hat E_2\;E_6 \pe
\label{a13}\ee
\es
There is also a holomorphic covariant derivative on forms of weight
$d$:
the quantity
\be
F_{d+2}=\left({i\over \pi}\partial_{\t}+{d\over 6}E_2\right)F_{d}
\equiv \hat D_d\;F_d
\label{a15}\ee
is a modular form of weight $d+2$.
It satisfies a distributive property similar to that in (\ref{a8}).
For the difference between the two covariant derivatives, we obtain: 
\be
\hat D_d-D_d={d\over 6}\hat E_2
\pe \ee

We also list a number of identities involving modular forms and
covariant derivatives, which are used in Appendices \ref{1li} and
\ref{lte}.
In these expressions the
quantity $D^s$ always stands for $D_{-2} D_{-4} \cdots D_{-2s}$.  \nl
1) Expansion formula
\bs
\be
(D^s \hE_2^{\n-s} \Phi_\n ) (\t)
= \sum_{r=0}^s \sum_{m=0}^{\n-s} a^{\n,s}_{r,m}
{1 \over (\p \t_2)^{s+m-r} } (q \pa_{q})^r
E_2^{\n -s -m} \Phi_\n (\t)
\label{b13a}
\ee
\be
\eqalign{
a^{\n,s}_{r,m}
& = (-1)^{s+m} {3^m 4^r \over 2^s} {s! \over r!}
\ch{\n-s}{m} \ch{2s+m-r}{s+m} \cr
&   0 \leq s \leq \n \sp 0 \leq r \leq s
\sp 0 \leq m \leq \n -s \cr}
\ee
\label{bb13}
\label{ef1} \es
Two useful special cases are
\be
{\rm Re} D^s i \t^{2s+1} =
- { (-2)^s s! \over \p^{s} } \t_2^{s+1}
\sp
D^s 1 = {  (2s)! \over (-2)^s s! \p^s}    \frac{1}{\t_2^s}
\pe \ee
2) A special function and its derivatives.
The following combined polylogarithm functions
$L_{(s)}$ play a very special role
in the modular-invariant integrals of Appendix \ref{1li}.
Their definition is
\be
L_{(s)} (x) \equiv
   \sum_{r=0}^s { (s+r) ! \over r ! (s-r)! (4 \p)^r}
 ({\rm Im}\,x)^{s-r}
 Li_{s +r +1} (e^{2 \pi i x}  )
\co \label{d18}
\ee
where
\be
Li_s(x) = \sum_{p=1}^\infty  \frac{1}{p^s} x^p
\label{d15}\ee
are the polylogarithm functions.
They satisfy the interesting relations that,
\be
D_U^s D_T^s Li_{2s+1} (q_T^k q_U^l)
= \sum_{m=0}^s {(s+m)! \over m! (s-m)! }  (4kl)^{s-m}
{ L_{(m)}(Tk +Ul) \over (\p T_2 U_2)^m }
\label{re1} \ee
and their  inversion
\be
{ L_{(s)}(Tk +Ul) \over (\p T_2 U_2)^s }
= \sum_{m=0}^s {s ! \over  (s-m)! }{ 2m  +1  \over (s +m +1)! }
(-4kl)^{s-m}
D_U^m D_T^m Li_{2m+1} (q_T^k q_U^l)
\pe \label{re2} \ee

\subsection{Generalized  Jacobi forms and covariant derivatives
\label{modf2}}

We give in this appendix a generalization of Jacobi forms
\cite{ez} and
their associated modular-covariant derivatives, and give various
properties and
application to characters.

We define a generalized Jacobi form of type $(d,m)$\footnote{The
number $d$ is
also called the weight and $m$ the index.}  to be a
holomorphic function $F_{d,m}(y^i|\tau)$ ($i=1\ldots S$)
with the following transformation properties
\bs
\be
F_{d,m}(y^i+\epsilon^i|\tau)=F_{d,m}(y^i|\tau)
\label{5j}\ee
\be
F_{d,m}(y^i+\tau\epsilon^i|\tau)=e^{-i\pi m(\tau\epsilon\cdot
\epsilon+2\epsilon\cdot y)}F_{d,m}(y^i|\tau)
\label{6j}\ee
\be
F_{d,m}(y^i,\tau+1)=F_{d,m}(y^i,\tau)
\label{7j}\ee
\be
F_{d,m}(y^i/\tau|-1/\tau)=\tau^d~e^{i\pi m y\cdot
y/\tau}F_{d,m}(y^i| \tau)
\label{8j} \co \ee
\label{jfo}
\es
where $\e^i$ is a vector in the lattice $\Ga^{S,0}$. For our purposes,
this
will generally be one of the even self-dual Euclidean lattices, which
are the $E_8$ root lattice with $S=8$ or the root lattice
of $E_8 \ti E_8$ or weight lattice of ${\rm Spin}(32)/Z_2$ with $S=16$.

Then it can be explicitly verified that there exists the following
non-holomorphic covariant derivative
\be
\tilde{D}
= D_\t   + {i \over \p \t_2 } \by_2 \cdot \pa_{\bar y}
- m {\by_2 \cdot \by_2 \over \t_2^2}
\co \label{jde}
\ee
which is such that $\tilde{D}F_{d,m}$ is a Jacobi form of type
$(d+2,m)$.
Here $D_\t$ is the usual $SL(2,Z)$ covariant derivative (\ref{a9}) on a
weight $d$ modular function, and $\by_2$ stands for the imaginary part
of the $S$-dimensional vector $\by$.  The inner product on this space
is
taken with the metric $\et_{(S)}$ on $\Ga^{S,0}$.

The generators of $O(S+2,2,Z)$ transformations are
\bs
\be
U \ra U +1 \sp T \ra T \sp \bar{y} \ra \bar{y}
\label{c17a} \ee
\be
U \ra -1/U  \sp T \ra T - {\by \cdot \by \over 2 U}
\sp \bar{y} \ra \bar{y}/U
\label{uin} \ee
\be
U \ra U   \sp T \ra T + \bar{\e} \cdot \by   + \frac{1}{2} \bar{\e}^2 U
\sp \bar{y} \ra \bar{y} + \bar{\e} U
\label{c17f} \ee
\be
U \ra U   \sp T \ra T
\sp \bar{y} \ra \bar{y} + \bar{\e}
\ee
\be
U \ra T    \sp T \ra U
\sp \bar{y} \ra \bar{y}
\ee
\be
U \ra U \sp T \ra T +1 \sp \bar{y} \ra \bar{y}
\ee
\be
U \ra U- {\by \cdot \by \over 2 T}   \sp T \ra -1 /T
\sp \bar{y} \ra \bar{y}/T \pe 
\label{tin} \ee
\label{o2t} \es
Note that the first four of these transformations, which leave
the
variable $\cV = T_2 - {\by_2 \by_2 \over 2U_2}$ invariant, are the
ones used in (\ref{jfo}) (ignoring $T$).
A function $F_{d}(\by,T,U)$ is of weight $d$ in both $T$ and $U$ if it
transforms with a factor $U^d$ and $T^d$ under the transformations
(\ref{uin})
and (\ref{tin}) respectively, and is invariant under the remaining
 transformations in  (\ref{o2t}).

We introduce the following notation for the $O(S+2,2)$ moduli,
\be
y^a = (y^i,y^+,y^-)= (\by, T, U)
\sp \et^{ab} = \left(
\matrix{ \et_{(S)}^{ij} & 0 \cr 0 & -\et_{(2)} \cr }  \right)
\sp \et_{(2)} = \left( \matrix{0 & 1 \cr 1 & 0 \cr} \right)
\co \label{o2m} \ee
where $\et_{(S)}$  is the metric on the lattice, generally taken to
be unity. Inner products on this $(S+2)$-dimensional space are
taken with
the above metric and denoted by $(,)$, so that, for example
\be
(y_2,y_2) = \by_2 \cdot \by_2 - 2 T_2 U_2
\pe \ee
On the space of $O(S+2,2,Z)$ covariant functions, we define the
following
operator
\be
\square_d  =
\frac{1}{2\p^2} \left(
  \et^{ab} \pa_{y_a} \pa_{y_b}
+ {S -2 d   \over (y_2,y_2) } \left[  \frac{d}{2} + i y_2^a \pa_{y_a} \right]
\right)
\co \label{box} \ee
which satisfies the property that
when $F_{(d)} (\by,T,U)$ is a function of weight $d$ in $T$ and $U$,
then the function $\square F_{(d)}(\by,T,U)$ is of weight $d+2$ in both
$T$ and $U$.
Also note that for  $\by = 0$, $S=0$ the operator $\square$
 reduces to the double covariant derivative $D_U D_T$.

We also recall the definition of character and affine
character lattice sums
\be
\chi( \t) = \sum_{\bar r  } q^{\bar r\cdot \bar r/2}
\sp
\chi ( \by|\t)
=\sum_{\bar r} q^{\bar r\cdot \bar r/2}e^{2\pi i \bar r\cdot \by }
\label{ach} \ee
where $\br$ runs over the appropriate lattice.
For example, when $\br \in \Ga^{S,0}$, we have
for the two relevant cases, $S=8$ and $S=16$, the affine character
lattice sums
\bs
\be
\chi_{E_8} ( \by| \t )
={1\over 2}\sum_{a,b=0}^1\prod_{i=1}^{8}\theta[^a_b]( y^i| \t )
\ee
\be
\chi_{E_8 \ti E_8 }( \by| \t)
=  {1\over 2}\sum_{a,b =0}^1\prod_{i=1}^{8}\theta[^a_b]( y_{(1)}^i| \t
)
 {1\over 2}\sum_{a,b}\prod_{i=1}^{8}\theta[^a_b]( y_{(2)}^i| \t)
\ee
\be
\chi_{S0(32)} (\by|q)
={1\over 2}\sum_{a,b=0}^1 \prod_{i=1}^{16}\theta[^a_b]( y^i| \t )
\pe \ee
\es
Comparison of the transformation properties of these affine character
lattice sums and (\ref{jfo}) shows that that they are in fact
Jacobi forms of weight $(S/2,1)$.
The full affine characters are obtained from the lattice sums by
dividing by $\eta^{S}$.

Some identities satisfied by the operators $\tilde{D}$ and
$\square$ are given below. All of these expressions have been
explicitly
checked for $s \leq 2$, which covers the cases needed for this paper.
We conjecture, however, that they are valid generally, and as a
non-trivial
 check one may verify that they correctly reduce to the identities
given in
(\ref{re1}), (\ref{re2}) for $S=0$. Below, the quantities
$\tilde{D}^s$ stand  for
$\tilde{D}_{-2}  \tilde{D}{-4} \cdots \tilde{D}_{-2s} $ and
similarly for $\square^s$. \nl
1) Expansion formula
\bs
\be
\eqalign{
(\tilde{D}^s \hE_2 ^{\n-s} \chi(\by) \Phi_\n)(\t)
& = \sum_{r=0}^s \sum_{m=0}^{\n-s} \sum_{p=0}^{r} \sum_{n=0}^{s-r}
   \a^{\n,s}_{r,m,p,n} {1 \over (\p \t_2)^{s+m-r+p+n} }
\cr 
 & \ti \sum_{\br} ( \p^2 \by_2 \cdot \by_2)^{n}
( \p \br \cdot \by_2)^{p}
 [q \pa_{q} ]^{r-p}
q^{\frac{1}{2} \br \cdot \br }
e^{2 \p i \br \cdot \by }
( E_2^{\n-s} \Phi_\n ) ( \t  )
\cr}
\ee 
\be
\a^{\n,s}_{r,m,p,n}  =
(-1)^{s+m} {3^m 4^r 2^n \over 2^s} {s! \over r! n!}
\ch{\n-s}{m} \ch{2s+m-r-n}{s+m} \ch{r}{p}
\ee
$$
   0 \leq s \leq \n \sp  0 \leq r \leq s
\sp 0 \leq m \leq \n -s 
\sp 0 \leq p \leq r \sp 0 \leq n \leq s-r 
$$
\label{ef2} \es
\ni where $\chi (\by |\t)$ is the affine character (\ref{ach})
of weight $(S/2,1)$.\nl
2) Covariant derivatives of special functions.
The combined polylogarithm function defined in (\ref{d18}) also
satisfies
\be
 \square^s Li_{2s+1} ( e^{2 \pi i (r,y) } )
= \sum_{m=0}^s  \ch{s}{m} {(S/2+s+m) ! \over (S/2 +s)! } (-2)^s
(r^2)^{s-m}
 { L_{(m)}( (r,y) )  \over (\p (y_2,y_2))^m }
\co \ee
where $r =(\br,-l,-k)$ so that
$(r,y) = \br  \cdot \by + Tk +Ul$ and $r^2 = \br \cdot \br - 2 kl $.
The inverse of this relation is
\be
{  L_{(s)}( (r,y) )  \over (\p (y_2,y_2))^s }
= \sum_{m=0}^s \ch{s}{m}
{ (S/2+m)! (S/2+ 2m+1) \over (S/2 +s +m +1)! } { (-1)^s \over 2^m}
(r^2)^{s-m}
 \square^m Li_{2m+1} ( e^{2 \pi i (r,y) } )
\pe \label{idl} \ee
We also have the following identity
\be
\eqalign{
{\rm Re} \square^s i
 d^{(s)}_{a_1 \ldots a_{2s+1}} & y^{a_1} \cdots y^{a_{2s+1} }
  \cr
= & - \frac{1}{ (2\p^2)^s} \sum_{m=0}^s  \ch{s}{m} {(S/2+s+m) ! \over
(S/2 +s)! }
{ (-4)^m m! (2s+1)! \over (2m+1)! }
{ d^{(m)}_{a_1 \ldots a_{2m+1}}\over (y_2,y_2)^m }
y_2^{a_1} \cdots y_2^{a_{2m+1} }
\cr}
\ee 
where the tensors $d^{(m)}$ are totally symmetric and recursively
defined
from  $d^{(s)}$ by
\be
d^{(m-1)}_{a_1 \ldots a_{2m -1} }
=d^{(m)}_{a_1 \ldots a_{2m +1} }  \et^{a_{2m} a_{2m+1} }
\sp 1 \leq m \leq s
\pe \ee
The inverse relation reads
\be
{d^{(s)}_{a_1 \ldots a_{2s+1}} \over (y_2,y_2)^s}
  y_2^{a_1} \cdots y_2^{a_{2s+1} }
=  -  \sum_{m=0}^s \ch{s}{m}
{ (S/2+m)! (S/2+ 2m+1) \over (S/2 +s +m +1)! }
{ (-2 \p^2 )^m (2s+1)! \over 4^s  s ! (2m+1) ! }
\label{idd} \ee
$$
 \ti {\rm Re}\,  \square^s 
d^{(m)}_{a_1 \ldots a_{2m+1}} y^{a_1} \cdots y^{a_{2m+1} }
\pe 
$$
We finally note the relation
\be
 \square^s 1  =
{ (-1)^{s} \over  2^s \p^{2s} }
{ (S/2 +2s)! \over (S/2+ s)! }
{ (2s)! \over s! }
{1 \over (y_2,y_2)^s }
\pe \label{cd1} \ee

\subsection{Hecke operators \label{modf3}}

Consider a Jacobi form as defined in (\ref{jfo}).
Let $J_{d,m}$ be the space of Jacobi forms of type $(d,m)$.
We will define the following operators \cite{ez}
\bs
\be
V_N~[F_{d,m}](y^i|\tau)=\frac{1}{N} \sum_{k,p>0\atop kp=N}~\sum_{ 0 \leq j <
k }~p^d
F_{d,m} \left(py^i\left.\right|{p\tau+j\over k}\right)
\label{9}\ee
\be
U_N~[F_{d,m}](y^i|\tau)=F_{d,m}(N~y^i|\tau)
\label{10}\ee
\be
U_N  ~~ : ~~J_{d,m}\to J_{d,mN^2}\;\;\;,\;\;\;V_N~~~:~~~J_{d,m}\to
J_{d,mN}
\ee
\be
U_N\cdot U_M=U_{MN}\;\;\;,\;\;\;V_{M}\cdot V_{N}=V_{N}\cdot V_{M}=
\sum_{D|(M,N)} D^{d-1}~U_D\cdot V_{MN/D^2}
\pe \label{mid} \ee
\es
where $D$ in (\ref{mid}) runs over the common divisors of $(M,N)$.
The operator $V_N$ is the generalization of the Hecke operator $H_N$  
given in
(\ref{hec}) and one may check that $V_N [F_{d,m}]$ gives a Jacobi form
of type $(d,mN)$.

\boldmath
\section{Heterotic/type-I duality in $D<10$ dimensions \label{htd} }
\renewcommand{\theequation}{B.\arabic{equation}}
\setcounter{equation}{0}
\unboldmath

In this appendix we will derive the heterotic/type-I duality map
once we have compactified both theories on a torus to $D<10$ dimensions.

The heterotic string action in $D$ dimensions is
\be
S^{\rm het}_{D}=\int
d^D x\sqrt{-g}e^{-\phi}\left[R+\partial^{\m}\phi\partial_{\m}\phi
-{1\over 12}\hat H^{\m\n\r}\hat H_{\m\n\r}
\;\;\;\;\;\;\;\;\;\;\;\;\;\;\;  
\;\;\;\;\;\;\;\;\;\;\;\;\;\;\;  
\right.
\label{c1}\ee
$$
\;\;\;\;\;\;\;\;\;\;\;\;\;\;\;  
\;\;\;\;\;\;\;\;\;\;\;\;\;\;\;  
\left.
-{1\over 4}(M^{-1})_{IJ}
F^{I}_{\m\n}F^{J\m\n}+{1\over 8}Tr(\partial_{\m} M\partial^{\m}
M^{-1})\right]
 $$
where
\be
\hat H_{\m\n\r}=\partial_{\m}B_{\n\r}-{1\over
2}A^{I}_{\m}L_{IJ}F^{J}_{\n\r}+{\rm cyclic}
\pe \label{c2}\ee
The $(2d+16)\times(2d+16)$ moduli matrix $M_{IJ} $ is
\be
M=\left(\matrix{G^{-1}& -G^{-1}C &-G^{-1}Y^{t}\cr
-C^{t}G^{-1}&G+C^{t}G^{-1}C+Y^{t}Y&C^{t}G^{-1}Y^{t}+Y^{t}\cr
-YG^{-1}&YG^{-1}C+Y&{\bf 1}+YG^{-1}Y^{t}\cr}\right)
\co \label{c3}\ee
written in terms of the metric $G_{\alpha\beta}$ of the $d$-torus
$(d=10-D)$, the
antisymmetric tensor
$B_{\a\b}$, and gauge moduli $Y^{i}_{\a}$ with
\be
C_{\a\b}=B_{\a\b}-{1\over 2}Y^{i}_{\a}Y^{i}_{\b}
\co \label{c4}\ee
with $\a,\b=1,2,\cdots ,d$ and $i=1,2,\cdots,16$: 
\be
L=\left(\matrix{0&{\bf 1}_{d}&0\cr
{\bf 1}_{d}&0&0\cr
0&0&{\bf 1}_{16}\cr}\right) \pe 
\label{c5}\ee
Going to the Einstein frame we obtain
\be
S^{\rm het}_{D,{\rm E}}=\int
\sqrt{-g_{\rm E}}\left[R-{1\over D-2}\nabla^{\m}\phi\nabla_{\m}\phi
-{e^{-4\phi\over D-2}\over 12}\hat H^{\m\n\r}\hat H_{\m\n\r}\right.
\;\;\;\;\;\;\;\;\;\;\;\;\;\;\;  
\;\;\;\;\;\;\;\;  
\label{c6}\ee
$$
\;\;\;\;\;\;\;\;\;\;\;\;\;\;\;  
\;\;\;\;\;\;\;\;\;\;\;\;\;\;\;  
\left.-{e^{-2\phi\over D-2}\over 4}(M^{-1})_{IJ}
F^{I}_{\m\n}F^{J\m\n}+{1\over 8}Tr(\partial_{\m} M\partial^{\m}
M^{-1})\right]
\pe $$

The ten-dimensional lowest-order effective action of the type-I string
is
\be
S^{\rm I}_{10}=\int
\sqrt{-G_{10}}\left[e^{-\Phi}\left(R+\partial_{\m}\Phi
\partial^{\mu}\right)
-{1\over 4}e^{-{\Phi\over 2}}F^{i}_{\m\n}F^{i,\m\n}-{1\over
12}H_{\mu\nu\rho} H^{\mu\nu\rho}\right]
\pe \label{c7}\ee
Doing the standard toroidal reduction to $D$
dimensions, we obtain
\be
S^{\rm I}_{D}=\int\sqrt{-g}\left[e^{-\phi}\left[
R+(\partial\phi)^2+{1\over 4}\partial G_{\a\b}\partial G^{\a\b}-{1\over
4}G_{\a\b}F^{A,\a}_{\m\n}
F^{A,\b\m\n}\right]\right.
\;\;\;\;\;\;\;\;\;\;\;\;\;\;\;  
\;\;\;\;\;\;\;\;\;\;\;\;\;\;\;  
\label{c8}\ee
$$\left.
-{1\over 4}e^{-{\phi\over 2}}G^{1/4}\left[\tilde
F^i_{\m\n}\tilde F^{i\m\n}+2\tilde F^i_{\m\a}\tilde
F^{i\m\a}\right]-\sqrt{G}\left[{1\over
12}H_{\m\n\r}H^{\m\n\r}+{1\over 4}H_{\m\n\a}H^{\m\n\a}
+{1\over 4}H_{\m\a\b}H^{\m\a\b}\right]\right]
\co $$
where $G$ stands for the determinant of the metric $G_{\a\b}$ and
\bs
\be
\tilde F^{i}_{\m\n}=F^{i}_{\m\n}+Y^{i}_{\a}F^{A,\a}_{\m\n}
\;\;\;,\;\;\;
\tilde F^{i}_{\m\a}=\partial_{\m}Y^{i}_{\a}
\;\;\;,\;\;\;
F_{\m\n}^{i}=\partial_{\m}A^{i}_{\n}-\partial_{\n}A^{i}_{\m}
\label{c11}\ee
\be
C_{\a\b}\equiv B_{\a\b}-{1\over 2} \d_{ij}Y^{i}_{\a}Y^{j}_{\b}
\;\;\;,\;\;\;
H_{\m\a\b}=\partial_{\m}C_{\a\b}+\d_{ij}Y^{i}_{\a}
\partial_{\m}Y^{j}_{\b}
\label{c13}\ee
\be
B_{\m,\a}\equiv\hat B_{\m\a}+B_{\a\b}A_{\m}^{\b}+{1\over
2}\d_{ij}Y^{i}_{\a}A^{j}_{\m}
\;\;\;,\;\;\;
F^{B}_{\a,\m\n}=\partial_{\m}B_{\a,\n}-\partial_{\n}B_{\a,\m}
\label{c15}\ee
\be
H_{\m\n\a}=F_{\a,\m\n}^{B}-C_{\a\b}F^{A,\b}_{\m\n}-\d_{ij}Y^{i}_{\a}
F_{\m\n}^{j}
\sp
F^{A,\a}_{\m\n} = \pa_\m A^\a_\n  - \pa_\n A^\a_\m
\label{c16}\ee
\be
B_{\m\n}=\hat B_{\m\n}+{1\over
2}\left[A^{\a}_{\m}B_{\n\a}+\d_{ij}A^{i}_{\m}
A^{\a}_{\n}Y^{j}_{\a}-(\m\leftrightarrow
\n)\right]-A^{\a}_{\m}A^{\b}_{\n}B_{\a\b}
\label{c17}\ee
\be
H_{\m\n\r}=\partial_{\m}B_{\n\r}-{1\over
2}\left[B_{\m\a}F^{A,\a}_{\n\r}+A^{\a}_{\m}
F^{B}_{\a,\n\r}+ \d_{ij}A^{i}_{\mu}F^{j}_{\n\r}\right]+{\rm cyclic}
\label{c18}
\ee
$$\equiv \partial_{\m}B_{\n\r}-{1\over 2}
A^{I}_{\m} L_{IJ} F^{J}_{\n\r}+{\rm cyclic}
\pe $$
\es
Here we have extended the index $i=1,2,\cdots ,16$ to
$I=1,2,\cdots,2d+16$
to incorporate the $2d$ extra gauge fields $A_\m^\a$, $B_{\a,\m}$
coming from the metric and the antisymmetric tensor respectively.
The hat over the $B $ in (\ref{c15}), (\ref{c17}) indicates the original
components
of the ten-dimensional antisymmetric tensor.
Furthermore
\be
\phi=\Phi-{1\over 2}\log({\rm det}G)
\pe \label{c19}\ee
We will go to the Einstein frame $g=e^{2\phi/(D-2)}g_{\rm E}$ to obtain
\be
S^{\rm I}_{D,{\rm E}}=\int\sqrt{-g_{\rm E} }\left[
R-{(\partial\phi)^2\over D-2}-{e^{(D-6)\phi\over D-2}\over
12}\sqrt{G}H_{\m\n\r}H^{\m\n\r}
+{1\over 4}\partial G_{\a\b}\partial G^{\a\b}-{e^{\phi\over 2}\over 2}
G^{1\over 4}\tilde F^i_{\m\a}\tilde F^{i\m\a}
\right.
\label{c20}\ee
$$\left.
-{1\over 4}\sqrt{G}e^{\phi}H_{\m\a\b}H^{\m\a\b}
-{1\over 4}e^{-2\phi\over D-2}G_{\a\b}F^{A,\a}_{\m\n}
F^{A,\b\m\n}-{1\over 4}e^{{(D-6)\phi\over 2(D-2)}}G^{1/4}\tilde
F^i_{\m\n}
\tilde F^{i\m\n}-{1\over 4}e^{(D-4)\phi\over
(D-2)}\sqrt{G}H_{\m\n\a}H^{\m\n\a}\right]
\pe $$

Define now in the type-I context
\bs
\be
\tilde G_{\a\b}=({\rm det}~G)^{-{1\over 4}}e^{-{\phi\over 2}}~G_{\a\b}
\label{c21}\ee
\be
\tilde \phi={6-D\over 4}\phi+{2-D\over 8}\log({\rm det}~G)
\pe \label{c22}\ee
\es
Then the type-I Einstein frame action becomes identical to the
heterotic one.
Thus, the duality dictionary in $D$ dimensions is
\bs
\be
{g_{\rm E}}'=g_{\rm E}\;\;\;,\;\;\;{Y^I_{\a}}'=Y^I_{\a}\;\;\;,\;\;\;
B_{\a\b}'=B_{\a\b}
\;\;\;,\;\;\;{A^i_{\m}}'=A^i_{\m}\;\;\;,\;\;\;B_{\m\n}'=B_{\m\n}
\label{c23}\ee
\be
G'_{\a\b}=({\rm det}~G)^{-{1\over 4}}e^{-{\phi\over 2}}~G_{\a\b}
\;\;\;,\;\;\;
\phi'={6-D\over 4}\phi+{2-D\over 8}\log({\rm det}~G)
\pe \label{c24}\ee
\es
where primed indices refer to the heterotic side.

{}From now on we will set the Wilson lines to zero.
In $D=9$ we will parametrize the metric $G_{\a\b}$ in terms of the
circle
length,
$G=R^2$.
Then (\ref{c24}) implies
\be
R^2_{\rm het}={R^2_{\rm I}\over \lambda_{\rm I}}
\co \label{c25}\ee
where $\lambda_{\rm I}=e^{\Phi/2}$ is the ten-dimensional type-I coupling
constant that organizes the genus expansion.
In $D=8$ we will use the $T,U$ basis for the moduli $G_{\a\b},
B_{\a\b}$
\be
G_{\a\b}={T_2\over U_2}\left(\matrix{1&U_1\cr
U_1&|U|^2\cr}\right)\;\;\;,\;\;\;
B_{\a\b}=T_1\left(\matrix{0&1\cr -1&0\cr}\right)
\pe \label{c26}\ee
Then
\be
\left.\left.T_1\right|_{\rm het}=T_1\right|_{\rm I}
\;\;\;,\;\;\;U_{\rm het}=U_{\rm I}
\;\;\;,\;\;\;
\left.\left.T_2\right|_{\rm het}={T_2\over \l}\right|_{\rm I}
\pe \label{c28}\ee
Finally, we note that  in any dimension we have 
\be
G_{\a\b}'=e^{-\Phi/2}G_{\a\b}
\pe \label{c29}\ee

\boldmath 
\section{Elliptic genera for general $N=4$ ground states \label{elg4}}
\renewcommand{\theequation}{C.\arabic{equation}}
\setcounter{equation}{0}
\unboldmath

We will consider  here $N=4$ heterotic ground states.
The simplest case, which is considered in the text, is the one
obtained
from toroidal compactification of the $O(32)$ ten-dimensional heterotic
string.
There are, however, more general ground states with maximal supersymmetry
once
we are in less than ten dimensions. Such ground states can be
constructed as
freely acting orbifolds of the toroidally compactified theory.
In order not to reduce the supersymmetry, the orbifold group must
contain rotations that act only on the (non-supersymmetric)
right-movers
and arbitrary lattice translations.
Such $N=4$ ground states have reduced rank and can contain current
algebras
with higher levels.

In all such ground states the one-loop corrections to the $F^4$ and
$R^4$
terms
can be obtained from
\be
{\cal A}_D=t_8\int_{\cal F} {d^2\tau\over \tau_2^2}\tau_2^{(10-D)/2}
\cA(\tau,\bar\tau,F_I,R)
\co 
\label{B1}\ee
where $t_8$ is the standard tensor \cite{GSW}.
In the above formula, $\cA$ is the elliptic genus of the internal CFT, 
which
has $(c,\bar c)=(15-3D/2,26-D)$ in the presence of background gauge
fields and curvature.
The left-moving internal CFT is free (toroidal).
The elliptic genus is defined as a trace in the internal CFT
\be
\left.\cA(\tau,\bar\tau, F_I,R)\right|_{R=F=0}=Tr[(-1)^F
q^{L_0-c/24}\bar
q^{\bar L_0-\bar c/24}]_{R}
\label{B2}\ee
in the Ramond sector.
In (\ref{B1}) we are supposed to keep the terms that are fourth order
in
$R,F_I$\footnote{The index $I$ runs over all Abelian and non-Abelian
factors of the gauge group.}.
The elliptic genus obtains contributions only from ground states in the
left-moving (supersymmetric) sector. The only $\tau$ dependence comes
from the lattice sum.

In order to calculate the dependence of the elliptic genus on the
background fields we have to calculate the appropriate integrated
correlation functions
of vertex operators.
The $R$ dependence of the elliptic genus does not depend on the details
of
the $N=4$ ground state (apart from the overall normalization).
It was calculated in \cite{Schellekens} with the result
\be
{\cA(R,0)\over \cA(0,0)}=\exp\left[-{\hat E_2\over
48}tr\left({iR\over 2\pi}\right)^2-\sum_{k=2}^{\infty}{B_{2k}\over
4k(2k!)}tr\left({iR\over 2\pi}\right)^{2k}E_{2k}\right]
\co \label{B3}\ee
where $B_{2k}$ are the Bernoulli numbers.

The dependence on the field strengths can be obtained from the
associated characters of the right-moving affine algebra.
Consider the characters $\chi^a(v_i| \tau)$ of the $I$-th component of
the gauge group $G_I$,  where $a$ labels the integrable affine
representations, $i=1,2,\cdots,\,$rank $G_I$, and $v_i$ are the skew
eigenvalues of $F_{I}/4\pi^2$,
\be
tr\left({iF\over 2\pi}\right)^2=2(2\pi i)^2\sum_i v_i^2\quad,\quad
tr\left({iF\over 2\pi}\right)^4=2(2\pi i)^4\sum_i v_i^4
\pe \label{B4}\ee

The characters transform homogeneously under the modular group. In
particular (see for example \cite{GW})
\be
\chi^a({v_i/ \tau}| -1/ \tau)
=e^{i\pi k\sum_i v_i^2/\tau}\sum_{b}~S_{ab}~\chi^b(v_i | \tau  )
\co \label{B5}\ee
where $k$ is the level (a positive integer) of the associated current
algebra.
To obtain the associated traces we expand the characters as
\be
\eqalign{
\chi^a(v_i|\t) = \chi^a(\tau) & + {(2\pi i)^2\over 2!} \left(\sum_i
v_i^2\right) \chi^a_2(\tau) +
{(2\pi i)^4\over 4!} \left(\sum_i v_i^4\right) \chi_4^a(\tau) 
\cr 
& + {(2\pi i)^4\over
(2!)^2} \left(\sum_i v_i^2\right)^2 \chi_{2,2}^a(\tau) + {\cal O}(v^6)
\pe \cr } \label{B6}\ee
The above transformations imply the following behaviour for the
traces
\bs
\be
\chi^a\left(-{1\over \tau}\right)=\sum_{b}~S_{ab}~\chi^b(\tau)
\quad,\quad
\chi^a_4\left(-{1\over
\tau}\right)=\tau^4~\sum_{b}~S_{ab}~\chi^b_4(\tau)
\label{B7}\ee
\be
\chi^a_2\left(-{1\over
\tau}\right)=\sum_{b}S_{ab}\left[\tau^2\chi^b_2(\tau)
+{k~\tau\over 2\pi i}~\chi^b(\tau)\right]
\label{B8}\ee
\be
\chi^a_{2,2}\left(-{1\over
\tau}\right)=\sum_{b}S_{ab}\left[\tau^4\chi^b_{2,2}(\tau)
+{k~\tau^3\over 2\pi i} \chi^b_{2}(\tau)-{k^2~\tau^2\over
8\pi^2}~\chi^b(\tau)\right]
\pe \label{B9}\ee
\es
Thus, the $F^2$ and the $(F^2)^2$ traces are not modular-covariant.
Modifications by non-holomorphic pieces are needed. These arise in the
straightforward  evaluation of the thresholds by integrating the
singular terms in the correlator of four currents on the torus.
Another way to see their presence without invoking the regularization
prescription is to compute them in an IR-regulated background where
they come from the gravitational back-reaction \cite{KK}.
We will denote $\chi_2$ by $Q^2\chi$, $\chi_4$ by $Q^4\chi$ and
$\chi_{2,2}$ by $[Q^2]^2\chi$.
Then,
\bs
\be
Q^2\chi\to Q^2\chi-{ k\over 4\pi\tau_2}\chi
\label{B10}\ee
\be
[Q^2]^2\chi\to [Q^2]^2\chi-{k\over 4\pi\tau_2}Q^2\chi+{k^2\over
8\pi^2\tau_2^2}\chi
\pe \label{B11}\ee
\es
We also need
\bs
\be
\int {d^2z\over \tau_2}\langle \bar J^a_I(\bar z)\bar J^b_I(0)\rangle
={1\over 4}tr_I[T^aT^b]Tr\left[Q_I^2-{k_I\over 4\pi\tau_2}\right]
\label{B12}\ee
\be
\eqalign{
\int \prod_{i=1}^3{d^2z_i\over \tau_2}\langle \bar J^a_I(\bar z_1)\bar
J^b_I(0)\rangle & \langle \bar J^c_J(\bar z_2)\bar J^d_J(\bar z_3)\rangle
\cr
= & {1\over 8}tr_I[T^aT^b]tr_J[T^cT^d] 
Tr\left[\left(Q_I^2-{k_I\over 4\pi\tau_2}\right)
\left(Q_J^2-{k_J\over 4\pi\tau_2}\right)\right]
\sp I \neq J
\cr}
\label{B122} \ee  
$$
\int \prod_{i=1}^3{d^2z_i\over \tau_2}\langle \bar J^a_I(\bar z_1)\bar
J^b_I(\bar z_2)\bar J^c_I(\bar z_3)\bar J^d_I(0)\rangle=
{1\over 2}tr_I[T^aT^bT^cT^d]Tr[Q_I^4]
\;\;\;\;\;\;\;\;\;\;\;\;\;\;\;\;\; 
\;\;\;\;\;\;\;\;\;\;\;\;\;\;\;\;\; $$
\be
+{1\over 48}\left(tr_I[T^aT^b]tr_I[T^cT^d]+tr_I[T^aT^c]tr_I[T^bT^d]+
tr_I[T^aT^d]tr_I[T^bT^c]\right)
\label{B13}\ee
$$\times Tr\left[
(Q_I^2)^2-{k_I\over 4\pi\tau_2}Q_I^2+{k_I^2\over
8\pi^2\tau_2^2}\right]
\co $$
\es
where $I$ labels the gauge-group factors,
and $T^a$ are matrices in the adjoint of the gauge-group factor
$G_I$.

Putting everything together we obtain
$$
{\cA(\tau,R,F_I)\over \cA(\tau,R,0)}=1+{1\over 4}\sum_I
tr\left({iF_I\over 2\pi}\right)^2~Tr\left[Q_I^2-{ k_I\over 4\pi
\tau_2}\right]
\;\;\;\;\;\;\;\;\;\;\;\;\;\;\;\;\; 
\;\;\;\;\;\;\;\;\;\;\;\;\;\;\;\;\; 
\;\;\;\;\;\;\;\;\;\;\;\;\;\;\;\;\; 
$$
\be+
{1\over 8}\sum_{I<J}tr\left({iF_I\over
2\pi}\right)^2~tr\left({iF_J\over
2\pi}\right)^2~Tr\left[\left(Q_I^2-{k_I\over 4\pi\tau_2}\right)
\left(Q_J^2-{k_J\over 4\pi\tau_2}\right)\right]
\;\;\;\;\;\;\;\;\;\;\;\;\;\;\;\;\; 
\label{B14}\ee
$$+{1\over 16}\sum_Itr\left[\left({iF_I\over
2\pi}\right)^2\right]^2~Tr\left[
(Q_I^2)^2-{k_I\over 4\pi\tau_2}Q_I^2+{k_I^2\over
8\pi^2\tau_2^2}\right]
+{1\over 2}\sum_I tr\left({iF_I\over 2\pi}\right)^4~Tr[Q_I^4]
\co $$
where $tr$ stands for the group trace and $Tr$ stands for the
(normalized) trace in the Hilbert space relevant to the elliptic
genus.

In ten dimensions there are two choices
for the gauge group, $O(32)$ and $E_8\times E_8$ both at level one.
For the case of $O(32)_1$ the elliptic genus was calculated
in \cite{Schellekens} with the result
\be
\cA^{O(32)}(\tau,R,F)=tr\left({iF\over 2\pi}\right)^4+{1\over 2^7\cdot
3^2\cdot
5}
{E_4^3\over \eta^{24}}tr\left({iR\over 2\pi}\right)^4
\;\;\;\;\;\;\;\;\;\;\;\;\;\;\;\;\; 
\;\;\;\;\;\;\;\;\;\;\;\;\;\;\;\;\; 
\;\;\;\;\;\;\;\;\;\;\;\;\;\;\;\;\; 
\label{2222}\ee
$$+{1\over 2^9\cdot 3^2}\left[{E_4^3\over \eta^{24}}+{\hat E^2_2
E_4^2\over \eta^{24}}
-2{\hat E_2E_4E_6\over \eta^{24}}-2^7\cdot
3^2\right]\left(tr\left({iF\over 2\pi}\right)^2\right)^2+{1\over
2^9\cdot 3^2}{\hat E^2_2E_4^2\over \eta^{24}}
\left(tr\left({iR\over 2\pi}\right)^2\right)^2
$$
$$
+{1\over 2^8\cdot 3^2}\left[{\hat E_2E_4E_6\over \eta^{24}}
-{\hat E^2_2E_4^2\over \eta^{24}}\right]tr\left({iR\over
2\pi}\right)^2tr\left({iF\over 2\pi}\right)^2
\co $$
while, for $E_8\times E_8$, a direct evaluation gives
\be
\cA^{E_8\times E_8}(\tau,R,F)={1\over 2^7\cdot 3^2\cdot
5}{E_4^3\over \eta^{24}}tr\left({iR\over 2\pi}\right)^4+{1\over
2^9\cdot 3^2}{\hat E^2_2E_4^2\over \eta^{24}}
\left(tr\left({iR\over 2\pi}\right)^2\right)^2
\label{2999}\ee
$$+{1\over 2^9\cdot 3^2}\left({(\hat E_2^2E_4-2\hat E_2
E_6+E_4^2)E_4\over \eta^{24}}\right) \left[\left(tr\left({iF_1\over
2\pi}\right)^2\right)^2+\left(tr\left({iF_2\over
2\pi}\right)^2\right)^2\right]
$$
$$-{\hat E_2 E_4(\hat E_2E_4-E_6)\over 2^8\cdot
3^2~\eta^{24}}tr\left({iR\over 2\pi}\right)^2\left[tr\left({iF_1\over
2\pi}\right)^2+tr\left({iF_2\over 2\pi}\right)^2\right]
+{(\hat E_2E_4-E_6)^2\over 2^8\cdot 3^2~\eta^{24}}tr\left({iF_1\over
2\pi}\right)^2tr\left({iF_2\over 2\pi}\right)^2
\co $$
where $F_{1,2}$ are the field strengths of the first, respectively
second
$E_8$.

Upon toroidal compactification to $D$ dimensions
the above formulae have to be multiplied by the $10-D$ toroidal
lattice
sum.

\section{Properties of the (2,2) lattice \label{lpr} }
\renewcommand{\theequation}{D.\arabic{equation}}
\setcounter{equation}{0}

The (2,2) lattice sum can be written as
\be
\Gamma_{2,2}(T,U)=\sum_{m_1,m_2,n_1,n_2\in Z}~q^{p_l^2/2}~\bar
q^{p_r^2/2}
\label{z13}\ee
where
\be
{1\over 2}p_r^2={|-m_1U+m_2+T(n_1+n_2U)|^2\over 4T_2U_2}
\sp 
{1\over 2}p_l^2={1\over 2}p_r^2+m_1n_1+m_2n_2
\pe \label{z15}\ee
Define the following ``momenta"
\be
p=m_2+Tn_1+U(-m_1+Tn_2)\;\;\;,\;\;\;q=m_2+Tn_1+\bar U(-m_1+Tn_2)
\pe \ee
Then we can write the lattice sum as
\be
\Gamma_{2,2}(T,U)=\sum e^{2\pi i\tau(\vec m\cdot\vec n)-
{\pi \tau_2\over T_2U_2}|p|^2}=\sum e^{2\pi i\bar
\tau(\vec m\cdot\vec n)-{\pi \tau_2\over T_2U_2}|q|^2}
\pe \ee
We also define the generalized lattice sums
\be
\langle p^{M_1}\bar p^{M_2}q^{N_1}\bar q^{N_2}\rangle\equiv
\sum
p^{M_1}\bar p^{M_2}q^{N_1}\bar q^{N_2}e^{2\pi i\tau
(\vec m\cdot\vec n)-{\pi \tau_2\over T_2U_2}|p|^2}
\pe \ee
In this notation, $\Gamma_{2,2}=\langle 1\rangle$.
Finally we define the (rescaled) covariant derivatives
\be
D^a_{u}=\partial_u-{ia\over u_2}\;\;\;,\;\;\;
D^a_{\bar u}=\partial_{\bar u}+{ia\over u_2}
\pe \ee
\def\ll{\langle 1\rangle}

Then, we can derive the following identities
\be
\square_T (\tau_2 \Gamma_{2,2}(T,U))=\square_U (\tau_2
 \Gamma_{2,2}(T,U))=
\tau_2^2\pa_{\tau}\pa_{\bar\tau}(\tau_2\Gamma_{2,2}(T,U))
\label{z10}\ee
with $\square_T\equiv T_2^2\partial_T\pa_{\bar T}$.
We also have
\bs
\be
D^{N-1}_UD^{N-2}_U\cdots D^0_U~\ll=\left({\pi\tau_2\over
2iT_2U_2^2}\right)^N
\langle \bar p^N q^N\rangle
\ee
\be
D^{N-1}_{\bar U}D^{N-2}_{\bar U}\cdots D^0_{\bar U}~\ll=
\left({\pi\tau_2\over -2iT_2U_2^2}\right)^N
\langle p^N \bar q^N\rangle
\ee
\be
D^{N-1}_TD^{N-2}_T\cdots D^0_T~\ll=\left({\pi\tau_2\over
2iT_2^2U_2}\right)^N
\langle \bar p^N \bar q^N\rangle
\ee
\be
D^{N-1}_{\bar T}D^{N-2}_{\bar T}\cdots D^0_{\bar T}~\ll=
\left({\pi\tau_2\over -2iT_2^2U_2}\right)^N
\langle p^N q^N\rangle
\ee
\be
(D^{N-1}_UD^{N-2}_U\cdots D^0_U)~(D^{N-1}_TD^{N-2}_T
\cdots D^0_T)~(\tau_2\ll)=
\left({i\pi\over 2T_2^2U_2^2}\right)^N(\tau_2^2
\partial_{\tau})^N\left(\tau_2
\langle \bar p^{2N}\rangle\right)
\ee
\be
(D^{N-1}_UD^{N-2}_U\cdots D^0_U)~(D^{N-1}_{\bar T}
D^{N-2}_{\bar T}\cdots D^0_{\bar T})~(\tau_2\ll)=
\left({i\pi\over 2T_2^2U_2^2}\right)^N(\tau_2^2
\partial_{\bar\tau})^N
\left(\tau_2
\langle q^{2N}\rangle\right)
\ee
\be
(D^{N-1}_{\bar U}D^{N-2}_{\bar U}\cdots D^0_{\bar U})~
(D^{N-1}_TD^{N-2}_T
\cdots D^0_T)~(\tau_2\ll)=
\left({i\pi\over 2T_2^2U_2^2}\right)^N(\tau_2^2
\partial_{\bar \tau})^N\left(\tau_2
\langle \bar q^{2N}\rangle\right)
\label{z50}\ee
\be
(D^{N-1}_{\bar U}D^{N-2}_{\bar U}\cdots D^0_{\bar U})~
(D^{N-1}_{\bar T}D^{N-2}_{\bar T}\cdots D^0_{\bar T})~
(\tau_2\ll)=
\left({i\pi\over 2T_2^2U_2^2}\right)^N(\tau_2^2
\partial_{\tau})^N
\left(\tau_2
\langle p^{2N}\rangle\right)
\pe \ee
\es
Also note  that in the above
\be
(\tau_2^2\pa_{\bar\tau})^N=\tau_2^{2N}D^{N-1}_{\bar\tau}
D^{N-2}_{\bar\tau}
\cdots
D^0_{\bar\tau}
\ee
and finally we  give the identity
\be
\pa_{\bar\tau}(\tau^2\pa_{\bar\tau})^N~D^N_{\tau}~
\Phi_{N}(\tau)\sim D^{N+1}_{\bar\tau}D^N_{\tau}\Phi_N(\tau)=0
\pe \ee

\newpage
\section{One-loop threshold integrals \label{1li} }

\subsection{Calculation of one-loop threshold integrals \label{c1li} }
\renewcommand{\theequation}{E.\arabic{equation}}
\setcounter{equation}{0}

In this appendix we compute the following two families of fundamental
domain
integrals
\bs
\be
I_{\n} (T,U) =\int_{\cal{F}}
{\rd^2 \t \over \t_2 } \left(   \Gamma_{2,2}(T,U) \hat
E_2^{\n}(\t) \Phi_{\n}(q)- c_0^{(\n)}\right)
\;\;\;\;\;\;\;\;\;\;\;\;\;\;\;\;\;\; 
\; \;\;\;\;\; \;\;\;\;
\label{d1}
\ee
\be
I_{\n} (y) =\int_{\cal{F}}
{\rd^2 \t \over \t_2 } \left( \Ga_{S+2,2} (y)  \hat
E_2^{\n}(\t) \Phi_{\n} (q)- d_0^{(\n)}\right)
\sp S= 8, 16 \co \label{win}
\ee
\es
where $\t = \t_1 + i \t_2$ is the complex modulus of the torus,
$q = e^{ 2 \p i \t}$, ${\cal F}$ is
the fundamental domain of $SL(2,Z)$ and $\n$ is an arbitrary
non-negative integer. For $\n=0,1$ these integrals were computed in
Ref. \cite{hm2}, which we will closely follow in notation and
computational
method. In the case of the integral in (\ref{win}), we will first keep
the results general for all $S$, but be more specific in explicit
expressions for the case $S=16$ with the $SO(32)$ lattice,
which is the one that has applications to the body of the paper.

\ni \underline{(2,2) case}
\vs .1cm
We first present the calculation of (\ref{d1})
in some detail.
The integrand involves the (2,2) lattice sum
\bs
\be
\Gamma_{2,2} (T,U)  = \sum_{p_l,p_r}
q^{p_l^2/2} \bq^{p_r^2/2}
\;\;\;\;\;\;\;\;\;\;\;\;\;\;\;\; 
\;\;\;\;\;\;\;\;\;\;\;\;\;\;\;\;\;\; 
\;\;\;\;\;\;\;\; \;\;\;\;\; 
\ee 
\be
\;\;\;\;\;\; \;\;\;\;\;\;\;\;\;\;\;\;\;\;\;\; 
 = \frac{1}{\t_2} \sum_{A \in Mat_{2 \ti 2} }
e^{ - 2\pi i T {\rm det}A }
\exp[ - {\p T_2 \over \t_2 U_2} | (1 \;\; U) A \left( {\t \atop 1}
\right)|^2 ]
\label{d2}
\ee
\be
p_r^2={|-m_1U+m_2+T(n_1+n_2U)|^2\over 2T_2U_2}
\sp
p_l^2 - p_r^2= 2( m_1n_1+m_2n_2)
\co \ee
\es
for which we use the second (Poisson resummed) form (\ref{d2}) in the
computations below.
Here $\hE_2$ is  as defined in (\ref{a6}), and
$\Phi_{\n}$ is a modular form of weight $- 2 \n$, which is holomorphic
everywhere except for a first-order pole at infinity; its Laurent
series
is given by,
\be
\Phi_{\n} (q) =\sum_{n = -1}^\infty ~c_n~q^n \pe
\label{d3}\ee
We also define, for any non-negative integer $s$, the power series
\be
E_2^s  (q)
\Phi_{\n} (q)
=\sum_{n = -1}^\infty ~c_n^{(s)}~q^n
\co \label{d4} \ee
so that in particular $c_n^{(0)} = c_n$ and the second term
in (\ref{d1}) proportional to $c_0^{(\n)}$ is chosen as to cancel the
IR-divergent part of the first term.

To evaluate this integral we use the method of orbits \cite{DKL},
splitting up
the integral in the sum of three terms,
$I_\n = \sum_{i=1}^3 I_\n^{(i)}$, corresponding to the trivial,
non-degenerate
and degenerate orbits, respectively, for which we outline the
computations
below.

\ni {\it Trivial orbit}.
In this case $A=0$ in (\ref{d2}) and the result is known
\cite{Schellekens} to be
\be
\eqalign{
I_\n^{(1)}=
 T_2 \int_{\cal{F}}
{\rd^2 \t \over \t_2^{2} } {\hat E_2^{\n} (\t) \Phi_{\n}} (q)
& ={\pi~T_2\over 3(\n+1)}
[E_2^{\n+1} (q) \Phi_{\n} (q) ]\vert_{{\rm coeff.~ of~}q^0}
\cr
& ={ \p T_2 \over 3 (\n+1) } [ c_0 - 24 (\n+1) c_{-1} ]
\pe \cr}  \
 \label{d5}
\ee

\ni {\it Non-degenerate orbit}.
Here, the representative matrices are
\be
A_0 =\left( \matrix{ k & j \cr 0 & p \cr}\right)
\sp 0 \leq j < k  \sp p \neq 0
\label{d6}\ee
and the integral unfolds over the double cover of the upper half-plane.
Expanding $\hE_2^\n$ we have
\be
I_\n^{(2)} = \sum_{s=0}^\n \ch{\n}{s} \left( - {3 \over \p} \right)^s
I_{\n,s}^{(2)}
\co \label{d7}\ee
where
\be
I_{\n,s}^{(2)}
= 2 T_2 \sum_{n=-1}^{\infty}  \sum_{ j \leq 0 < k \atop p \neq 0 }
\int_{- \infty}^{\infty} \rd \t_1 \int_0^\infty {\rd \t_2 \over
\t_2^{2+s}}
e^{ - 2 \p i T k p }
 e^{ - {\p T_2 \over \t_2 U_2} | k \t + j + p U |^2 }
c_n^{(\n-s)}  e^{ 2  \p i \t n } \pe
\label{d77}\ee
We first do the Gaussian integral over $\t_1$, with the result: 
\be
\eqalign{
\;\;\;\; I_{\n,s}^{(2)} = 2 \sqrt{T_2 U_2}
\sum_{n=-1}^{\infty}  \sum_{ j \leq 0 < k \atop p \neq 0 }
 \int_0^\infty \rd \t_2
& {1 \over \t_2^{3/2+s}}  \frac{1}{k}
e^{ - 2 \p i T kp }
e^{- {\p T_2 \over \t_2 U_2} ( k\t_2 + p U_2 )^2 }
\cr
& \ti e^{ -{2 i \p \over k} (j + pU_1) n - \p n^2 {\t_2 U_2 \over
T_2 k^2} }
c_n^{(\n-s)} e^{- 2 \p \t_2 n }\pe \cr} \
\label{d8}\ee
Then we do the  $j$ summation, using the identity
\be
\sum_{n=-1}^{\infty} \sum_{j=0}^{k -1}
 e^{ -{2 \p i n j \over k} } f(n)
=
\sum_{l=-1}^{\infty} \sum_{b=0}^{k-1}  \sum_{j=0}^{k -1}
 e^{ -{2 \p i  b  j \over k} } f(kl +b )
=
\sum_{l=-1}^{\infty} \sum_{b=0}^{k-1}  k \d_{b,0}
 f(kl +b )
=
k \sum_{l=-1}^{\infty}  f( kl )
\co \label{d9}\ee
after which we use
$ \sum_{p \neq 0} g(p) = \sum_{p > 0} [ g(-p) + g (p) ] $ and
we find
\be
I^{(2)}_{\n,s} =
 4 {\rm Re}  \sqrt{T_2 U_2}
\sum_{l=-1}^\infty   \sum_{k,p =1 }^\infty
 \int_0^\infty \rd \t_2
\frac{1}{\t_2^{3/2+s}}
e^{  2 \p i (T k + U_1 l) p  }
e^{ - {\p T_2 \over \t_2 U_2} ( k\t_2 - p U_2 )^2 }
 e^{ - \p \t_2 (2 l k + { U_2 l^2   \over  T_2 }) }
   c^{(\n-s)}_{  kl}
\pe \label{d10}\ee

To do the $\t_2$ integral we use
\be
\int_0 ^{\infty} \rd x \frac{1}{x^{1-\l}} e^{- c x - b/x}  =
 2 \left( {b \over c} \right)^{\l/2}
K_\l ( 2 \sqrt{ b c })
\sp {\rm Re}\,b, {\rm Re}\,c > 0
\co \label{d11}\ee
where the Bessel function $K_\l$ is given by
\be
K_{n+1/2} ( x)
= \sqrt{ \p \over 2x} e^{-x}  \sum_{r=0}^n { (n+r) ! \over r !
(n-r)! (2x)^r }
\sp
K_{-n} ( x)
= K_{n} ( x) \pe
\label{d12}\ee
If the moduli of the torus are in the fundamental chamber
 $T_2 > U_2$, we then obtain the result
\be
I^{(2)}_{\n,s} = 4 {\rm Re} {1 \over (T_2 U_2)^s   }
\sum_{l=-1}^{\infty}  \sum_{k,p =1}^\infty
 (q_T^{k} q_U^l)^p
   \sum_{r=0}^s { (s+r) ! \over r ! (s-r)! (4 \p)^r}
 (T_2k +U_2 l)^{s-r} \frac{1}{ p^{s + r +1}}
c^{(\n-s)}_{ kl}
\label{d13} \ee
where we have defined
\be
q_T = e^{ 2 \pi i T } \quad \sp \quad  q_U = e^{ 2 \pi i U } \pe
\label{d14}\ee
On the other hand, when $U_2 > T_2$, we find the same result with
$T$ and $U$ interchanged. We will generally assume that the moduli are
in the fundamental chamber, unless specified otherwise.

To further simplify the expression (\ref{d13}), we evaluate the
$p$-sum by using the polylogarithm functions  defined in (\ref{d15}),
giving
\be
I^{(2)}_{\n,s} = 4 {\rm Re} {1 \over (T_2 U_2)^s   }
\sum_{l=-1}^{\infty}  \sum_{k > 0}
   \sum_{r=0}^s { (s+r) ! \over r ! (s-r)! (4 \p)^r}
 (T_2k +U_2 l)^{s-r} Li_{s + r +1}
(q_T^k q_U^l)
c^{(\n-s)}_{ kl }
\pe \label{d16} \ee
Finally, using the definition (\ref{d18}) of the combined
 polylogarithm function $L_{(s)}$, we
conclude that the total contribution of the non-degenerate
orbits
to the integral is  given by
\be
I_\n^{(2)} =
4 {\rm Re} \sum_{s=0}^\n \ch{\n}{s} \left(  {- 3 \over \p T_2 U_2 }
\right)^s
\sum_{l=-1}^{\infty}  \sum_{k =1}^\infty
L_{(s)} (Tk + Ul)
c^{(\n-s)}_{ kl }
\pe \label{d17}
\ee

\ni {\it Degenerate orbit}.
For the degenerate orbits the representative matrices are
\be
A_0 =\left( \matrix{ 0 & j \cr 0 & p \cr}\right)
\sp (j,p) \neq (0,0)
\co \label{d19}\ee
where $j,p$ run over both positive and negative integers to account for
the
double covering, and the integration extends over the strip.

In this case we need to compute
\be
I_\n^{(3)} = \sum_{s=0}^\n \ch{\n}{s} \left( - {3 \over \p} \right)^s
I_{\n,s}^{(3)}
\co \label{d20}
\ee
where
\be
I_{\n,s}^{(3)}
 =
\int_{- 1/2}^{1/2} \rd \t_1 \int_0^\infty {\rd \t_2 \over
\t_2^{2+s} } \left[
T_2  \sum_{n=-1}^\infty \sum_{(j,p)\neq(0,0)}
e^{ -  {\p  T_2 \over  \t_2 U_2 } | j + p U |^2 }
c_n^{(\n-s)} e^{2 \p i \t n} -
c_0^{(\n)}  \d_{s,0} \d (\t \in {\cal F}) \t_2 \right] \pe
\label{d21} \ee
For $s=0$ we need to regularize the integral, and following \cite{DKL}
we  multiply the integrand by the regulator  $(1-e^{-N/\t_2})$ in
this case, taking the limit $N \ra \infty$ after evaluation of the integral. 
To keep the computation below uniform for all $s$, we use the fact that
the above prescription effectively amounts to omitting the constant
term in the integrand and replacing in the end
\be
\sum_{p=1} \frac{2}{p} = 2 \zeta (1) \ra - [ \log T_2 U_2  + \K]
\sp \K \equiv \log {8 \p e^{1-\g_E} \over 3 \sqrt{3}}
\label{d22}
\ee
where $\g_E$ is the Euler--Mascheroni constant.

So we focus on the first term in (\ref{d21}) and, after performing
the trivial $\t_1$ integration and subsequently the standard
$\t_2$ integration, we arrive at
\bs
\be
 I_{\n,s}^{(3)}   =   c_0^{(\n-s)} T_2  s !
\left( {  U_2 \over \p T_2 } \right)^{s+1}
 \sum_{(j,p)\neq(0,0) }
{1 \over | j + p U |^{2 (1+s)} }
\;\;\;\;\;\;\;\;\;\;\;\;\;\;\;\;\;\;\;\;  
\;\;\;\;\;\;\;\;\;\;\;\;\;\;\;\;\;\;\;\;  \;\;\;\;\;\;\;\;\;\;\;  
\label{d23}
\ee
\be
\;\;\;\;\;\;\;\; 
 =   c_0^{(\n-s)} T_2  s ! \left( {  U_2 \over \p T_2 } \right)^{s+1}
 \left(2 \sum_{j =1}^\infty  \frac{1}{j^{2(1+s)}} +
\sum_{j = -\infty}^\infty \sum_{ p \neq 0 }
{1 \over [ (j + p U_1)^2   +(p U_2 )^2]^{(1+s)} } \right) \pe
\label{d24} \ee
\es

For the first term in (\ref{d24}) we use the standard identity
\be
  \sum_{j=1}^\infty \frac{1}{j^{2m}} = \zeta (2m )
\sp \zeta(2m) = (-)^{m +1} {2^{2m -1} \p^{2m} \over (2m)! } B_{2m}
\label{d25} \ee
where $B_m$ are the Bernoulli numbers. For the explicit examples in
the text, the relevant values are $B_2=1/6$, $B_4=-1/30$ and $B_6=1/42$.

To evaluate the second term in (\ref{d24}) we use the identities
\bs
\be
\sum_{j= - \infty}^{\infty} {1 \over (j+B)^2  + C^2  }
= {i \p \over 2 C} [ \cot  \p (B+iC) - \cot  \p (B-iC) ]
\label{d26}
\ee
\be
\sum_{p= 1}^\infty
 {1 \over p^s}
{ q_U^{p} \over 1 - q_U^{p} }
= \sum_{l =1}^\infty  Li_s(q_U^l)
\label{d27}
\ee
\be
\left( \frac{1}{U_2} {\pa \over \pa U_2} \right)^s
 \frac{1}{U_2}  Li_{m} (q_U^l)
= {(-)^s  (2 \p)^s \over U_2^{2\n+1} }
   \sum_{r=0}^s { (s+r) ! \over r ! (s-r)! (4 \p)^r}
 (U_2  l )^{s-r}
 Li_{m + r - s } (q_U^l )
\sp m \geq s
\pe \label{d28} \ee
\es
The identity (\ref{d28}) may be derived 
 by recursion, using in particular that
$ {\pa \over \pa U_2} Li_s (q_U^l)\  = - 2 \p l
Li_{s-1} (q_U^l)$.
Then we can rewrite
\be
\eqalign{
\sum_{j=-\infty}^\infty \sum_{p \neq 0 }
& {1 \over [ (j + p U_1)^2   +(p U_2 )^2]^{(1+s)} }
= \sum_{p \neq 0 }
{ (-1)^s \over s ! } {1 \over p^{2 s} }
\left( \frac{1}{2U_2} {\pa \over \pa U_2} \right)^s
\sum_{j=-\infty}^\infty
{1 \over  (j + p U_1)^2   +(p U_2 )^2 }
\cr
& = { (-1)^s \p \over s ! }
\left( \frac{1}{2U_2} {\pa \over \pa U_2} \right)^s
\sum_{p =1}^\infty
{1 \over p^{2 s} }
{2 \over U_2} {1 \over p}
\left[{ q_U^{p} \over 1 - q_U^{p} }
+ {\bq_U^{p} \over 1 - \bq_U^{p} } +1 \right]
\cr
 & = { (-1)^s \p \over 2^s s ! }
\left( \frac{1}{U_2} {\pa \over \pa U_2} \right)^s
\left( \frac{1}{U_2} 4 {\rm Re} \sum_{l=1}^{\infty} Li_{2s+ 1} (q_U^l)
+ \frac{2}{U_2} \zeta (2 s +1) \right)
\cr
& =
4 {\rm Re}
{\p^{s+1} \over U_2^{2s+1} } \sum_{l=1}^\infty
L_{s}(Ul) + {2\p (2s)! \over (s!)^2 4^s }
{\zeta (2s +1) \over U_2^{2s+1} } \pe 
\cr}
\label{d29} \ee
Here we have used (\ref{d26}) and some rearrangement in the second
step,
(\ref{d27}) in the  third step; the last step uses (\ref{d28}), 
along with the definition of $L_{(s)}$ in (\ref{d18}).

Inserting the results  (\ref{d29}), (\ref{d25}), (\ref{d24})
in (\ref{d23}), we can write the total result for (\ref{d21})
as
\be
\eqalign{
I_{\n}^{(3)}& =
 4 {\rm Re} \sum_{s=0}^\n
\ch{\n}{s}
\left({ -3 \over \p T_2 U_2 }\right)^s
\sum_{l =1}^{\infty}  L_{(s)} ( Ul  ) c_{0}^{(\n -s)}
 -c_0^{(\n)} [ \log T_2 U_2 + {\cal K} ] + { \p U_2 \over 3  } c_0^{(\n
)}
\cr
+& \sum_{s=1}^\n
c_0^{(\n-s)} \ch{\n}{s}
\left\{ { 4 \p (12)^{s} s ! B_{2s +2} \over (2s+2) ! }
{U_2^{s+1} \over T_2^s }
+ 2  {(2s)! \over s!} \left( {3 \over 4\p^2 T_2U_2 }\right)^s
\zeta (2 s +1)   \right\}
\co \cr }
\label{d30}
\ee
which completes the calculation of the degenerate orbit.

\ni {\it Total result}.
Adding the three expressions (\ref{d5}), (\ref{d17}) and (\ref{d30}), we
obtain our final result for the fundamental domain integral in
(\ref{d1}): 
\be
\eqalign{
I_\n (T,U) & =
4 {\rm Re} \sum_{s=0}^\n \ch{\n}{s} \left(  {- 3 \over \p T_2 U_2 }
\right)^s
\sum_{k,l}'
L_{(s)} (Tk + Ul)
c^{(\n-s)}_{ kl }
\cr
&
-c_0^{(\n)} [ \log T_2 U_2 + {\cal K} ]
+ { \p T_2 \over 3 (\n +1) } [c_0^{(0)} - 24 (\n +1) c_{-1}^{(0)} ]
+ { \p U_2 \over 3  } c_0^{(\n )}
\cr
& +\sum_{s=1}^\n
c_0^{(\n-s)} \ch{\n}{s}
\left\{ { 4 \p (12)^{s} s ! B_{2s +2} \over (2s+2) ! }
{U_2^{s+1} \over T_2^s }
+ 2  {(2s)! \over s!} \left( {-3 \over 4\p^2 T_2U_2 }\right)^s
\zeta (2 s +1)   \right\}
\cr}\
\label{d31}
\ee
where
\be
\sum_{k,l}'  \equiv
 \sum_{k,l=0 \atop (k,l) \neq (0,0)}^\infty
+ \;(\;\;) \vert_{(k,l)=(1,-1)}
\co \label{d32}\ee
and $L_{(s)}$ is the combined polylogarithm function defined in
(\ref{d18}). This expression is valid in the fundamental chamber
$T_2 > U_2$, while for $U_2 > T_2$ we obtain the same result with
$T$ and $U$ interchanged.
For the special cases $\n=0$ and 1, with  
$\Phi_0 = E_4^3/\et^{24} $ and $\Phi_1 = E_4 E_6/\et^{24}$, 
respectively, 
the expression (\ref{d31}) agrees with
that obtained in the appendix of \cite{hm2}.

\ni \un{$(S+2,2)$ case}
\vs .1cm

We next evaluate the integrals in (\ref{win}), which involve
the $(S+2,2)$ lattice sum: 
\be
\Ga_{S+2,2} (y)   =  \sum_{p_l,p_r }
 q^{p_l^2/2} \bq^{p_r^2/2}
\co \label{wil} \ee
where our notations and conventions are as follows.
The $(S+2,2)$ lattice is
obtained by an $SO(S+2,2)$ rotation of some standard lattice, which we
take
to be of the form $\Ga^{S,0} \oplus \Ga^{2,2}$. Here,  $\Ga^{S, 0}$ is
the $S=8$ or 16-dimensional, even self-dual Euclidean lattice, i.e.
either the
$E_8$ root lattice or the
$E_8 \times E_8$ root lattice or the ${\rm Spin} (32) /  \Z_2$
weight lattice. For $\Ga^{2,2}$ we use the conventions of the (2,2)
case
discussed above. A general lattice vector is denoted by
\be
\ell \in \Ga^{S+2,2}: \quad
\ell =  (\br,\vec{n},\vec{m}) \sp \br \in \Ga^{S,0}
\sp
(\vec{n},\vec{m})  \in \Ga^{2,2}
\ee
so that barred vectors are the components in $\Ga^{S,0}$.
The complex moduli $y$ are parametrized as in (\ref{o2m}) so that
\be
y = (\by, T, U )
\sp (y,y) = \by \cdot \by - 2 T U
\sp (y_2,y_2) = \bar{y}_2 \cdot \bar{y}_2 - 2 T_2 U_2
\co \ee
with $\by$ an $S$-dimensional complex vector.
The subscript ``2'' on the moduli denotes the imaginary part as usual,
and
we have the restrictions that $U_2 > 0$ and $(y_2 , y_2) < 0 $.
In these coordinates, the left- and right-moving components of $p \in
\Gamma^{S+2, 2}$ are given by
\bs
\be
 p_r^2 = {1 \over - (y_2, y_2)} \left| \bar{r} \cdot \bar{y} +
m_1 U + n_1 T  - m_2 -  {1 \over 2} n_2 (y, y) \right|^2
\ee
\be
 p_l^2 -  p_r^2  =  \bar{r} \cdot \bar{r} - 2 m_1 n_1 - 2 m_2 n_2
\pe \ee
\es

After a Poisson resummation in $m_1,m_2$, the lattice sum (\ref{wil})
takes the alternate form
\be
\Ga_{S+2,2} ( y )
= {-(y_2, y_2) \over 2 \tau_2 U_2 }  \sum_{\bar{r} \in \Ga^{S,0} }
\sum_{A}
q^{{1 \over 2} \bar{r} \cdot \bar{r}} e^{{\cal  G}(A,\t) }
\label{wip} \ee
where
\bs
\be
\eqalign{
{\cal G}  (A,\t)
=  &  { \p (y_2, y_2) \over 2 (U_2)^2 \tau_2} | {\cal A} |^2 - 2 \pi
i T
\det A + {\p \over U_2} \left( \bar{r} \cdot \bar{y} \tilde{\cal A}  -
\bar{r} \cdot \bar{y}^* {\cal A} \right) \cr
& - {\p n_2 \over 2 U_2}  \left( \bar{y} \cdot \bar{y} \tilde{\cal A} -
\bar{y}^* \cdot \bar{y}^* {\cal A} \right) +
{i \p  \bar{y}_2 \cdot \bar{y}_2
\over
(U_2)^2} \left( n_1 + n_2  U^*  \right) {\cal A}
}
\label{c9a} \ee
\be
A = \pmatrix{n_1   & m_1   \cr n_2  & m_2  \cr}
\sp
 {\cal A}  = (1  \;\,U) A \pmatrix{\tau \cr  1}
\sp \tilde{\cal A}  = (1  \;\,U^*) A \pmatrix{\tau \cr  1}
\pe \label{c9b} \ee
\es

For completeness we also give an alternative form of the expression
(\ref{wip}): 
\be
\label{c10}
\eqalign{
\Ga_{S+2,2} (G,B,Y)
& 
= { \sqrt{ {\rm det} G} \over \t_2}
\sum_{\vec{m},\vec{n} }
e^{ - {\p \over \t_2} (m^I +n^I \t) (G+B)_{IJ} (m^J +n^J \bar{\t} ) }
\;\;\;\;\;\;\;\;\;\;\; 
 \cr & 
\;\;\;\;\;\;\;\;\;\;\; 
\ti \frac{1}{2} \sum_{a,b=0,1}
\prod_{i=1}^{S}
e^{ - i \p [ n^I Y_I^i Y_J^i m^J + b  n^IY_I^i] }
\th \left[ {}^{ a + 2 n^I Y_I^i  }_{ b + 2 m^I Y_I^i}\right]
\co \cr }
\ee
where the $\th$-function is defined in (\ref{the}).
Here, the connection between the real moduli $G,B,Y$
in the form (\ref{c10}) and the complex moduli $y= (\by,T,U)$ in
(\ref{wip}) is as follows,
\bs
\be
G =  {-(y_2,y_2) \over 2 U_2^2 } \pmatrix{ 1 & U_1 \cr U_1 & |U|^2 \cr}
\sp B_{12} = T_1 - { \by_1 \cdot \by_2 \over 2 U_2}
\ee
\be
y^i= (y_1 + i y_2)^i   = - Y_2^i + U Y_1^i
\pe \ee
\label{c12} \es
To check the equivalence between the expressions (\ref{wip}) and
(\ref{c10}), one
uses eq. (\ref{the}) and the relations in (\ref{c12}).

The modular properties under $\t$  are most easily derived from
(\ref{wip}) or
 (\ref{c10})
and we find that  $ \Ga_{S+2,2}$ is of weight $S/2$.
The lattice sum is also properly invariant
under the $O(S+2,2,Z)$ transformations (\ref{o2t}) of  the moduli.
For modular invariance of the integrand in (\ref{win}), the function
$\Phi_\n $
transforms
with weight $-S/2-2\n$, and we assume the same expansion as in
(\ref{d3}),
(\ref{d4}) for this function.

It turns out that since the lattice is even self-dual,
the
contribution to (\ref{wip}) from two matrices  $A$ that are related by
a modular
transformation is again, as in the (2,2) case, given by a modular
transformation on $\t$. As a consequence we can use the method of
orbits as above. We omit the details of the calculations, which
are similar to the ones given for the (2,2) case, 
 and generalize those in \cite{hm2}, but
only give the final result.

To write the total result we introduce the following notation \cite{hm2}.
The triplet $r = (\bar{r}, -l, -k)$ is positive if
\be
 k > 0 \;\;\;\;\; {\rm or}
\;\;\;\;\;
 k = 0, \;\; l > 0 \;\;\;\;\; {\rm or}
\;\;\;\;\;  k = l = 0, \;\; \bar{r} > 0 .
\ee
and we use the  definition
\be
d^{(s)} ( r ) \equiv   c^{(s)}_ {- {1 \over 2} (r, r) }
\sp
(r,r) =  \br \cdot \br - 2 kl
\co \label{dde}
\ee
where the coefficients $c_n^{(s)} $ are as in (\ref{d4}).
We will also use the functions in (\ref{dde}) with argument $\bar{r}$
instead of $r$, meaning that $k = l = 0$.
For example, the coefficient of the second term in (\ref{win})
(which subtracts the divergent part) is
\be
d_0^{(\n)} =
 \sum_{\bar{r}  \atop \bar{r} \cdot \bar{y} = 0} d^{(\n)} (\bar{r} )
\pe \ee
We define the product $(r ; y)$ as
\be
(r ; y) = \cases{
\bar{r} \cdot \bar{y}_1 + l U_1 + k T_1 + i \left| \bar{r} \cdot
\bar{y}_2
+ l U_2 + k T_2 \right| & for $k > 0$ \cr
\bar{r} \cdot \bar{y} + l U - \left[ {\bar{r} \cdot \bar{y}_2 \over
U_2}
\right] U & for $k = 0$, $\bar{r} \geq 0$ \cr
\bar{r} \cdot \bar{y} + l U + \left[ -{\bar{r} \cdot \bar{y}_2 \over
U_2}
\right] U & for $k = 0$, $\bar{r} < 0$ , \cr
}
\ee
where $[x]$ is the greatest integer smaller than or equal to $x$.

Then, we have the following result for the threshold including
Wilson lines
\be
\label{hmr}
\eqalign{
I_\n (y)
 =  4 {\rm Re} \sum_{s=0}^\n & \ch{\n}{s} \left( { 6 \over \p (y_2,y_2) }
\right)^s
\sum_{r > 0}'
L_{(s)} \left( (r;y) \right) d^{(\n-s)} (r)
\cr 
+ d_0^{(\n)} & 
\left( - \log \bigl(-(y_2, y_2) \bigr) - {\cal K} \right)
- { (y_2,y_2) \over 2 U_2} {\p \over 3 (\n +1) } [ E_2^{\n+1} \chi
\Phi_\n ]
\vert_{q^0}
\cr 
 + & \sum_{s=1}^\n   d_0^{(\n-s)}   \ch{\n}{s}
 {2 (2s)! \over s!} \left( {3 \over 2 \p^2 (y_2,y_2) } \right)^s
 \zeta (2s +1)
\cr 
& +  2 {\rm Re} \sum_{s=0}^\n \ch{\n}{s}
\left( { 6 \over \p  (y_2,y_2) } \right)^s
 {  U_2^{2s+1} s! \over \p^{s+1}   }
\sum_{\br }
Li_{2s+2} \left( e^{ 2 \pi i \br \cdot \by_2 / U_2 } \right)
d^{(\n-s)} (\br)
\co 
\cr} 
\ee  
where $L_{(s)}$ is defined as in (\ref{d18}), and
${\cal K}$ is given in (\ref{d22}).
 The prime on the sum over $r > 0$
 indicates that terms with $k = l = 0$ and
$\bar{r} \cdot \bar{y} = 0$ for generic values of the moduli are
omitted.

Further simplifications of this expression occur when the moduli are in
the (generalized) fundamental Weyl chamber \cite{hm2}
\bs
\be
0 < { \br \cdot \by_2 \over U_2 } < 1 \sp {\rm for}\;\,
\br > 0 \sp \br \cdot \br \leq 2
\ee
\be
0 < U_2 < T_2
\co \ee
\es
which means that $(r;y)=(r,y) = \br \cdot \by + l U + k T$ for all
$r$ such that $ -\frac{1}{2} (r,r) \geq -1$.

For generic moduli we also have $\br \cdot \by =0 $, which implies $\br =0$,  
and since $c _{n < 1} =0 $ the $\br$ sum in the last line of (\ref{hmr})
restricts to the subset $\br^2 = 2$ only.
Hence, we have in the generalized fundamental Weyl chamber
 the simplified expression: 
\be
\label{hmr2}
\eqalign{
I_\n (y)
= & 4 {\rm Re} \sum_{s=0}^\n \ch{\n}{s} \left( { 6 \over \p (y_2,y_2) }
\right)^s
\sum_{r > 0}'
L_{(s)} \left( (r,y) \right) c^{(\n-s)}_{kl - \br^2/2}
\cr 
 + & c^{(\n)}_0
\left( - \log \bigl(-(y_2, y_2) \bigr) - {\cal K} \right)
- { (y_2,y_2) \over 2 U_2} {\p \over 3 (\n +1) } [ E_2^{\n+1} \chi
\Phi_\n ]
\vert_{q^0}
\cr 
 + & \sum_{s=0}^\n
c^{(\n-s)}_0
\ch{\n}{s} \left\{
{ 4 \p (-24)^s s! B_{2s+2} \over (2s+2)! } {U_2^{2s+1} \over
(y_2,y_2)^s }
+ \th(s \geq 1) {2 (2s)! \over s!} \left( {3 \over 2 \p^2 (y_2,y_2) }
\right)^s
 \zeta (2s +1) \right\}
\cr 
 + & 2 {\rm Re} \sum_{s=0}^\n
\ch{\n}{s}
\left( { 6 \over \p  (y_2,y_2) } \right)^s
 {  U_2^{2s+1} s! \over \p^{s+1}   }
\sum_{\br^2 =2 }
Li_{2s+2} \left( e^{ 2 \pi i \br \cdot \by_2/ U_2 } \right)
c^{(\n-s)}_{-1}
\cr }
\ee 
where we also used that $Li_{s} (1) =  \zeta (s)$ and eq. (\ref{d25}).

\ni {\it Simplification of rational terms}.
For the calculation (and existence) of the generalized prepotentials
in Appendix \ref{gpp}, it is necessary to simplify the rational terms,
which
are defined as follows
\be
I_\n^{\rm rat} (y) =
- { (y_2,y_2) \over 2 U_2} {\p \over 3 (\n +1) } [ E_2^{\n+1} \chi
\Phi_\n ]
\vert_{q^0}
+\sum_{s=0}^\n
c^{(\n-s)}_0
\ch{\n}{s}
{ 4 \p (-24)^s s! B_{2s+2} \over (2s+2)! } {U_2^{2s+1} \over
(y_2,y_2)^s }
\label{rat} \ee
$$
 + 2 {\rm Re} \sum_{s=0}^\n
\ch{\n}{s}
\left( { 6 \over \p  (y_2,y_2) } \right)^s
 {  U_2^{2s+1} s! \over \p^{s+1}   }
\sum_{\br^2 =2 }
Li_{2s+2} \left( e^{2 \pi i \br \cdot \by_2 / U_2 } \right)
c^{(\n-s)}_{-1}
\pe $$
In the fundamental Weyl chamber we can use the following identities
(relevant for $\n \leq 2$) on the even polylogarithms,
\bs
\be
{\rm Re} Li_2(e^{2\pi i x}) = \p^2 \left( \frac{1}{6} - |x| +x^2
\right)
\;\;\;\;\;\;\;\;\;\;\;\; 
\;\;\;\;\;\;\;\;\;\;\;\; \;\;\;\;\; \;\;\;\; \;\;\;\; 
\ee
\be
{\rm Re} Li_4(e^{2\pi i x}) = \p^4 \left( \frac{1}{90} - \frac{1}{3}
x^2
+ \frac{2}{3} |x|^3 - \frac{1}{3} x^4 \right)
\;\;\;\;\;\;\;\;\;\;\;\; 
\;\;\;\;\;\;\; 
\ee
\be
{\rm Re} Li_6(e^{2\pi i x}) = \p^6 \left( \frac{1}{945} - \frac{1}{45}
x^2
+ \frac{1}{9} x^4 - \frac{2}{15} |x|^5
+ \frac{2}{45} x^6 \right)
\co \ee
\es
which hold for $|x| < 1$.
To simplify the expression in (\ref{rat}) for $\n \leq 2$, we will also
use the fact that  $\chi (q)  = 1 + 2 D q + \cO (q^2)$, 
where $D$ is the number of positive roots (for $E_8$ this is 120, while
for $E_8 \ti E_8$ or $SO(32)$ this is 240). 
Moreover, we use the explicit functions  $\Phi_\n$, which are
\bs
\be
S=8: \;\;\;\;\;  \Phi_0 =  {E_4^2 \over \et^{24} } \sp
\Phi_1 =  {E_6 \over \et^{24} } \sp
\Phi_2 =  {E_4 \over \et^{24} }
\;\;\;\;\;\;\;\;\;\;\;\;\;\;  \ee
\be
S=16: \;\;\;\;\;  \Phi_0 =  {E_4 \over \et^{24} } \sp
\Phi_1 = \mbox{non-existent} \sp
\Phi_2 =  {1  \over \et^{24} }
\pe \ee
\es

We will also need to define the following completely symmetric
Lie algebra tensors
\be
\sum_{\br^2 = 2\atop \br > 0} r_{i_1} r_{i_2} \ldots r_{i_n}
= \a^{(n)}_{i_1 i_2 \ldots i_n}
\pe \label{lat} \ee
In particular, by definition, $\a^{(1)}_i = 2 \r_i$, where $\r$ is the
Weyl vector, while it is also known for any simply-laced group (we take
$\br^2=2$) that
$ \a^{(2)}_{ij} =  \tilde{h} \d_{ij} $
where $\tilde{h}$ is the dual Coxeter number (equal to 30 for $E_8$ and
$SO(32)$).  We also have for $E_8$ and $SO(32)$ the identities
\bs
\be
\sum_{ijkl} \a^{(4)}_{ijkl} v^i v^j v^k v^l  =
\left\{ \matrix{ 18 (\bv \cdot \bv)^2 & \sp & E_8 \cr
 6 (\bv \cdot \bv)^2 + 24 \sum_i v_i^4 & \sp & SO(32)  \cr}
 \right.
\ee
\be
\sum_{ijklmn} \a^{(6)}_{ijklmn } v^i v^j v^k v^l v^m v^n  =
\left\{ \matrix{ 15 (\bv \cdot \bv)^3 & \sp & E_8 \cr
 30 (\bv \cdot \bv) \sum_i v_i^4 & \sp & SO(32)  \cr}
 \right.
\ee
\es
The tensors in (\ref{lat})  satisfy the contraction property
\be
\a^{(n)}_{i_1 i_2 \ldots i_n} \et^{i_{n-1} i_n}
= 2 \a^{(n-2)}_{i_1 i_2 \ldots i_{n-2}}
\pe \ee

Then, after some algebra, we find the following results (for $\n \leq
2$):
in eq. (\ref{rat}) there appear {\it a priori} terms of the form
$(y_2^i)^{2\n +2} \over U_2 (y_2,y_2)^\n$; however,
these  vanish because of non-trivial root identities. The vanishing of these
terms is essential
for the integrability of the thresholds in terms of (generalized)
prepotentials
as discussed in Appendix \ref{gpp}.
The final simplification for the rational terms can be summarized in
terms of
a set of symmetric tensors as follows: 
\be
I_\n^{\rm rat} (y) = -  {8 \p \over (y_2,y_2)^\n }
d^{(\n,\n)}_{a_1 \ldots a_{2\n +1} } y_2^{a_1} \cdots y_2^{a_{2\n+1} }
\pe \label{rat2} \ee
In particular, for the case $\n=0$ we have the explicit result
\be
S=8,16: \;\;\;\; d^{(0,0)}_a = (\bar{\r}, -30, -31) \equiv - \et_{ab}
\r^b
\pe \label{wvec}
\ee
As pointed out in Ref. \cite{hm2}, for the case $S=8$ we have
 $\r^a = - \et_{ab}  d^{(0)}_b  = -(\bar{\r}_{E_8} , 31, 30)$,  
which is the Weyl vector of the $E_{10}$ KM algebra.
  For the case $\n=1$ and $S=8$ the result agrees with \cite{hm2}, and
will not be given explicitly here.
 Finally,  we give the corresponding expressions for the case $\n=2$.
For $E_8$ we have
\be
\eqalign{
E_8 : \;\;\;\;  I_2^{\rm rat} (y) =  - & {8 \p \over (y_2,y_2)^2} \left(
[ \bar{\r} \cdot \by - 30 T_2 - 31 U_2] (y_2,y_2)^2 
\right. 
\cr 
+ & [8U_2^3 - 168 U_2^2 T_2 - 144 U_2 T_2^2  - 4
 \a^{(3)}_{i_1 i_2 i_3 } y_2^{i_1} y_2^{i_2}  y_2^{i_3} ] (y_2,y_2)
\cr 
& \left. - \frac{48}{5} U_2^5 + 48 U_2^4 T_2 -288 U_2^3 T_2^2
- 192 U_2^2 T_2^3 + \frac{24}{5}
\a^{(5)}_{i_1 i_2 i_3 i_4 i_5} y_2^{i_1} y_2^{i_2}  y_2^{i_3}y_2^{i_4}
y_2^{i_5}
\right)
\cr} 
\ee 
where $i=1, \ldots ,8$, while for $E_8 \ti E_8$ we find
\be
\eqalign{
E_8 \ti E_8  : \;\;\;\;
I_2^{\rm rat} (y) = - & {8 \p \over (y_2,y_2)^2} \left(
[ \bar{\r} \cdot \by - 30 T_2 - 31 U_2] (y_2,y_2)^2 
\right.
\cr 
+ & [8U_2^3 - 168 U_2^2 T_2 - 144 U_2 T_2^2  - 4
 \a^{(3)}_{i_1 i_2 i_3 } y_2^{i_1} y_2^{i_2}  y_2^{i_3} ] (y_2,y_2)
\cr 
 - & \frac{48}{5} U_2^5 + 48 U_2^4 T_2 -288 U_2^3 T_2^2
- 192 U_2^2 T_2^3 
\cr 
&  \left. + 144 (U_2+T_2)  (\by_2 \cdot \by_2)_1 (\by_2 \cdot
\by_2)_2
 + \frac{24}{5}
\a^{(5)}_{i_1 i_2 i_3 i_4 i_5} y_2^{i_1} y_2^{i_2}  y_2^{i_3}y_2^{i_4}
y_2^{i_5}
\right)
\co \cr}
\ee 
where now
 $i=1, \ldots ,16$, and $ (\by_2 \cdot \by_2)_1$ and $ (\by_2 \cdot
\by_2)_2$
refer to the two $E_8$ factors respectively.
For $SO(32)$ we find
\be
\eqalign{
SO(32): \;\;\;\; I_2^{\rm rat}(y) = - & {8 \p \over (y_2,y_2)^2} \left(
[ \bar{\r} \cdot \by +18 T_2 +17  U_2] (y_2,y_2)^2 
\right. 
\cr 
+ & [8U_2^3 +24  U_2^2 T_2 +48 U_2 T_2^2 - 4
 \a^{(3)}_{i_1 i_2 i_3 } y_2^{i_1} y_2^{i_2}  y_2^{i_3} ] (y_2,y_2)
\cr
 & \
 -  \frac{48}{5} U_2^5 + 48 U_2^4 T_2 -96 U_2^3 T_2^2
- 96 (U_2 +T_2) \sum_i (y_2^i)^4
\cr
 & \;\;\;\; + \left.  \frac{24}{5}
\a^{(5)}_{i_1 i_2 i_3 i_4 i_5} y_2^{i_1} y_2^{i_2}  y_2^{i_3}y_2^{i_4}
y_2^{i_5}
\right)
\pe \cr}
\ee
We also note that the corresponding tensors satisfy the
identities
\be
d^{(2,2)}_{abcde} \et^{bc} \et^{de} =
\left\{
\matrix{ - \frac{24}{5} \r_a & \sp & S=8 \cr
 \frac{8}{3} \r_a & \sp & S=16 \cr}
\right.
\ee
where $\r_a$ is the generalized Weyl vector defined in (\ref{wvec}).

\subsection{Generalized prepotentials \label{gpp} }

\ni \un{(2,2) case}
\vs .1cm

Using identities (\ref{re2}) and (\ref{bb13}),
it can be shown that the result (\ref{d31}) for the one-loop threshold
integral (\ref{d1}) can be written in terms of $\nu$ ``prepotentials"
$f_{(\n,s)}(T,U)$ in the following way
\be
I_{\n} (T,U) =
-  c_0^{(\n)} [ \log T_2 U_2 + {\cal K} ]
- 2 \log | \tf_{(\n,0)}(T,U)  |^2
+ {\rm Re} \sum_{s=1}^\n \ch{\n}{s}  D_T^s D_U^s f_{(\n,s)} (T,U )
\co \label{d33}\ee
where
\bs
\be
\tf_{(\n,0)}(T,U)
=
q_T^{ [c_0^{(0)}/24 - (\n +1) c_{-1}^{(0)} ]/(\n+1) } q_U^{ c_0^{(\n
)}/24}
\prod_{m=0}^\n \prod_{k,l}' (1 - q_T^k q_U^l)^{ c_\n(m;k,l) }
\label{d35}\ee
\be
 c_\n(m;k,l) \equiv  \ch{\n}{m} {(-3)^m \over (m+1)! }  (-4 kl)^m
c^{(\n-m)}_{ kl }
\ee
\be
f_{(\n,s\geq 1)} = 4   \sum_{m=0}^{\n -s} \ch{\n-s}{m} {s!  (2s+1)
(-3)^{s+m}\over (2s+m+1)!}  \sum_{kl}'
(-4 kl)^m Li_{2s+1} (q_T^k q_U^l)c^{(\n-s-m)}_{ kl }
\ee
$$
 - c_0^{(\n-s)} 4 i \p^{2s+1}
 {  (12)^{s} s ! B_{2s +2} \over (2s)! (2s+2) ! } U^{2s+1}
+
c_0^{(\n-s)}
 2  {s! (-3)^s  \over (2s)!}  \zeta (2 s +1) \pe
$$
\label{gp1}
\es
The function $f_{(\n,s)}$ is an (almost) modular function of $T$ and
$U$, of
weight $-2s$, and the appropriate covariant derivatives are defined in
Appendix A. In particular, under modular transformations the
functions $f_{(\n,s)}$ transform with an additive piece.
In the case of $N=2$ threshold integrals, $\nu=1$ and $f_{(1,1)}$ is the
one-loop prepotential of the $N=2$ effective supergravity.
Writing the integral in this form suggests that in $N=1$ supergravity in
eight dimensions, the four-derivative terms can be written in terms of
holomorphic prepotentials.

\ni \un{$(S+2,2)$ case}
\vs .1cm
Using the identities (\ref{idl}), (\ref{idd}) and (\ref{cd1}),
it can be shown that the
result (\ref{hmr2}) of the thresholds (\ref{win})
 can also be rewritten in terms of $\n$
``prepotentials'', whose form in the generalized fundamental chamber
is as follows
\be
I_\n (y) = -c_0^{(\n)}[ \log -(y_2,y_2) + {\cal K}]
- 2 \log | \tf_{(\n,0)}(y)  |^2
+ {\rm Re} \sum_{s=1}^\n \ch{\n}{s} \square^s f_{(\n,s)}  (y)
\co \ee
where the second-order operator $\square$ is defined in eq. (\ref{box})
and
\bs
\be
\tf_{(\n,0)} (y)
=   e^{2 \pi i (\s_\n, y)} \prod_{m=0}^\n \prod_{r>0}'
(1  -  e^{2 \pi i (r, y)} )^{ c_\n(m,r) }
\ee
\be
\s_\n^a \equiv  - { (S/2+1) ! \over (S/2 + \n +1)!} { (2\n+1) ! \over
4^\n \n !}
\et^{ab} d^{(\n,0)}_b
\ee
\be
c_\n(m,r) \equiv \ch{\n}{m} { (S/2+1) ! \over (S/2 + m +1)!}
(-6r^2)^m   c^{(\n-m)}_{-r^2/2}
\ee
\be
f_{(\n,s\geq 1)} (y) = 4 \sum_{m=0}^{\n-s} \ch{\n-s}{m}
{ (S/2+s) ! (S/2 +2s +1) \over (S/2 + 2s +m +1) ! } (-3)^{s+m}
\sum_{r>0}' (2r^2)^m Li_{2s+1}(  e^{2 \pi i (r, y) } )
c^{(\n-s-m)}_{-r^2/2}
\ee
$$
 + 8 \p^{2s+1} { (S/2+s) ! (S/2 +2s +1) \over (S/2 + \n +s  +1) ! }
{ (-2)^s \over 4^\n} { (2 \n +1) ! \over \n ! (2s+1) ! }
 d^{(\n,s)}_{a_1 \ldots  a_{2s+1}} y^{a_1} \cdots   y^{a_{2s+1}}
$$
$$
 + 2 (-3)^s { (S/2+s) !  \over (S/2 + 2s ) ! } \zeta (2s+1)
c_0^{(\n-s)}
\co $$
\label{gp2} \es
where we have used the simplified form (\ref{rat2})  of the rational
terms
in the generalized fundamental Weyl chamber and used the
recursive definition
\be
d^{(\n,s-1)}_{a_1 \ldots a_{2s -1} }
=d^{(\n,s)}_{a_1 \ldots a_{2s +1} }  \et^{a_{2s} a_{2s+1} }
\sp 1 \leq s \leq \n
\pe \ee
For the case $\n=1$, $S=8$, we have checked the agreement with the
one-loop prepotential given in \cite{hm2}. We note here again that
the above expressions have only been proved for $\n \leq 2$. For higher
$\n$, these expressions remain true if the conjectured relations
(\ref{idl}), (\ref{idd}), (\ref{cd1})
and the form (\ref{rat2}) for the rational terms remain valid. We
strongly
believe this to be the case, and also note that
the expressions in (\ref{gp2}) correctly
reduce to those in (\ref{gp1}) for $S=0$.

\boldmath
\section{Large $T_2$ expansion of heterotic
one-loop integrals \label{lte} }
\unboldmath
\renewcommand{\theequation}{F.\arabic{equation}}
\setcounter{equation}{0}

The main results (\ref{d31}) and (\ref{hmr2})  of  Appendix \ref{1li}
are the general
form
 of the elliptic-genus contributions
to the one-loop free energy of the
heterotic string compactified on a two-torus, without and with Wilson
lines.
In this appendix we compute
the large $T_2$ expansion of these two expressions, by re-expanding the
result
in a double power series in the variables
$T_2$ and $q_T =e^{2 \pi i T}$. Using heterotic/type-I duality the
resulting expansion can be decomposed into  the perturbative part
(powers
of $T_2$ only) and the non-perturbative part
(powers of $q_T$) from the type-I point of view. We
will use this terminology below, in accordance with
the physical interpretation discussed in the text.

\ni \un{(2,2) case}
\vs .1cm
Our aim is to use the large $T_2$ expansion to rewrite the expression
in
(\ref{d31}) in the form
\be
I_\n (T,U) = I_\n^{({\rm p})} (T_2,U)  + I_\n^{({\rm n.p})} (T_2,q_T,U)
+ I^{({\rm d})}  (T_2)
\co \label{d36}
\ee
where $I_\n^{({\rm p})}$ and
 $I_\n^{({\rm n.p})}$ stand for the perturbative and non-perturbative
parts, respectively, and
$ I^{({\rm d})}  (T_2)$
 collects logarithmically divergent and constant pieces.

In fact, by examining the separate contributions $I_\n^{(i=1,2,3)}$ in
(\ref{d5}), (\ref{d17}) and (\ref{d30})
of the trivial, non-degenerate and degenerate orbits, respectively,
it is not difficult to see that
\bs
\be
I_\n^{({\rm p})} (T_2,U) +
 I^{({\rm d})}  (T_2)
= I_\n^{(1)} +I_\n^{(3)}
\label{d37}
\ee
\be
I_\n^{({\rm n.p})} (T_2,q_T,U) = I_\n^{(2)}
\co \label{d38}
\ee
\es
so that the perturbative contributions are included in the trivial
and degenerate orbit, while
the non-degenerate orbits generate
non-perturbative terms.

In further detail, it follows from (\ref{d5}) and
 (\ref{d30}) that the perturbative terms are
\be
I_\n^{({\rm p})} (T_2,U)  =
 { \p T_2 \over 3 (\n +1) } [c_0^{(0)} - 24 (\n +1) c_{-1}^{(0)} ]
+ \sum_{s=0}^\n \ch{\n}{s} \left( \frac{-3}{\p T_2} \right)^s
Y_{(s)}(U) c_0^{(\n-s)}
\co \label{d39}\ee
where the functions $Y_{(s)}$ are given by
\bs
\be
Y_{(0)} (U) =   4 {\rm Re}
\sum_{l =1}^\infty L_{(0)} ( Ul  ) - \log U_2 + \frac{\p U_2}{3}
= -\log U_2 | \et (U) |^4
\ee
\be
Y_{(s \geq 1)} (U) =   4 {\rm Re}
{ 1 \over  U_2^s}
\sum_{l =1}^\infty L_{(s)} ( Ul  )
+ { (-1)^s (4 \p )^{1+s} s ! B_{2s +2} \over (2s+2) ! }
U_2^{s+1}
+ 2  {(2s)! \over s!} \left( {1  \over 4\p U_2 }\right)^s
\zeta (2 s +1)  \pe
\ee
\es
Note that  modular invariance in the $T$ and $U$
moduli of the total integral (\ref{d1}) implies that these functions
are
modular-invariant in the $U$ modulus. For $Y_{(0)}$ this fact
corresponds to 
 the usual modular transformation of the $\et$-function.
For $s \geq 1$, however, this fact implies highly non-trivial modular
properties of the polylogarithms, which in some sense generalize those
of the $\et$-function. Similar identities were noted in \cite{hm2}.

For the logarithmically divergent/constant pieces, we easily read off
\be
 I^{({\rm d})}  (T_2) = -c_0^{(\n)} [ \log T_2 + {\cal K} ] \pe
\label{d40}
\ee
Using the intermediate result (\ref{d23}) in the degenerate orbit
, we can also write down the following alternative 
form of the perturbative terms: 
\be
\eqalign{
\;\;\;\; I_\n^{({\rm p})} (T_2,U)
& = { \p T_2 \over 3 (\n +1) } [c_0^{(0)} - 24 (\n +1) c_{-1}^{(0)} ]
 -I^{({\rm d})}  (T_2)
\cr
& +U_2 \sum_{s=0}^\n \ch{\n}{s} \left( {- 3U_2 \over \p^2 T_2}
\right)^s
 c_0^{(\n-s)} \sum_{(j,p)\neq(0,0) }
{1 \over | j + p U |^{2 (1+s)} }
\co \cr}
\label{d41}\ee
which expresses the contributions at order $1/T_2^s$ as a sum over
inverse
powers $1/P^{2(1+s)}$ of the internal momenta of the type-I string.

Moving on to the non-perturbative terms, we start with
the expression (\ref{d17}) for the
non-degenerate orbit and rewrite it as follows. First, we substitute
the
explicit form of $L_{(s)}$ in (\ref{d18}), yielding
\bs
\be
I_\n^{({\rm n.p})} =
4 {\rm Re} \sum_{s=0}^\n
\left(  {- 3 \over \p T_2 U_2 } \right)^s
   \sum_{r=0}^s
\ch{\n}{s}
{ (s+r) ! \over r ! (s-r)! (4 \p)^r}
\sum_{k,l \atop k\neq 0 }'
 (T_2 k + U_2 l)^{s-r}
 Li_{s +r +1} (q_T^k q_U^l  )
c^{(\n-s)}_{ kl }
\label{d42}\ee
\be
=
4 {\rm Re} \sum_{s=0}^\n
   \sum_{r=0}^s
   \sum_{m=0}^{s-r}
{ (-3)^s \over 4 ^r}
\ch{\n}{s}
{ (s+r) ! \over r ! (s-r)! }
\ch{s-r}{m}
 \frac{1}{(\p T_2)^{s-m} } \frac{1}{(\p U_2)^{r+m} }
\label{d43} \ee
$$
\ti
\sum_{l,k \atop k\neq 0 }'
\sum_{p=1}^\infty
k^m l^{s-r-m}
 \frac{1}{p^{s +r +1} } q_T^{k p} q_U^{l p}
c^{(\n-s)}_{ kl }
\co $$
\es
where in the second step we have also used the summed form of the
polylogarithms and expanded the $(T_2k +U_2l)^{r-s}$ factor.

Next we do the $l$ summation, by rewriting
\be
\label{d44}
\eqalign{
  \sum_{l=-1 }^\infty    l^{s-r-m} q_U^{lp} c_{kl}^{(\n-s)}
&   = \sum_{l=-1 }^\infty   l^{s-r-m} q_{pU/k}^{kl} c_{kl}^{\n-s} \cr
 &  = \sum_{l=-1 }^\infty   \frac{1}{k^{s-r-m}}  [q_{u'} \pa_{q_{u'}
}]^{s-r-m}
q_{u'}^{kl}  c^{(\n-s)}_{ kl }
\cr
 &
  = \sum_{j=0}^{k-1}   \frac{1}{k^{s-r-m+1}}  [q_{u'} \pa_{q_{u'}
}]^{s-r-m}
  ( E_2^{\n-s} \Phi_\n ) ( u' +{j\over k}  ) \cr
 &   = \sum_{j=0}^{k-1}   \frac{1}{k^{s-r-m+1}}  [q_{u} \pa_{q_{u}
}]^{s-r-m}
  ( E_2^{\n-s} \Phi_\n ) ( u  )
\cr}\
\ee
where we have introduced, in the second step, $u'=pU/k$ and, in the
last step: 
\be
 u = {p U + j \over k }
\co \label{inm} \ee
which is identified with the complex modulus of the world-volume of the
D1-brane.
In the third step  we also used the identity
\be
\sum_{l=-1}^\infty  q_U^{ kl } C_{kl}
= \frac{1}{k} \sum_{j=0}^{k-1}  F (U + \frac{j}{k} )
\sp
F (U) = \sum_{n=-1}^\infty  C_n q_U^n
\label{sid} \ee
and the definition (\ref{a6}).

Substituting the result (\ref{d44}) in (\ref{d43}) we obtain
\bs
\be
\eqalign{
 I_\n^{({\rm n.p})} =
4 {\rm Re} \sum_{s=0}^\n
   & \sum_{r=0}^s
   \sum_{m=0}^{s-r}
{ (-3)^s \over 4 ^r}
\ch{\n}{s}
{ (s+r) ! \over r ! (s-r)! }
\ch{s-r}{m}
 \frac{1}{(\p T_2)^{s-m} }
\cr
 & \ti
\sum_{k,p =1}^\infty  \sum_{j=0}^{k-1}
\frac{1}{(\p pU_2/k)^{r+m} }
 \frac{1}{(k p)^{s -m +1} } q_T^{k p}
 [q_{u} \pa_{q_{u} }]^{s-r-m}
  ( E_2^{\n-s} \Phi_\n ) ( u  )
\cr}\
\ee
\be
 \eqalign{
\;\;\;\;\;\;\;\;\;\; =
4 {\rm Re} \sum_{s=0}^\n \ch{\n}{s} &
\left( { 3 \over 2 \p T_2} \right)^s
   \sum_{r=0}^s
   \sum_{m=0}^{\n-s}
{ (-)^{s+m} 3^m 4^r \over 2^s} {s! \over r!}
\ch{\n-s}{m}
\ch{2s+m-r}{s+m}
\cr
&
\ti
\sum_{k,p =1}^\infty  \sum_{j=0}^{k-1}
  \frac{1}{(\p u_2)^{s+m-r} }
 \frac{1}{(k p)^{s +1} } q_T^{k p}
 [q_{u} \pa_{q_{u} }]^{r}
  ( E_2^{\n-s-m} \Phi_\n ) ( u  )
\cr}\
\label{d45}
\ee
\es
where, in the second step, we used $u_2 = pU_2/k$, the summation identity
\be
 \sum_{s=0}^\n \sum_{r=0}^s \sum_{m=0}^{s-r} f(s,r,m)
= \sum_{s=0}^\n \sum_{r=0}^s \sum_{m=0}^{\n-s} f(s+m,s-r,m)
\co \label{sid2} \ee
and performed some regrouping of terms.

Finally, we use the identity (\ref{ef1})
to obtain
the interesting result that
the non-perturbative part of the integral over the elliptic genus,
\be
\eqalign{
 I_\n^{({\rm n.p})} (T_2,q_T,U)
= & \int_{\cal{F}}
{\rd^2 \t \over \t_2 }  \left(   \Gamma_{2,2}(T,U) \hat
E_2^{\n}(\t) \Phi_{\n}(q)- c_0^{(\n)}\right) \vert_{\rm non-pert.}
\cr
& = 4 {\rm Re} \sum_{s=0}^\n \ch{\n}{s}
\left(\frac{3}{2\p T_2} \right)^s
\sum_{p,k=1 }^\infty
\frac{1}{(kp)^{s+1} }  q_T^{kp}   \sum_{j=0}^{k-1}
   (D^s  \hE_2^{\n-s} \Phi_\n ) \left({p U + j \over k }\right)
\co \cr}\
\label{b13n} \ee
depends again on the elliptic genus and covariant derivatives thereof.

We continue to simplify this by noting that (\ref{b13n})
has the form
\bs
\be
 I_\n^{({\rm n.p})} (T_2,q_T,U)
 = 4 {\rm Re} \sum_{s=0}^\n \ch{\n}{s}
\left(\frac{3}{2\p T_2} \right)^s
   \sum_{N=1}^\infty \frac{1}{N^{s}} q_T^N 
g_{(s,N)} (U)
\label{brut}
\ee
\be
g_{(s,N)} (U)
\equiv
 \frac{1}{N} \sum_{p,k=1 \atop p k = N }^\infty   \sum_{j=0}^{k-1}
   (D^s  \hE_2^{\n-s} \Phi_\n ) \left({p U + j \over k }\right)
\co \ee
\es
where the functions $g_{(s,N)}(U)$ entering at the $N$-th instanton
contribution $q_T^N$ are modular functions of $U$, given the
fact that $\Phi_\n$ are modular functions of weight $-2\n$.
In terms of the Hecke operator (\ref{hec}), 
we have $g_{(s,N )} (U) = H_N [ D^s \hE_2^{\n-s} \Phi_\n ] (U)$.
 
Although for a given instanton number $N$ (and any $s$)
the function $g_{(s,N)}$
is modular-invariant in $U$, the sum in this expression is
reducible when $N=n m^2$ for some $m > 1$, in the sense that the
function
can then be split up into more than one part, each of which is
separately
modular-invariant.
Here we will do the reduction using the modular invariance of the
result on $U$.  An algebraic explanation can also be given,  
by looking at the classifications of the mappings
between the lattice characterizing the instanton world-sheet and the
torus.

When the sum cannot be further reduced
into separate modular-invariants, we will call the resulting sum an
irreducible  modular-invariant.
In particular, when $N = n m^2=p k $, there will be one or more
 triplets of numbers $(p,k,j)$, which  have a greatest common divisor
${\rm g.c.d.}(p,k,j) = m >1$, and it is not difficult to see that the
corresponding subset of terms have already appeared as the 
modular-invariant
$g_{(s,n)}$, i.e. at lower instanton number $n<N$.
Hence, the irreducible modular
invariants are characterized by the $n$-instanton modular function
\be
G_{(s,n)}(U) \equiv \frac{1}{n}  \sum_{p,k=1 \atop p k =n}^\infty
 \sum_{j=0}^{k-1}
\d({\rm g.c.d.}(p,k,j) = 1 )
    (D^s  \hE_2^{\n-s} \Phi_\n ) \left({p U + j \over k }\right)
\co \label{b13} \ee
which is the minimal modular-invariant completion of
$(D^s  \hE_2^{\n-s} \Phi_\n ) (n U)$, in that all  terms in the sum
of (\ref{b13}) are necessary and sufficient to make the entire function
$G_{(s,n)}(U)$ modular-invariant.

Using the definition (\ref{b13}) in (\ref{b13n}) we can rearrange
the non-perturbative contributions as follows: 
\be
\eqalign{
 I_\n^{({\rm n.p})} (T_2,q_T,U)
 & = 4 {\rm Re} \sum_{s=0}^\n \ch{\n}{s}
\left(\frac{3}{2\p T_2} \right)^s
   \sum_{n=1}^\infty \sum_{m=1}^\infty \frac{1}{(n m^2)^ {s+1}}
q_T^{n m^2}
\cr
& \;\;\;\;\;\; \;\;\;\;\;
\ti \sum_{p,k=1 \atop p k =n }^\infty   \sum_{j=0}^{k-1}
 \d({\rm g.c.d.}(p,k,j) = 1)
(D^s  \hE_2^{\n-s} \Phi_\n ) ({p U + j \over k })
\cr
& = 4 {\rm Re} \sum_{s=0}^\n \ch{\n}{s}
\left(\frac{3}{2\p T_2} \right)^s
   \sum_{n=1}^\infty \frac{1}{n^{s}}
G_{(s,n)} (U) \sum_{m=1}^{\infty} \frac{1}{m^{2(s+1)}}
(q_T^n)^{m^2} \pe
\cr }
\ee
We finally write the result as
\be
 I_\n^{({\rm n.p})} (T_2,q_T,U)
 = 4 {\rm Re} \sum_{s=0}^\n \ch{\n}{s}
\left(\frac{3}{2\p T_2} \right)^s
   \sum_{N=1}^\infty \frac{1}{N^{s}}
G_{(s,N)} (U) \Theta_{(s)} (q_T^N)
\co \label{fwr}
\ee
where we have introduced the function
\be
\Theta_{(s)} (q)=
 \sum_{m=1}^{\infty} \frac{1}{m^{2(s+1)}}
q^{m^2}
\ee
and we recall that $G_{(s,N)}$ are the irreducible 
modular-invariants defined in (\ref{b13}).

The sum over the integer $m$ in the above formulae is interpreted as a
sum of
multiple D1-branes wrapped around the torus.
We can argue that from the type-I point of view they must be included, 
otherwise the $SL(2,Z)_T$ invariance will be broken.
This can be seen most easily for the $Tr F^4$ threshold, where the
result is given by $\log(T_2|\eta(T)|^4)$. Decomposing this threshold
as
above,
it is obvious that on the one hand, the logarithmic divergence plus
$SL(2,Z)_T$ invariance uniquely specifies that only the $\eta$-function
can appear.
If on the other hand we drop in the instanton expansion of the
threshold the terms corresponding to the multiply wrapped branes, then
the $SL(2,Z)_T$ symmetry will be broken.
We conclude that $SL(2,Z)_T$ symmetry forces the inclusion of multiply
wrapped D-instantons.

Similarly, we have computed the large $T_2$ limit of the generalized
prepotentials in (\ref{gp1}), where, for uniformity with $f_{(\n,s \geq 1)}$, 
we will use $f_{(\n,0)} \equiv - 4 \log \tf_{(\n,0)}$ below.
They exhibit a structure similar to that in (\ref{d36}): 
\be
f_{(\n,s)}(T,U) =
f_{(\n,s)}^{({\rm p})} (U)  +
f_{(\n,s)}^{({\rm n.p})} (U,q_T) +
f_{(\n,s)}^{({\rm d})} (T)
\co \ee
where we have separated perturbative, non-perturbative and
divergent parts. Here, the divergent and perturbative parts are given
by
\bs
\be
f_{(\n,s)}^{({\rm d})}  (T)
= -\d_{s,0} { [c_0^{(0)} - 24 (\n +1) c_{-1}^{(0)} ] \over 6(\n+1) } 2
\pi i T
\;\;\;\;\;\;\;\;\;\;\;\;\;\;\;\;\;\;
\;\;\;\;\;\;\;\;\;\;\;\;\;\;\;\;\;\;
\;\;\;\;\;\;\;
\ee
\be
f_{(\n,0)}^{({\rm p})} (U)
= - {  c_0^{(\n )} \over 6} 2 \pi i U
- \sum_{r=0}^\n
 4 \ch{\n}{r} {(-3)^r \over (r+1)! }
\sum_{l=1}^\infty   \log (1 -  q_U^l) c_{0}^{(\n-r)}
\;\;\;\;\;\;\;\;
\ee
\be
\eqalign{
\;\;\;\;\;\;\;\;\;\;\;\; f_{(\n,s \geq 1 )}^{({\rm p})} (U)  & =
\sum_{r=0}^{\n -s} \ch{\n-s}{r} 4s!
{2s+1 \over (2s +r +1) !}   (-3)^{s+r} \sum_{l=1}^\infty
 Li_{2s+1} (q_U^l)c^{(\n-s-r)}_{0 }
\cr
 & - c_0^{(\n-s)} 4 i \p^{2s+1}
 {  (12)^{s} s ! B_{2s +2} \over (2s)! (2s+2) ! } U^{2s+1}
+
c_0^{(\n-s)}
 2  {s! (-3)^s  \over (2s)!}  \zeta (2 s +1)
\co \cr}
\ee
\es
and we note that the functions $f_{(\n,s)}^{({\rm p})} $ are almost
modular functions of weight $-2s$, transforming with additional pieces
that
are annihilated by the covariant derivatives.

For the non-perturbative part we find the instanton expansion
\bs
\be
f_{(\n,s)}^{({\rm n.p})} (U,q_T)
=\sum_{n=1}^\infty q_T^N f^{(N)}_{(\n,s)} (U)
\;\;\;\;\;\;\;\;\;\;\;\;\; 
\;\;\;\;\;\;\;\;\;\;\;\;\; 
\;\;\;\;\;\;\;\;
\ee
\be
f^{(N)}_{(\n,s)} (U)
\equiv \frac{1}{N} \sum_{k,p = 1 \atop kp =N}^\infty \sum_{j=0}^{k-1}
\frac{1}{p^{2s}} F_{(\n,s)} \left( { p U + j \over k} \right)
\pe \label{hfu} \ee
\label{npp} \es
Here $F_{(\n,s)} (u) $ is given by
\be
F_{(\n,s)} (u)
= 4 s!  (-3)^s \sum_{m=0}^{\n-s} \ch{\n-s}{m} { 2s  +1  \over (2s+m+1)
! }
[ 12 q_u \pa_{q_u} ]^m   (E_2^{\n-s-m} \Phi_\n) (u)
\pe \label{b50} \ee
We conjecture that this is a holomorphic modular function in $u$
of weight $-2s$, which implies that the function is of the form
\be
F_{(\n,s)} (u)
= {4 s! (2s+1) (-6)^\n \over (\n+s+1) ! 2^s}
 \sum_{p,q,r=0 \atop p + 2q + 3r = \n-s}^\infty
b_{p,q,r}^{\n,s} E_6^r E_4^q \hat{D}^p \Phi_\n (u)
\co \ee
where $\hat{D}$ is the holomorphic covariant derivative in (\ref{a15})
and
the coefficients $b_{p,q,r}^{\n,s}$ are computable in principle by
comparison
with (\ref{b50}) and use of eqs. (\ref{a12})--(\ref{a15}).

We have checked the conjecture for $\n -s \leq 3 $, obtaining the
coefficients
\bs
\be
b_{0,0,0}^{\n,\n} = 1
\sp
b_{1,0,0}^{\n,\n-1} = 1
\sp b_{2,0,0}^{\n,\n-2} = 1
\sp
b_{0,1,0}^{\n,\n-2} = - { (\n- 1) \over 18}
\ee
\be
b_{3,0,0}^{\n,\n-3} = 1
\sp b_{1,1,0}^{\n,\n-3} = - { (3 \n- 5) \over 18}
\sp b_{0,0,1}^{\n,\n-3} =   - { (2 \n- 3) \over 27}
\pe \ee
\label{bex} \es
Moreover, additional evidence in support of the conjecture is the fact
that when $F_{(\n,s)}(u)$ is of weight $-2s$ in $u$, it follows that
the function $f_{(\n,s)}^{(N)} (U)$ in (\ref{hfu}) 
is of weight $-2s$ in $U$ as it should.
In particular, this function can be rewritten in the form
$ f_{(\n,s)}^{(N)} (U) = H_N[ F_{(\n,s)}] (U) $ where $H_N$ is the Hecke 
operator defined in (\ref{hec}). 

\ni \un{$(S+2,2)$ case}

In this case, for brevity, we restrict ourselves to the
non-perturbative
contributions, which clearly come from the $k>0$ sum in the first
term of (\ref{hmr}) only.  Moreover, in this case with non-zero Wilson
lines,
we need to employ the following loop counting parameter,
\be
\cV  \equiv G^{1/2}  \sp G^{1/2} = T_2  - \frac{1}{2 U_2}
\by_2 \cdot \by_2
\pe \ee
We omit the details of the resulting calculation, in which we closely
follow
the steps taken in the (2,2) case. We list, however, some of the
main identities that are used: the analogue of (\ref{sid}) is here: 
\be
\sum_{l=-1}^\infty  q_U^{ kl - \frac{1}{2} \br \cdot \br }
C_{kl - \frac{1}{2} \br \cdot \br  }
= \frac{1}{k} \sum_{j=0}^{k-1}  F \left(U + \frac{j}{k} \right)
e^{\pi i \br \cdot \br j/k}
\sp
F (U) = \sum_{n=-1}^\infty  C_n q_U^n
\pe \ee
We also need to define as in (\ref{inm}) the complex modulus $u$ of the
world-volume of the D1-brane, along with the induced D1-brane Wilson
lines,
\be
\bw = p \by
\pe \label{inw} \ee
Finally, we now need the expansion formula (\ref{ef2}),
which involves the Jacobi covariant derivative $\tilde{D}$ of
(\ref{jde})
 and the affine characters $\chi (\by | t)$ in (\ref{ach}).

The final result is
\be
\eqalign{
 I_\n^{({\rm n.p})} (\cV,q_T,\by,U)
= & \int_{\cal{F}}
{\rd^2 \t \over \t_2 }  \left(   \Gamma_{S+2,2}(y) \hat
E_2^{\n}(\t) \Phi_{\n}(q)- d_0^{(\n)}\right) \vert_{\rm non-pert.}
\cr
&
 = 4 {\rm Re} \sum_{s=0}^\n \ch{\n}{s} \left( { 3 \over 2 \p \cV}
\right)^s
\sum_{p,k=1 }^\infty
\frac{1}{(kp)^{s+1} }  q_T^{kp}   \sum_{j=0}^{k-1}
 (\tilde{D}^s \hE_2 ^{\n-s} \chi(\bw) \Phi_\n)(u) \pe \cr}
 \label{ltw} \ee
Since we used the conjectured identity (\ref{ef2}), we emphasize here
again
that this has only been explicitly checked up to $\n=2$, but we note
the correct reduction for zero Wilson lines to the result (\ref{b13n}),
as well as the fact that
(\ref{ltw}) has the correct transformation properties. We strongly
believe
the above result to be generally valid.

We also give the large $T_2$ expansion of the non-perturbative part of
the generalized prepotentials in (\ref{gp2}). This is exactly of the
form (\ref{npp}), but with $F_{(\n,s)} (u) \ra F_{(\n,s)} (\bw|u)$
given
by
\be
F_{(\n,s)} (\bw|u)
= 4 (S/2+s)!  (-3)^s \sum_{m=0}^{\n-s} \ch{\n-s}{m}
{ (S/2+ 2s  +1)  \over (S/2+ 2s+m+1) ! }
\label{b50p}
\ee
$$
\ti \chi (\bw |u)
[ 12 q_u \pa_{q_u} ]^m   (E_2^{\n-s-m} \Phi_\n) (u)
\pe
$$
We conjecture that this is a holomorphic Jacobi form of type
$(-2s,1)$ in $(u,\bw)$, which implies that the function is of the form
\be
F_{(\n,s)} (\bw|u)
= {4 (S/2+s)! (S/2+ 2s+1) (-6)^\n \over (S/2+ \n+s+1) ! 2^s}
 \sum_{p,q,r=0 \atop p + 2q + 3r = \n-s}^\infty
b_{p,q,r}^{\n,s} \chi (\bw |u) E_6^r E_4^q \hat{D}^p \Phi_\n (u)
\co \ee
where $\hat{D}$ is the holomorphic covariant derivative in (\ref{a15})
and the coefficients $b_{p,q,r}^{\n,s}$ are obtained from (\ref{bex})
using the replacement $\n \ra S/4 + \n$.

\section{Recursion relations and prepotentials \label{rrp} }
\def\ca{{\cal A}}
\renewcommand{\theequation}{G.\arabic{equation}}
\setcounter{equation}{0}

Let us consider the following integrals
\be
\Psi_s=\int_{\cal F}{d^2\t\over
\tau_2}\left[\Gamma_{2,2}(T,U){\cal A}_s -C\delta_{s,0}\right]
\co \label{z1}\ee
where $s=0,1,2,\cdots ,\nu_{\rm max}$ and ${\cal A}_s$ are the
relative elliptic genera defined in (\ref{ell2}); 
$C$ is the coefficient of the $q^0$ term in $\ca_0$ given in (\ref{ell1}), 
and  $\Psi_s$ is real.
The relatives of the elliptic genus satisfy the following recursion
relations
\be
\tau_2^2\pa_{\tau}\pa_{\bar\tau}{\cal A}_s={s(s+1)\over 4}{\cal A}_s+
{3\over 2}(s+1){\cal A}_{s+1}
\co \label{z12}\ee
with ${\cal A}_{\nu_{\rm max}+1}=0$.
They also satisfy
\be
(\tau_2^2\pa_{\bar\tau})^{\nu_{\rm max}+1}{\cal
A}_s=0\;\;\;,\;\;s=0,1,\cdots,\nu_{\rm max}
\co \label{z11}\ee
which will be useful as well.

We first analyze the cases $\nu_{\rm max}=0,1,2$ separately
and then describe the general case.

\bigskip
\ni $\underline{\nu_{\rm max}=0}$
\bigskip

Using (\ref{z12}), (\ref{z11}), (\ref{z10}), (\ref{z50}) on the integral
representation
and doing some integration by parts, keeping boundary terms, we
obtain the following equations
\be
\square_T\Psi_0=\square_U\Psi_0={C\over 4}\;\;\;,\;\;\;
\pa_T\pa_{\bar U}\Psi_0
=0 \pe \ee
The most general solution to the above equations is
\be
\Psi_0=-C\log(T_2U_2)+\left[f(T,U)+cc\right]
\co \ee
which concludes the analysis.

\bigskip
\ni $\underline{\nu_{\rm max}=1}$
\bigskip

Using (\ref{z12}), (\ref{z11}), (\ref{z10}), (\ref{z50}) we obtain the
following equations: 
\bs
\be
\square_T\Psi_0={C\over 4}+{3\over 2}\Psi_1\;\;\;
\left(\square_T-{1\over 2}\right)\Psi_1=0
\label{z23}\ee
\be
D^1_TD^0_TD^1_{\bar U}D^0_{\bar U}\Psi_0=D^1_TD^0_TD^1_{\bar U}
D^0_{\bar U}\Psi_0=0
\label{z24}\ee
\es
as well as those that are obtained by $T\leftrightarrow U$.

The second equation in (\ref{z23}) for $\Psi_1$ has as general
solution
\be
\Psi_1^*={1\over 3}\left[D_TD_U f_1(T,U)+D_TD_{\bar U}
\tilde f_1(T,\bar U)\right]+cc
\ee
while (\ref{z24}) implies that $\tilde f_1(T,\bar U)$
can be set to zero.
Thus we find that 
\be
\Psi_1={1\over 3}D_TD_U f_1(T,U)+cc
\pe \ee
Then the general solution to the equations for
$\Psi_0$ is 
\be
\Psi_0=-C\log(T_2U_2)+\left[f_0(T,U)+D_TD_U f_1(T,U)+
cc\right]
\pe \ee

\bigskip
\ni $\underline{\nu_{\rm max}=2}$
\bigskip

Using the above, we can now derive 
the following recursion relations
\bs
\be
\square_T\Psi_0={3\over 2}\Psi_1+{C_0\over 4}
\label{z2}\ee
\be
\square_T\Psi_1={1\over 2}\Psi_1+3\Psi_2
\label{z3}\ee
\be
\square_T\Psi_2={3\over 2}\Psi_2
\label{z4}\ee
\be
(D^2_TD^1_TD^0_T)(D^2_{\bar U}D^1_{\bar U}D^0_{\bar U})
\Psi_s\;\;\;,\;\;\;
s=0,1,2
\label{z49}\ee
\es
and similarly for $U$.
The simplest equation to solve is (\ref{z4}).
Its general solution is
\be
\Psi_2={1\over 3}\left(D^2_T D^2_{U}f_2(T,U)+D^2_T
D^2_{\bar U}\tilde f_2(T,\bar U)+cc\right)
\co \label{z38}\ee
where as usual $D^2=D_{-2}D_{-4}$.
The kernel of $D^2_T$ are functions of the form
$A(\bar T,U,\bar
U)
(T-\bar T)^4+B(\bar T,U,\bar U)(T-\bar T)^3$.
Using (\ref{z49}) on the general solution (\ref{z38})
we obtain
\be
D^3_{\bar U}D^3_{T}\Psi_2\sim \pa_{\bar U}^5\pa_T^5
\tilde f_2(T,\bar U)=0
\pe \ee
Thus, $\tilde f_2$ must satisfy this equation, so it is
a polynomial of degree at most 4 in $T,\bar U$.
In this case the function vanishes, when acted on by the covariant derivatives
in (\ref{z38}), so, without loss of generality,  it can be taken to be zero.
Thus, we have shown that
\be
\Psi_2={1\over 3}\left(D^2_T D^2_{U}f_2(T,U)+cc\right)
\pe \label{z6}\ee
Let us now solve the next equation, (\ref{z3}), which reads
\be
\left(\square_T-{1\over 2}\right)\Psi_1=D^2_T D^2_{U}f_2(T,U)
+cc
\pe \label{z7}\ee
The general solution is
\be
\Psi_1=D^2_T D^2_{U}f_2(T,U)+{1\over 3}\left(D_TD_Uf_1(T,U)+
D_TD_{\bar U}\tilde f_1(T,\bar U)\right)+cc
\pe \label{z8}\ee
Moreover, (\ref{z48}) implies that $\tilde f_1(T,\bar U)$
must be set to zero
so that
\be
\Psi_1=D^2_T D^2_{U}f_2(T,U)+{1\over 3}D_TD_Uf_1(T,U)+cc
\pe \label{z9}\ee
Finally, the general solution to (\ref{z2}) is
\be
\Psi_0=-C_0\log(T_2U_2)+\left[D^2_T D^2_{U}f_2(T,U)+
D_TD_Uf_1(T,U)
+f_0(T,U)+cc\right]
\pe \ee

The general $\nu_{\rm max}$ case is now transparent.
We have the following differential equations
\be
\left(\square_T-{s(s+1)\over 4}\right)\Psi_s={3\over 2}(s+1)
\Psi_{s+1}+{C\over 4}\delta_{s,0}
\label{z122}\ee
and
\be
(D^{\nu_{\rm max}}_{\bar U}D^{\nu_{\rm max}-1}_{\bar U}\cdots
D^0_{\bar U})~(D^{\nu_{\rm max}}_TD^{\nu_{\rm max}-1}_T\cdots
D^0_T)~\Psi_s=0\;\;\;,\;\;s=0,1,\cdots,\nu_{\rm max}
\pe \label{z48}\ee
The general solution is
\be
\Psi_s=-C\delta_{s,0}\log(T_2U_2)+\sum_{\nu=s}^{\nu_{\rm max}}
{(\nu+s)!\over 6^s(\nu-s)!s!}\left[D_T^{\nu}D_U^{\nu}
f_{\nu}(T,U)+cc\right]
\co \ee
which establishes the existence of generalized holomorphic
prepotentials.

\section{Heterotic threshold integrals for \\ general toroidal
compactification
\label{htt} }
\renewcommand{\theequation}{H.\arabic{equation}}
\setcounter{equation}{0}

We wish to compute the  integrals relevant for the heterotic
thresholds in toroidal compactifications,
\be
I_{\n} (G,B) =\int_{\cal{F}}
\rd^2 \t \; \t_2^{d/2 -2}    \Gamma_{d,d}(G,B) \hat
E_2^{\n}(\t) \Phi_{\n}(q)
\pe \label{ddc}
\ee
The integrand involves the $(d,d)$ lattice sum
\bs
\be
\Gamma_{d,d}(G,B) = \sum_{m_i,n^i}
q^{p_l^2/2} \bq^{p_r^2/2}
\sp p_{l,r}^2 = p_{l,r}^i G_{ij} p_{l,r}^j
\ee
\be
p_l^i  = { 1\over \sqrt{2} } (G^{-1} ( m + (G-B) n )^i
\sp
p_r^i  = { 1\over \sqrt{2} } (G^{-1} ( m - (G+ B) n )^i \sp i = 1
\ldots d
\co \ee
\es
where $G$ and $B$ are the $d$-dimensional metric and antisymmetric
tensor
of the $d$-torus respectively.
The integral is IR-divergent and can be regulated  by removing the
massless contribution. For the function $\Phi_\n$ we assume the
same expansion as in (\ref{d4}). 

For the computations and result
described
below, it will be useful to introduce the pull back of the $G$ and $B$
field 
\bs
\be
\hG_{IJ} = M_I^i G_{ij} M_J^j \sp
\hB_{IJ} = M_I^i B_{ij} M_J^j \sp I,J = 1,2
\ee
\be
M_I^i= (n^i,m^i)
\co \label{mma} \ee
\label{gbd} \es
and the corresponding induced K\"ahler form and complex structure
\be
\eqalign{
T^{(m,n)} & =T_1+ iT_2=- \hB_{12} + i \sqrt{\hG_{11} \hG_{22} -
\hG_{12}^2 } \cr
U^{(m,n)}  & = U_1 + i U_2 = \left(- \hG_{12} + i
 \sqrt{\hG_{11} \hG_{22} - \hG_{12}^2 } \right)/\hG_{11} \pe  \cr }
 \label{inm2} \ee
Below, we omit the superscripts $(m,n)$ on these induced moduli,
for simplicity.  
Then, we may write the lattice sum after a Poisson
resummation on
$m_i$ in the form (\ref{ddt}), which can be recast as 
\bs
\be
\Gamma_{d,d}(G,B) = \frac{1}{\t_2^{d/2} } \sqrt{G} \sum_{A \in
Mat_{d \ti 2 }    }
e^{  2\pi i \bar{T} }
\exp[ - {\p T_2 \over \t_2 U_2} | \t - \bar{U} | ^2 ]
\label{prf}
\ee
\be
A^T=  M = \left( \matrix{
n_1 & n_2 & \ldots & n_d  \cr
m_1 & m_2 &  \ldots &  m_d  \cr
} \right)
\co \ee
\es
where $T,U$ depend on the entries of $A$ through the definitions in
(\ref{gbd}), (\ref{inm2}). Note also that we used here the $ 2 \ti d$  
matrix
$M$ defined in (\ref{mma}) and that its transpose $A=M^T$ coincides with 
the matrix $A$ in (\ref{d2}) for $d=2$. In particular, $SL(2,Z)$
transformations
on $\t$ act on the right of $A$ as  $SL(2,Z)$ transformations on the  
lattice.
Hence, we can use the method of orbits to evaluate the integral.

\newpage
The orbits of $SL(2,Z)$ in the set of $2\ti d$ matrices with
integer entries are as follows: 
\bs
\be
\mbox{trivial orbit} : \;\;\;\; A^T = 0
\;\;\;\;\;\;\;\;\;\;\;\;\;\;\;\;\;\; 
\;\;\;\;\;\;\;\;\;\;\;\;\;\;\;\;\;\; 
\;\;\;\;\;\;\;\;\;\;\;\;\;\;\;\;\;\; 
\;\;\;\;\;\;\;\;\;\;\;\;\;\;\;\;\;\; 
\;\;\;\;\;\;\;\;\;\;\;\;\; 
\ee
\be
\mbox{degenerate orbit} : \;\;\;\; A^T =
 \left( \matrix{
0 & 0 & \ldots & 0  \cr
m_1 & m_2 &  \ldots &  m_d  \cr
} \right)
\sp (m_1,m_2,\ldots,m_d) \neq (0,0,\ldots,0)
\ee
\be
\mbox{non-degenerate orbit} : \;\;\;\; A^T =
 \left( \matrix{n_1 &  \ldots & n_k & 0 &  \ldots &  0  \cr
 m_1 &  \ldots & m_k & m_{k+1} &  \ldots  & m_d   \cr } \right)
\;\;\;\;\;\;\;\;\;\;\;\;\;\;\;\;\;\; 
\;\;\;\;\;\;\;\;\;\;\;\;\;\;
\ee
\be
\;\;\;\;\;\;\;\;\;\;\;\;\;\;\;\;\;
\;\;\;\;\;\;\;\;\;\;\;\;\;\;\;\;\;
 1 \leq k < d   \sp n_k > m_k \geq 0 \sp (m_{k+1},\ldots,m_d)  \neq
(0,\ldots,0)
 \ee
\es
The stabilizer group in each of these three cases is the same as for
the $d=2$ case, so again we split up the integral into
three separate parts, for which we give the results below.
Here, we will denote the
degenerate and non-degenerate orbits by $\sum_{m}'$ and $\sum_{m,n}'$.

\ni {\it Trivial orbit}.
The result is identical to the one given in (\ref{d5}).

\ni {\it Non-degenerate orbit}. Performing first the Gaussian $\t_1$
integration and subsequently using (\ref{d11}), (\ref{d12}) to evaluate
the $\t_2$ integration, we find
\be
I_\n^{(2)} =
2
\sum_{s=0}^\n \ch{\n}{s}
 \sum_{m,n}' { \sqrt{G} \over T_2}
\left(  {- 3 \over \p T_2 U_2 } \right)^s q_T
\sum_{l=-1}^{\infty}
   \sum_{r=0}^s { (s+r) ! \over r ! (s-r)! (4 \p)^r}
 (T_2 +U_2 l)^{s-r}
q_U^l c^{(\n-s)}_{l}
\pe \ee
Using the summation identity (\ref{sid2}) and the covariant derivative
identity (\ref{b13a}), it is not difficult to see that
this can be re-expressed in terms of the original function, as
\be
I_\n^{(2)} =2
\sum_{s=0}^\n \ch{\n}{s}
 \sum_{m,n}' { \sqrt{G} \over T_2}
 \left(  { 3 \over 2 \p T_2 } \right)^s q_T
(D^s \hat E_2^{\n-s}  \Phi_\n )(U)
\co \ee
where we remind the reader again that the induced moduli $T,U$ defined
in (\ref{inm2}) are $m,n$-dependent.

\ni {\it Degenerate orbit}. In this case we need to regulate the 
IR divergence. Since we do not need the
exact regulated result for this paper, we confine ourselves here to
giving the unregulated result  for the degenerate orbit
\be
I^{(3)}_\n
\simeq  \sum_{s=0}^\n c_0^{(\n-s)} \ch{\n}{s} s!  \sum_{m}'
\sqrt{G} \left( {- 3 \over \p} \right)^s
 \left( {U_2 \over \p T_2} \right)^{s+1}  {1 \over |U|^{2(1+s)}  }
\ee
where $T_2 |U|^2/U_2 = m G m $ since $n=0$.

\ed
\newpage
\begin{thebibliography}{6666}


\bibtem{Po} M. Green and J. Polchinski, Phys. Lett. {\bf B335} (1994)
377,
hep-th/9406012;\\
J. Polchinski, Phys. Rev. {\bf D50} (1994) 6041,
hep-th/9407031;\\
M.B. Green, Phys. Lett. {\bf B354} (1995) 271,  hep-th/9504108.

\bibtem{GG} M.B. Green and M. Gutperle, hep-th/9612127;
Nucl. Phys. {\bf B498} (1997) 195, hep-th/9701093.

\bibtem{B} J.L.F. Barbon, Phys. Lett. {\bf B404} (1997) 33,
hep-th/9701075;\\
J.L.F. Barbon and M.A. Vasquez-Mozo, Nucl. Phys. {\bf B497} (1997) 236,
hep-th/9701142.

\bibtem{GV} M.B. Green and P. Vanhove, hep-th/9704145;\\
M.B. Green, M. Gutperle and P. Vanhove,  hep-th/9706175.

\bibtem{kp} E. Kiritsis and B. Pioline, hep-th/9707018.

\bibtem{anto} I. Antoniadis, B. Pioline and T.R. Taylor,
hep-th/9707222.

\bibtem{rt} J. Russo and A.A. Tseytlin, hep-th/9707134.

\bibtem{BBS} K. Becker, M. Becker and A. Strominger, Nucl. Phys. {\bf
B456} (1995) 130,  hep-th/9507158.

\bibtem{OV} H. Ooguri and C. Vafa, Phys. Rev. Lett. {\bf 77} (1996)
3296, hep-th/9608079;\\
B. Greene, D. Morrison and C. Vafa, Nucl. Phys. {\bf B481} (1996) 513,
hep-th/9608039.

\bibtem{O} M. O'Loughlin, Phys. Lett. {\bf B385} (1996),
hep-th/9601179.

\bibtem{W} E. Witten, Nucl. Phys. {\bf B474} (1996) 343,
hep-th/9604030;\\
R. Donagi, A. Grassi and E. Witten, Mod. Phys. Lett. {\bf A11} (1996)
2199, hep-th/9607091;\\
P. Mayr, Nucl. Phys. {\bf B494} (1997) 489, hep-th/9610162.

\bibtem{KSS} S. Kachru, E. Silverstein and N. Seiberg, Nucl. Phys.
{\bf B480} (1996) 170,
hep-th/9605036;\\
S. Kachru and E. Silverstein, Nucl. Phys. {\bf B462} (1996) 92,
hep-th/9608194.

\bibtem{HM} J. Harvey and G. Moore, hep-th/9610237.

\bibtem{6} A. Gregori, E. Kiritsis, C. Kounnas, N. Obers, M.
Petropoulos and B. Pioline, hep-th/9708062.

\bibtem{fg}  M. Bershadsky, S. Cecotti, H. Ooguri and C. Vafa, Commun.
Math. Phys. {\bf 165} (1994) 311,  hep-th/9309140; \\
I. Antoniadis, E. Gava, K.S. Narain and T.R. Taylor, Nucl. Phys. {\bf
B413} (1994) 162,  hep-th/9307158.

\bibtem{BaKi} C. Bachas and E. Kiritsis, hep-th/9611205.

\bibtem{BV}   N. Berkovits and  C. Vafa, Nucl. Phys. {\bf B433} (1995)
123,  hep-th/9407190;\\
H. Ooguri and C. Vafa, Nucl. Phys. {\bf B451} (1995) 121,
hep-th/9505183.

\bibtem{bk2} C. Bachas, C. Fabre, E. Kiritsis, N. Obers and P.
Vanhove,
hep-th/9707126.

\bibitem{talk} C. Bachas, talks given at Strings 97, Amsterdam and at the
Euroconference on Advanced Quantum Field Theory, Lalonde les Maures.

\bibtem{lec} E. Kiritsis, hep-th/9708130.

\bibtem{Windey} O. Alvarez, T. P. Killingback, M. Mangano and P.
Windey,
 in the Proceedings of the Irvine Conf. on Non-Perturbative Methods in
 Physics, Irvine, Calif., Jan 5-9, 1987 Nucl. Phys. {\bf B }(Proc.
Suppl.)
 1A;\\
O. Alvarez, T. P. Killingback, M. Mangano and P. Windey,
Commun. Math. Phys, {\bf 111} (1987) 1.

\bibtem{ellwit} E. Witten, Commun. Math. Phys. {\bf 109} (1987) 525;\\
E. Witten, {\it The Dirac Operator in Loop Space}, in P. Landweber, {\it
Elliptic curves and modular forms in algebraic topology\/}, Lecture
Notes in Mathematics,
Springer Verlag, 1988.

\bibtem{5d} M. Douglas, J. Polchinski and A. Strominger,
hep-th/9703031.

\bibtem{ffg}   I. Antoniadis,  E. Gava,  K. S. Narain and  T. R.
Taylor,
 Nucl. Phys. {\bf B455} (1995) 109,  hep-th/9507115.

\bibtem{roo}   E. Bergshoeff and M. de Roo, Nucl. Phys. {\bf B328}
(1989)
439;\\
M. de Roo, H. Suelmann and A. Wiedemann, Phys. Lett. {\bf B280} (1992)
39;
Nucl. Phys. {\bf B405} (1993) 326,  hep-th/9210099.

\bibtem{ABFPT} I. Antoniadis, C. Bachas, C. Fabre, H. Partouche and
T. Taylor, Nucl. Phys {\bf B489} (1997) 160, hep-th/9608012. 

\bibtem{BF}  C. Bachas and C. Fabre, Nucl. Phys. {\bf B476} (1996)
418, hep-th/9605028.

\bibtem{FI} I. Antoniadis, H. Partouche and T.R. Taylor, Nucl. Phys.
{\bf B499} (1997) 29, hep-th/9703076;\\
M. Serone, Phys. Lett. {\bf B395} (1997) 42; Erratum-ibid {\bf B401}  
(1997) 363, hep-th/9611017;\\
M. Serone and J.F. Morales, hep-th/9703049.

\bibtem{Tseytlin} A. Tseytlin,
 Phys. Lett. {\bf B367} (1996) 84, hep-th/9510173;
Nucl. Phys. {\bf B467} (1996) 383, hep-th/9512081

\bibtem{GSW} M.B. Green, J.H. Schwarz and E. Witten, {\it Superstring
Theory}, Cambridge U. Press, 1987. 

\bibtem{pert} E. Verlinde and H. Verlinde, Phys. Lett. {\bf B192}
(1987) 95; \\
J. Atick, G. Moore and A. Sen, Nucl. Phys. {\bf B308} (1988) 1.

\bibtem{cato} J. Atick, G. Moore and A. Sen, Nucl. Phys. {\bf B307}
(1988) 221.

\bibtem{D} J. Atick and A. Sen, Nucl. Phys. {\bf B296} (1988) 157.

\bibtem{ya} O. Yasuda, Phys. Lett. {\bf B215} (1988) 306;
Phys. Lett. {\bf B218} (1989) 455.

\bibtem{chs} C. Callan, J. Harvey and A. Strominger,
Nucl. Phys. {\bf B359} (1991) 611.

\bibtem{ds} M. Dine and N. Seiberg,  hep-th/9705057.

\bibtem{dab} A. Dabholkar, Phys. Lett. {\bf B357} (1995) 307,
hep-th/9506160.

\bibtem{PW} J. Polchinski and E. Witten,  Nucl. Phys. {\bf B460}
(1996) 525,
 hep-th/9510169.

\bibtem{bpst} A. Belavin, A. Polyakov, A. Schwartz and Y. Tyupkin,
Phys. Lett. {\bf B59} (1975) 85.

\bibtem{poly} A. Polyakov, Nucl. Phys. {\bf B120} (1977) 429.

\bibtem{KT} J. Kosterlitz and D. Thouless, J. Phys. {\bf C6} (1973)
1181.

\bibtem{mer} V. DeAlfaro, S. Fubini and G. Furlan, Phys. Lett. {\bf
B65}
(1976) 163;\\
C. Callan, R. Dashen and D. Gross, Phys. Rev. {\bf D17} (1978) 2717;\\
A. Actor, Rev. Mod. Phys. {\bf 51} (1979) 461.

\bibtem{zerosize} E. Witten,  Nucl. Phys.  {\bf B460} (1996) 541,
hep-th/9511030.

\bibtem{bk} C. Bachas and E. Kiritsis, Phys. Lett. {\bf B325}
 (1994) 103, hep-th/9311185.

\bibtem{Schellekens}  W. Lerche, B. Nilsson and  A. Schellekens,
Nucl. Phys. {\bf B289} (1987) 609;\\
W. Lerche, B.E.W. Nilson, A.N. Schellekens and
N.P. Warner, Nucl. Phys. {\bf B299} (1988) 91;\\
W. Lerche, A. Schellekens and  N. Warner, Phys. Rep. {\bf 177} (1989)
1.

\bibtem{Lerche} W. Lerche, Nucl. Phys. {\bf B308} (1988) 102


\bibtem{AS} J. Atick and A. Sen, Phys. Lett. {\bf B186} (1987) 339.

\bibtem{slo} D. Gross and J. Sloan, Nucl. Phys. {\bf B291} (1987) 41.

\bibtem{KK} E. Kiritsis and C. Kounnas, Nucl. Phys. {\bf B442} (1995)
472,  hep-th/9501020; Nucl. Phys. [Proc. Suppl.] {\bf 41} (1995) 331,
 hep-th/9410212; hep-th/9507051.

\bibtem{chem} M. Chemtob, hep-th/9703206.

\bibtem{DKL}L.J. Dixon, V.S. Kaplunovsky and J. Louis, Nucl. Phys.
{\bf B355} (1991) 649.

\bibtem{Serre} J.P. Serre, {\it Cours d' Arithm\'etique}, PUF  Paris
1970.

\bibtem{KK1} E. Kiritsis, C. Kounnas, M. Petropoulos and J. Rizos,
hep-th/9605011;
Nucl. Phys. {\bf B483} (1997) 141,  hep-th/9608034.

\bibtem{agnt} I. Antoniadis, E. Gava, K.S. Narain and T.R. Taylor,
 Nucl. Phys. {\bf B407} (1993) 706; hep-th/9212045.

\bibtem{stie} K. Foerger and S. Stieberger,  hep-th/9709004.

\bibitem{sing} B. de Wit, V. Kaplunovsky, J. Louis and  D. L\"ust,
Nucl. Phys. {\bf B451} (1995) 53; hep-th/9504006;\\
 I. Antoniadis, S. Ferrara, E. Gava, K. S. Narain and T. R. Taylor,
 Nucl. Phys. {\bf B447} (1995) 35; hep-th/9504034.

\bibtem{tasi} J. Polchinski, S. Chaudhuri and  C. Johnson,
hep-th/9602052;\\
J. Polchinski, hep-th/9611050

\bibtem{Polch} J. Polchinski, Phys. Rev. Lett. {\bf 75} (1995) 4724,
hep-th/9510017.

\bibtem{F} C. Vafa,  Nucl. Phys. {\bf B469} (1996) 403,
hep-th/9602022.

\bibtem{rey} N. Kim and S. J. Rey, hep-th/9701139;\\
S. J. Rey,  hep-th/9704158.

\bibitem{matrix}  R. Dijkgraaf, E. Verlinde and H. Verlinde,
hep-th/9703030.

\bibtem{DMVV} R. Dijkgraaf, G. Moore, E. Verlinde and H. Verlinde,
hep-th/9608096.

\bibtem{ez} M. Eichler and D. Zagier, {\it The Theory of Jacobi Forms},  
Birkh\"auser, 1985.

\bibtem{KK5} E. Kiritsis and C. Kounnas,  Nucl. Phys. [Proc. Suppl.]
 {\bf 45BC} (1996) 207, hep-th/9509017; hep-th/9507051.

\bibtem{GW} D. Gepner and E. Witten,  Nucl. Phys. {\bf B278} (1986)
493.

\bibtem{hm2} J. Harvey and G. Moore, Nucl. Phys. {\bf B463} (1996)
315, hep-th/9510182.


\end{thebibliography}
